\documentclass[%
superscriptaddress,
 amsmath,amssymb,
 aps,
]{revtex4-2}

\usepackage[left]{lineno}

\usepackage{physics}
\usepackage{amsmath}
\usepackage{amsthm}
\usepackage{}
\usepackage{subfigure}
\usepackage{diagbox}
\makeatletter
\usepackage{graphicx}
\graphicspath{{figures/}}
\usepackage{booktabs}
\usepackage[colorlinks,linkcolor=blue,citecolor=blue]{hyperref}
\usepackage[capitalize]{cleveref}

\usepackage[most]{tcolorbox}

\tcbuselibrary{theorems}

\newtcbtheorem[number within=section]{mybox}{Box}{
    halign title=left, 
    halign=left, 
}{bx}

\usepackage{enumitem}

\newtheorem{proposition}{Proposition}
\newtheorem{assumption}{Assumption}
\newtheorem{lemma}{Lemma}
\newtheorem{definition}{Definition}

\begin{document}




\title{Secure quantum key distribution against correlated leakage source}


\author{Jia-Xuan Li}
\author{Yang-Guang Shan}
\affiliation{Laboratory of Quantum Information, University of Science and Technology of China, Hefei 230026, China}
\affiliation{CAS Center for Excellence in Quantum Information and Quantum Physics, University of Science and Technology of China, Hefei, Anhui 230026, China}
\affiliation{Anhui Province Key Laboratory of Quantum Network, University of Science and Technology of China, Hefei 230026, China}
\author{Rong Wang}
\affiliation{School of Cyberspace, Hangzhou Dianzi University, Hangzhou 310018, China}
\author{Feng-Yu Lu}
\author{Zhen-Qiang Yin}
\email[]{yinzq@ustc.edu.cn}
\author{Shuang Wang}
\author{Wei Chen}
\author{De-Yong He}
\author{Guang-Can Guo}
\author{Zheng-Fu Han}
\affiliation{Laboratory of Quantum Information, University of Science and Technology of China, Hefei 230026, China}
\affiliation{CAS Center for Excellence in Quantum Information and Quantum Physics, University of Science and Technology of China, Hefei, Anhui 230026, China}
\affiliation{Anhui Province Key Laboratory of Quantum Network, University of Science and Technology of China, Hefei 230026, China}
\affiliation{Hefei National Laboratory, University of Science and Technology of China, Hefei 230088, China}


\date{\today}

\begin{abstract}

 
Quantum key distribution (QKD) provides information theoretic security based on quantum mechanics,
however, its practical deployment  is challenged by imperfections of source devices. 
Among various source loopholes, correlations between transmitted pulses pose a significant yet underexplored security risk, 
potentially compromising QKD's theoretical guarantees. 
In this work, we propose a security analysis framework for QKD under correlations, 
enabling finite-key analysis for the first time by extending and rearranging QKD rounds and leveraging the generalized chain rule. 
Based on this framework, 
and inspired by the idea of side-channel-secure QKD,
we develop a secure QKD against 
correlated leakage source
only need the characterization of correlation range and the lower bound on the vacuum component of the prepared states.
Additionally, our framework can be extended to other QKD protocols, 
offering a general approach to consider correlation induced security vulnerabilities.
The simulation results demonstrate the effectiveness of our protocol and its significantly superior tolerance to imperfect parameters compared to existing protocols.
This work provides a crucial step toward closing security loopholes in QKD, enhancing its practicality, and ensuring long-distance, 
high-performance secure communication under real-world constraints.

\end{abstract}


\maketitle

\noindent{\bf INTRODUCTION} 

\hfill

With the rapid development and widespread adoption of the Internet, 
information security has become one of the most critical issues in modern society. 
The modern cryptography, as the most widely used approach, 
ensures data privacy and security, but its security relies on computational complexity~\cite{6773024},
which is threatened by advancing computing power, especially quantum computing~\cite{365700,cryptoeprint:2017/190}.
As a result, there is an urgent need to explore cryptographic methods that are not based on computational hardness but instead rely on the fundamental principles of physics. 
Quantum key distribution (QKD)~\cite{BB84} has gained significant attention as a promising solution in this context.

QKD use the principles of quantum mechanics to achieve information theoretic security in key distribution~\cite{doi:10.1126/science.283.5410.2050, PhysRevLett.85.441, RevModPhys.81.1301, doi:10.1142/S0219749908003256},
which means that theoretically, even an eavesdropper with unlimited computational power cannot break its security.
However, the theoretical security of QKD still faces challenges in practical deployments, 
especially when considering imperfections in real world devices. 
Fortunately, all vulnerabilities related to measurement devices can be addressed 
using measurement-device-independent QKD (MDI QKD)~\cite{PhysRevLett.108.130502,PhysRevLett.108.130503,Curty2014}, 
as well as variants such as twin-field QKD (TF QKD)~\cite{Lucamarini2018,PhysRevX.8.031043,PhysRevA.98.062323,PhysRevApplied.11.034053} 
and mode-pairing QKD (MP QKD)~\cite{Zeng2022}. 
However, the imperfections of the source still introduce potential security loopholes in practical QKD implementations, 
which require further resolution.

The security loopholes caused by imperfect sources primarily stem from three aspects~\cite{aaz4487}: 
state preparation flaws (SPFs) due to the limited modulation accuracy of the device, 
information leakage caused by side channels, 
and information leakage resulting from classical correlations between pulses.
Here, we refer to a source exhibiting these three types of imperfections as correlated leakage source.
Fortunately, several effective solutions for the SPF problem already exist~\cite{PhysRevA.90.052314,PhysRevA.90.052319}. 
Moreover, based on the resolution of SPF, 
multiple approaches have also been proposed to address the side-channel issue~\cite{1365172,10.5555/2011832.2011834,Koashi_2009,Coles2016,Wang2019,Winick2018reliablenumerical,Pereira2019,sixto2025quantumkeydistributionimperfectly}.
In particular, a recently proposed protocol, 
known as side-channel-secure QKD (SCS QKD)~\cite{PhysRevApplied.12.054034,PhysRevApplied.19.064003,PhysRevResearch.6.013266,shan2024improvedpostselectionsecurityanalysis}, 
is based on the sending-or-not-sending QKD (SNS QKD)~\cite{PhysRevA.98.062323} scheme. 
By imposing a lower bound constraint on the vacuum component of the transmitted states, 
SCS QKD is immune to all information leakage caused by any unknown side channels.
Compared to device-independent QKD (DI QKD)~\cite{PhysRevLett.98.230501}, SCS QKD is considered the only protocol secured under side-channel 
that can be implemented with commercial devices while achieving long-distance transmission~\cite{PhysRevLett.128.190503}.

However, compared to the well-studied side-channel problem, the impact of correlations on security remains less thoroughly investigated,
leaving the QKD systems not completely secure under 
correlated leakage source.
Currently, the primary approaches for handling correlations include the post-selection methods~\cite{PhysRevA.93.042325,Mizutani2019,Yoshino2018} 
and enhancing protocols~\cite{aaz4487,zapatero2021security,PhysRevApplied.18.044069,PhysRevResearch.5.023065,Currs-Lorenzo_2024,li2025quantumkeydistributionovercoming,currslorenzo2024securityframeworkquantumkey},
using mathematical techniques such as 
reference techniques~\cite{aaz4487}, quantum coin~\cite{currslorenzo2024securityframeworkquantumkey} and so on.
Nevertheless, these methods have notable limitations, such as the inability of some protocols to handle high-order correlations, 
the need for complex modeling of the magnitude of correlation, and poor tolerance to existing device parameters.
These factors make handling correlations in practice a challenging task, 
and only a few experiments have addressed the correlation problem 
for certain parameters~\cite{li2025quantumkeydistributionovercoming}.
Moreover, 
none of these existing protocols provide a complete finite-key analysis, limiting their applicability in real world QKD systems.

To address these challenges, this paper makes two main contributions. 
First, we propose a security analysis framework for QKD under correlation. 
By extending and rearranging QKD rounds, 
using the generalized chain-rule result~\cite{6408179},
we establish security constraints that allow for finite-key analysis in the presence of correlations. 
Second, we introduce a secure QKD against 
correlated leakage source
based on the two-state SNS QKD~\cite{PhysRevApplied.12.054034,PhysRevApplied.19.064003,PhysRevResearch.6.013266,shan2024improvedpostselectionsecurityanalysis}. 
This protocol only assumes a bounded correlation range and a lower bound on the vacuum component of the prepared states, 
enabling secure key generation even in the presence of both SPF, side-channels and correlations.
Compared to existing protocols, 
our security analysis framework is the first to enable finite-length analysis under correlation conditions,
and this framework is not only applicable to our protocol but can also be easily extended to any protocol involving correlations.
Moreover, our protocol does not require any characterization of the magnitude or the specific form of side channels and correlations. 
It can tolerate practical device imperfections and demonstrates strong robustness against high-order correlations. 
Simulation results show that our protocol can efficiently generate keys under realistic device parameters.
When the correlation range is $5$, it only loses $10{\rm dB}$ of the maximum attenuation, 
and it can tolerate a correlation range as high as $1000$, 
far exceeding the maximum correlation range of $6$ observed in existing experiments~\cite{trefilov2024intensitycorrelationsdecoystatebb84}.
This significantly enhances the security and practical feasibility of QKD systems under realistic physical constraints,
marks a crucial step toward achieving loophole-free and high-performance QKD.

\hfill

\noindent{\bf RESULTS} 

\hfill

\noindent{\bf Security analysis framework addressing correlated sources.}
Due to correlation, one of the most critical assumptions in QKD security proofs, independent distributed state preparation, is violated. 
This significantly complicates the security analysis, 
especially under finite-key conditions. 
In this work, we
employ the generalized chain-rule result~\cite{6408179} to prove the security equivalence 
between correlated sources and uncorrelated sources, 
effectively establishing a reduction from non-independent scenarios to the independent case. 
Furthermore, through additional discussion, this equivalence can be extended 
to
prove that any protocol with 
correlated leakage source, including correlations, side channels and SPFs, 
can find a security bound to an i.i.d protocol. 

First we give our basic assumption.
We assume that the range of correlation is limited~\cite{Yoshino2018,aaz4487,zapatero2021security,PhysRevApplied.18.044069,PhysRevResearch.5.023065,Currs-Lorenzo_2024,li2025quantumkeydistributionovercoming}. 
The correlation range refers to the maximum number of previous rounds whose settings can influence the state preparation in a given round.
This assumption has been validated as reasonable by some experiments~\cite{Yoshino2018,Roberts:18,Lu2021,10050030,9907821,marcomini2024characterisinghigherorderphasecorrelations,li2025quantumkeydistributionovercoming,trefilov2024intensitycorrelationsdecoystatebb84}.
Thus, we introduce the first Assumption.
\begin{assumption}
    \label{assu:range}
    The correlation is constrained within a maximum range $\xi$.
\end{assumption}

\begin{figure*}[htbp]
    \centering
    \subfigure[
        Schematic diagram of the original protocol. 
        Each square represents a single round, with yellow, green, ..., and blue indicating rounds 
        $\mod (\xi + 1) = 1, 2, \ldots, (\xi + 1)({\rm or\,}0)$; the rightward arrow represents the process of filtering raw key bits. 
        The protocol proceeds from left to right for a total of 
        $N$ rounds, with each category being executed 
        $N_i$ rounds.
        The definitions of $\mathbf{D}$ and $\mathbf{Z}$ are detailed in 
        section `Security analysis framework addressing correlated sources'.
    ]{
        \includegraphics[width=0.8\textwidth]{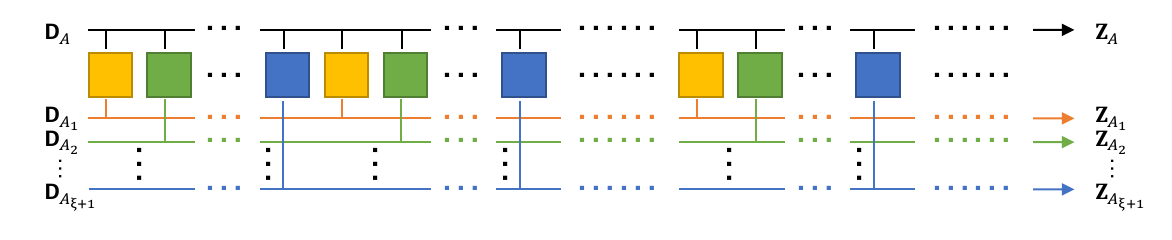}
        \label{fig:oriANDnewPROT_ori}
    }
    \\
    \subfigure[
        Schematic diagram of the new protocol. 
        Each square represents a single round, where colored squares indicate ``key generation rounds" 
        and uncolored squares represent rounds where data is disclosed. 
        Each pair of adjacent colored rounds is separated by $\xi$ data disclosure rounds. 
        Yellow, green, ..., and blue correspond one-to-one with the rounds of the same colors in Sub. ~\ref{fig:oriANDnewPROT_ori}. 
        The protocol proceeds from left to right, moving to the next row after completing one. 
        A total of $N'=N$ key generation rounds are executed, which corresponds to $(\xi + 1)N'$ rounds in total, with each category being executed $N_i'=N_i$ rounds.
        The definitions of $\mathbf{D}'$ and $\mathbf{Z}'$ are detailed in 
        section `Security analysis framework addressing correlated sources'.
    ]{
        \includegraphics[width=0.8\textwidth]{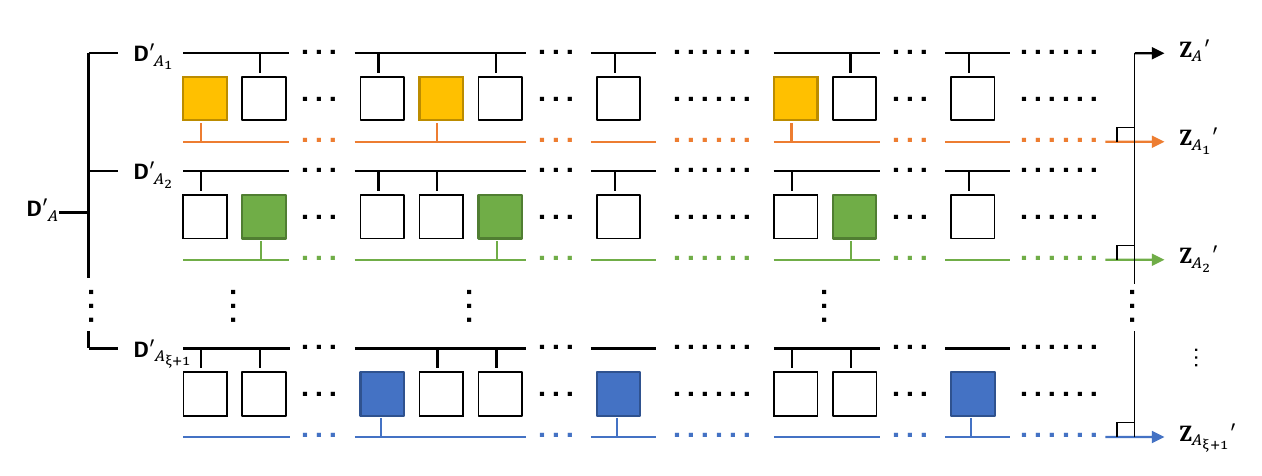}
        \label{fig:oriANDnewPROT_new}
    }
    \caption{
        Schematic diagrams of the original and new protocols.
    }
    \label{fig:oriANDnewPROT}
\end{figure*}

The security of a QKD protocol 
can be characterized by the conditional smooth min-entropy
$H^{\epsilon}_{\rm min}(\mathbf{Z}_A|\mathbf{E}')_{\rho}$, where $\mathbf{Z}_A$ refers to the raw key of Alice,
$\mathbf{E}'$ refers to the uncertainty system of the eavesdropper Eve
and $\rho$ represents the quantum state shared in the protocol, which includes Eve's optimal attack system.
Since the correlation range $\xi$ ensures that two keys separated by $\xi$ rounds do not influence each other, we divide 
$\mathbf{Z}_A$ into several subsets
$
        \mathbf{Z}_A
        =
        \mathbf{Z}_{A_1}
        \mathbf{Z}_{A_2}
        \ldots
        \mathbf{Z}_{A_{\xi+1}}
$,
where $\mathbf{Z}_{A_i}$ represents the key sequence generated in the 
$k$-th rounds of the protocol,
where $k \in \{i+n(\xi+1)| n \in \mathbb{N}_0, i+n(\xi+1)\leq N \}$.
By the generalized chain-rule result~\cite{6408179}
and data processing inequality, 
can use 
$
            H^{\epsilon_i}_{\rm min}\left(\mathbf{Z}_{A_i}|
            \left(
                \complement_{\mathbf{Z}_{A}} \mathbf{Z}_{A_i}
            \right)
            \mathbf{E}'
            \right)_{\rho}
$ to estimate a lower bound for 
$H^{\epsilon}_{\rm min}(\mathbf{Z}_A|\mathbf{E}')_{\rho}$, 
where
$\left(
    \complement_{\mathbf{Z}_{A}} \mathbf{Z}_{A_i}
\right)$
represents the part of set 
$\mathbf{Z}_{A}$ excluding 
$\mathbf{Z}_{A_i}$, denoted as 
$\mathbf{Z}_{A_1}\mathbf{Z}_{A_2}\ldots\mathbf{Z}_{A_{i-1}}\mathbf{Z}_{A_{i+1}}\ldots\mathbf{Z}_{A_{\xi+1}}$.

Furthermore, we can relax the condition $\complement_{\mathbf{Z}_{A}} \mathbf{Z}_{A_i}$ in 
$
            H^{\epsilon_i}_{\rm min}\left(\mathbf{Z}_{A_i}|
            \left(
                \complement_{\mathbf{Z}_{A}} \mathbf{Z}_{A_i}
            \right)
            \mathbf{E}'
            \right)_{\rho}
$ 
even more. Define all of Alice's local systems (including the raw key, (potential) basis selection, intensity selection, as well as ancillas control random drifts and fluctuations, etc.) 
as $\mathbf{D}_A$, and 
similar to
$\mathbf{Z}_A$, we divide it into
$
        \mathbf{D}_A
        =
        \mathbf{D}_{A_1}
        \mathbf{D}_{A_2}
        \ldots
        \mathbf{D}_{A_{\xi+1}}
$,
where $\mathbf{D}_{A_i}$ represents the data in the 
$k$-th rounds of the protocol, 
where $k \in \{i+n(\xi+1)| n \in \mathbb{N}_0, i+n(\xi+1)\leq N \}$.
From the definitions, there is 
$\mathbf{Z}_{A_i} \in \mathbf{D}_{A_i}$,
and further we can calculate that
$
        \left(
            \complement_{\mathbf{Z}_{A}} \mathbf{Z}_{A_i}
        \right)
        \in
        \left(
            \complement_{\mathbf{D}_{A}} \mathbf{D}_{A_i}
        \right)
$,
$\left(
    \complement_{\mathbf{D}_{A}} \mathbf{D}_{A_i}
\right)$
represents the part of set 
$\mathbf{D}_{A}$ excluding 
$\mathbf{D}_{A_i}$, denoted as 
$\mathbf{D}_{A_1}\mathbf{D}_{A_2}\ldots\mathbf{D}_{A_{i-1}}\mathbf{D}_{A_{i+1}}\ldots\mathbf{D}_{A_{\xi+1}}$.
Using data processing inequality, we can calculate that
$
H^{\epsilon_i}_{\rm min}\left(\mathbf{Z}_{A_i}|
\left(
    \complement_{\mathbf{Z}_{A}} \mathbf{Z}_{A_i}
\right)
\mathbf{E}'
\right)_{\rho'_i}
\geq
H^{\epsilon_i}_{\rm min}\left(\mathbf{Z}_{A_i}|
    \left(
        \complement_{\mathbf{D}_{A}} \mathbf{D}_{A_i}
    \right)
    \mathbf{E}'
    \right)_{\rho'_i}
$,
where $\rho'_i$ denotes the quantum state in the protocol that include the space
$\mathbf{Z}_{A_i}$, $
\left(
    \complement_{\mathbf{D}_{A}} \mathbf{D}_{A_i}
\right)
$.
The schematic diagram of the process is shown in
\cref{fig:oriANDnewPROT_ori},
and this we can give our first lemma.
\begin{lemma}
    \label{lemma:correlated-source-devide}
    The lower bound of the smooth min-entropy $H^{\epsilon}_{\rm min}(\mathbf{Z}_A|\mathbf{E}')_{\rho}$ of an original protocol with a maximum correlation range $\xi$ can be
    bound by 
    the sum of the smooth min-entropies $H^{\epsilon_i}_{\rm min}\left(\mathbf{Z}_{A_i}|
    \left(
        \complement_{\mathbf{D}_{A}} \mathbf{D}_{A_i}
    \right)
    \mathbf{E}'
    \right)_{\rho'_i}$,
    where each $H^{\epsilon_i}_{\rm min}\left(\mathbf{Z}_{A_i}|
    \left(
        \complement_{\mathbf{D}_{A}} \mathbf{D}_{A_i}
    \right)
    \mathbf{E}'
    \right)_{\rho'_i}$ corresponds to the rounds in the original protocol that share the same modulo-$(\xi+1)$ remainder, under the condition that all other rounds are made public.
    This relation satisfies
    \begin{equation}
        \label{equ:lemma-1-needddd}
        \begin{aligned}
            &
            H^{\epsilon}_{\rm min}(\mathbf{Z}_A|\mathbf{E}')_{\rho}
            \\
            \geq
            &
            \sum_{i=1}^{\xi+1} \left[
                H^{\epsilon_i}_{\rm min}\left(\mathbf{Z}_{A_i}|
                \left(
                    \complement_{\mathbf{D}_{A}} \mathbf{D}_{A_i}
                \right)
                \mathbf{E}'
                \right)_{\rho'_i}
            \right]
            -
            \sum_{i=1}^{\xi} (
                f_i
            )
            ,
        \end{aligned}
    \end{equation}
    where
$
        f_1 
        = 
        2
        \log_2 \frac{1}{\epsilon-2\epsilon_1-\epsilon_1'}
$
        ,
$
        f_i
        = 
        2
        \log_2 \frac{1}{\epsilon_{i-1}'-2\epsilon_i-\epsilon_i'}
$
if $i \in [2,\xi]$,
$
        \epsilon_{\xi+1} 
        = \epsilon_{\xi}'
$.

    \begin{proof}
        See section `Method' for detail.
    \end{proof}
\end{lemma}


\cref{lemma:correlated-source-devide}
shows that the smooth min-entropy of a protocol with correlation can be bounded by the sum of the smooth min-entropies of its uncorrelated subcomponents. 
However, \cref{lemma:correlated-source-devide} cannot be directly used to compute a lower bound on the smooth min-entropy of the original protocol, 
because the subcomponents are required to be conditioned on the disclosure of the other parts. To address this, it is necessary 
to construct an uncorrelated protocol such that a certain part of it possesses the same smooth min-entropy as
$
H^{\epsilon_i}_{\rm min}\left(\mathbf{Z}_{A_i}|
    \left(
        \complement_{\mathbf{D}_{A}} \mathbf{D}_{A_i}
    \right)
    \mathbf{E}'
    \right)_{\rho'_i}
$, and for this we repeat the original protocol $\xi+1$ times to form a new protocol.
Therefore, the new protocol actually executes 
$(1+\xi)N$ rounds, which we refer to as \textit{physical rounds}. 
We stipulate that the new protocol uses only $k$-th \textit{physical rounds} for the QKD process, which we call \textit{key generation rounds},
where
$k \in \left\{
    (i-1) N + i + n (\xi+1)
    |
    i \in \mathbb{N},
    n \in \mathbb{N}_0,
    i + n (\xi+1)\leq N
\right\}$,
to keep the number of \textit{key generation rounds} remain $N$, the same as the the original protocol.
While the left $\xi N$ \textit{physical rounds} not only send the prepared quantum states through the channel but also publicly disclose all their data
(all local ancillas or their measurement results, due to the requirements of the specific protocol),
which we call \textit{leakage rounds}.
Similar to
$\mathbf{Z}_A$,
the raw key bits of Alice in the new protocol, denoted as 
$\mathbf{Z}_A'$, can categorize into 
$\xi$ types, represented as 
$
        \mathbf{Z}_A'
        =
        \mathbf{Z}_{A_1}'
        \mathbf{Z}_{A_2}'
        \ldots
        \mathbf{Z}_{A_{\xi+1}}'
$,
with 
$\mathbf{Z}_{A_i}'$ represents the key sequence generated in the 
$k$-th \textit{key generation round} of the protocol,
where
$k \in \left\{
    (i-1) N + i + n (\xi+1)
    |
    n \in \mathbb{N}_0,
    i + n (\xi+1)\leq N
\right\}$.
From the above discussion, we can also obtain that $\mathbf{Z}_{A_i}$ and $\mathbf{Z}_{A_i}'$ corresponding one-to-one.
This correspondence is intuitively illustrated in Fig.~\ref{fig:oriANDnewPROT}, 
where the rounds of the original protocol are marked with the same color as the \textit{key generation rounds} in the new protocol, 
while the \textit{leakage rounds} are marked in white.
For clarity, in summary, we give the definition of the \textit{new protocol}.
\begin{definition}
    \label{def:new-protocol}
    For any original protocol with a maximum correlation range $\xi$, 
    we define a corresponding new protocol by repeating the original protocol $ \xi + 1 $ times. 
    In the $i$-th repetition, only the rounds whose indices satisfy modulo $ \xi + 1 $ congruent to $ i $ (with the remainder $ \xi + 1 $ interpreted as $0$) 
    are used for key generation, while all other rounds are disclosed. 
    The raw key of original protocol is denoted by $\mathbf{Z}_A$, the raw key of the $i$-th repetition of the new protocol is denoted by $\mathbf{Z}_{A_i}'$ and the the raw key of the whole new protocol is denoted by $\mathbf{Z}_{A}'$.
    Similarly, define $\mathbf{D}_{A}' =\mathbf{D}_{A_1}'\mathbf{D}_{A_2}'\ldots \mathbf{D}_{A_{\xi+1}}'$ as the data from all disclosed rounds, 
    where $\mathbf{D}_{A_i}'$ denotes the data revealed in the public rounds during the $i$-th repetition of the original protocol.
\end{definition}

In the new protocol, there exists an attack by Eve such that the joint density matrix of the raw key of the new protocol, the system of the \textit{leakage rounds} and Eve's system, 
satisfies $\bigotimes_{i=1}^{\xi+1} \rho_i''$,
where
$\rho_i''$ is the density matrix of the rounds that generalize $\mathbf{Z}_{A_i}'$
and satisfies $\rho_i''=\rho_i'$.
Thus we have
$
H^{\tilde{\epsilon}_i}_{\rm min}\left(\mathbf{Z}_{A_i}|
        \left(
            \complement_{\mathbf{D}_{A}} \mathbf{D}_{A_i}
        \right)
        \mathbf{E}'
        \right)_{\rho'_i}
        =
        H^{\tilde{\epsilon}_i}_{\rm min}(\mathbf{Z}_{A_i}'|\mathbf{D}_{A_i}'\mathbf{E}')_{\rho_i''}
$.
Recall that the new protocol to publicly disclose all ancillas (or their measurement results) in \textit{leakage rounds} and 
Eve's optimal attack will also yield a smaller smooth min-entropy.
Thus, we can calculate that
$
H^{\tilde{\epsilon}_i}_{\rm min}(\mathbf{Z}_{A_i}'|\mathbf{D}_{A_i}'\mathbf{E}')_{\rho_i''}
        \geq
        H^{\tilde{\epsilon}_i}_{\rm min}(\mathbf{Z}_{A_i}'|\mathbf{E}')_{\rho'}
$, where $\rho'$ denote the total quantum state of the raw key of the new protocol and Eve's system.
Thus, we give our second Lemma.
\begin{lemma}
    \label{lemma:equ-between-have-or-not-corr}
    The lower bound of the smooth min-entropy $H^{\epsilon}_{\rm min}(\mathbf{Z}_A|\mathbf{E}')_{\rho}$ of an original protocol with a maximum correlation range $\xi$ can be
    bound by the sum of the smooth min-entropy $ H^{\tilde{\epsilon}_i}_{\rm min}(\mathbf{Z}_{A_i}'|\mathbf{E}')_{\rho'}$ of parts of the new protocol,
    where the definition of new protocol is in \cref{def:new-protocol}.
    This relation satisfies
    \begin{equation}
        \label{equ:lemma1}
        \begin{aligned}
            &
            H^{\epsilon}_{\rm min}(\mathbf{Z}_A|\mathbf{E}')_{\rho}
            \geq
            &
            \sum_{i=1}^{\xi+1} \left[
                H^{{\epsilon}_i}_{\rm min}(\mathbf{Z}_{A_i}'|\mathbf{E}')_{\rho'}
            \right]
            -
            \sum_{i=1}^{\xi} (
                f_i
            )
            ,
        \end{aligned}
    \end{equation}
where
$
        f_1 
        = 
        2
        \log_2 \frac{1}{\epsilon-2\epsilon_1-\epsilon_1'}
$
        ,
$
        f_i
        = 
        2
        \log_2 \frac{1}{\epsilon_{i-1}'-2\epsilon_i-\epsilon_i'}
$
if $i \in [2,\xi]$,
$
        \epsilon_{\xi+1} 
        = \epsilon_{\xi}'
$.

    \begin{proof}
        See section `Method' for detail.
    \end{proof}
\end{lemma}

\cref{lemma:equ-between-have-or-not-corr} has already established a security constraint that connects a correlated protocol to an uncorrelated one. 
However, \cref{lemma:equ-between-have-or-not-corr} requires estimating the smooth min-entropy of several subcomponents of the new protocol individually, 
which undoubtedly increases both the complexity and the impact of statistical fluctuations in the data. 
Therefore, a more effective approach is to further aggregate these parts and estimate them using the overall phase error rate.
For most protocols, the smooth min-entropy can be
estimated only for specific events, such as single-
photon events in typical BB84 and MDI protocols, while
multi-photon events cannot be directly estimated.
Thus, it is necessary to apply the chain-rule result~\cite{6408179} once again to bound $H^{\tilde{\epsilon}_i}_{\rm min}(\mathbf{Z}_{A_i}'|\mathbf{E}')_{\rho'}$,
using $ H^{\tilde{\epsilon}_i'}_{\rm min}(\mathbf{Z}_{\mathcal{Z},A_i}'|\mathbf{E}')_{\rho'}$,
where 
$ \mathbf{Z}_{A_i}' = \mathbf{Z}_{\mathcal{Z},A_i}' \mathbf{Z}_{\mathcal{X},A_i}'$,
$ \mathbf{Z}_{\mathcal{Z},A_i}'$ denote the parts that can estimated,
$ \mathbf{Z}_{\mathcal{X},A_i}'$ denote the parts that can not.
Further, from the uncertainty relation and the parameter estimation of the QKD process~\cite{Tomamichel2012}, we can bound the 
$ H^{\tilde{\epsilon}_i'}_{\rm min}(\mathbf{Z}_{\mathcal{Z},A_i}'|\mathbf{E}')_{\rho'}$ using 
$h\left(\overline{e_i}^{\rm U}_{{\epsilon}_i}\right)$, where $\overline{e_i}^{\rm U}_{{\epsilon}_i}$ the upper bound of the phase error rate $e_i$
and $h(x) = -x {\rm log}_2 (x)-(1-x) {\rm log}_2 (1-x)$.
Since the smooth min-entropy is estimated using the phase error rate, its physical meaning lies in estimating based on a given phase error rate, 
with the failure probability bounded by the square of the smoothing parameter~\cite{6408179}. 
Given the fact that if each $\mathbf{Z}_{\mathcal{Z},A_i}'$ estimation fails, then the estimation of $\mathbf{Z}_{\mathcal{Z},A}'$ as a whole must also fail (and likewise for success), 
and combining this with the convexity of $h(x)$, we can use the total phase error rate and the overall failure probability to estimate the sum of the smooth min-entropies of each part. 
Consequently, together with \cref{lemma:equ-between-have-or-not-corr}, we can estimate the smooth min-entropy of the original protocol using the phase error rate of the newly constructed protocol, which gives us another Lemma.
\begin{proposition}
    \label{Proposition:Proposition}
    The lower bound of the smooth min-entropy $H^{\epsilon}_{\rm min}(\mathbf{Z}_A|\mathbf{E}')_{\rho}$ of an original protocol with a maximum correlation range $\xi$ can be
    bound by the upper bound of the estimation of phase error rate $\overline{e}^{\rm U}_{\widehat{\epsilon}}$ of the new protocol,
    where the definition of new protocol is in \cref{def:new-protocol}.
    This relation satisfies
    \begin{equation}
        \label{equ:lemma2}
        \begin{aligned}
            H^{\epsilon}_{\rm min}(\mathbf{Z}_A|\mathbf{E}')_{\rho}
            \geq
            n
            \left(
                1 - h\left(
                    \overline{e}^{\rm U}_{\widehat{\epsilon}}
                \right)
            \right)
            -
            \xi f
            -
            \left(
                \xi+1
            \right)
            f'
            ,
        \end{aligned}
    \end{equation}
    where $\epsilon$, $\widehat{\epsilon}$ and $f$ satisfy
    $
    {\widehat{\epsilon}} =
            \left(
                \frac{
                    \epsilon - \xi \frac{1}{2^{f/2}}
                }{2 \xi +1}
                -
                \frac{1}{2^{f'/2}}
            \right)^{{\xi+1}}
    $.
    
    \begin{proof}
        See section `Method' for detail.
    \end{proof}
\end{proposition}


It is worth noting that if all components of $\mathbf{Z}_A$ can be used to estimate the phase error, 
then further scaling of 
$ \mathbf{Z}_{A_i}'$
to
$\mathbf{Z}_{\mathcal{Z},A_i}'$
is unnecessary. 
In this special case, we simply need to delete all terms containing $ f' $ from the conclusions in \cref{equ:lemma2}, which results in a more compact estimate of the smooth min-entropy.

Moreover, it is worth noting that although our security proof requires a rearrangement of rounds as illustrated in \cref{fig:oriANDnewPROT}, 
this rearrangement is not necessary in the actual data processing of the final protocol. 
Since the rounds are independent after the transformation, 
we are free to rearrangement them again to restore their original sequence. 

\hfill

\noindent{\bf Secure quantum key distribution against correlated leakage source.}
After completing the security analysis framework, we will construct a secure QKD against 
correlated leakage source based on the two-state SNS protocol~\cite{PhysRevApplied.12.054034,PhysRevApplied.19.064003,PhysRevResearch.6.013266,shan2024improvedpostselectionsecurityanalysis}.
To analyze the security, 
we impose 
a lower bound constraint on the vacuum component of the state sent by the source into the channel~\cite{PhysRevApplied.12.054034,PhysRevApplied.19.064003,PhysRevResearch.6.013266,shan2024improvedpostselectionsecurityanalysis}.
The same constraint can be achieved through pre-characterization, limit the upper bounds of pulse intensities or other methods, 
and it has already been experimentally implemented~\cite{PhysRevLett.128.190503}.
The specific assumption is as follows.
\begin{assumption}
    \label{assu:vacuum}
    For the two-state SNS-QKD protocol, the lower bound of the proportion of vacuum states in each round, under both the send and not send scenarios, is known. 
    Specifically, given the $i$-th round and its preceding $\xi$ rounds, the state sent into the channel during the current round 
    $\rho_{
        \mathbf{r}_{i-\xi}^i
        ,
        \mathbf{a}_{i-\xi}^i
    }^{\rm A(B)}$
    satisfies
    \begin{equation}
        \label{equ:ass2}
        \begin{aligned}
            \mathop{\rm min}\limits_{\mathbf{r}_{i-\xi}^{i-1}}
            \left(
                \mathop{\rm min}\limits_{\mathbf{a}_{i-\xi}^{i}}
                \left(
                \left|
                    \bra{0}
                    \rho_{
                        \mathbf{r}_{i-\xi}^i
                        ,
                        \mathbf{a}_{i-\xi}^i
                    }^{\rm A(B)}
                    \ket{0}
                \right|
                \right)
            \right)
            \geq
            V_{r_i}^{\rm A(B)}
            ,
        \end{aligned}
    \end{equation}
    where
    $r_i \in \{0,1\}$ denotes the encoding setting in $i$-th round,
    $a_i$ denotes the set of ancillas in the system that are potentially related to rounds and can influence the transmitted state, which includes controls over SPF, correlation, side-channel and so on,
    $ V_{r_i}^{\rm A(B)}$ denotes the lower bound of the proportion of vacuum states
and 
the sequence from $i$-th to the $j$-th round for  
$a$ and $r$ are defined as 
$
{a}_{i}^j 
:= a_{j}a_{j-1}\ldots a_{i}
$
and
$
{r}_{i}^j
:= r_{j}r_{j-1}\ldots r_{i}
$
respectively.
\end{assumption}
Following the above approach, we construct an \textit{equivalent protocol}.  
Furthermore, by applying a unitary mapping to the transmitted states, we establish the security equivalence between the \textit{equivalent protocol} and an independent and identically distributed (i.i.d.) protocol.  
Based on this, we can estimate the secret key rate.

\begin{figure}[htbp]
    \includegraphics[width=0.4\textwidth]{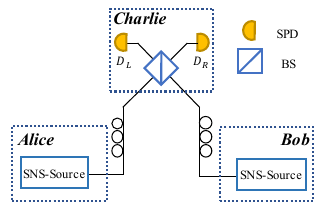}
    \caption{
        The schematic diagram of the protocol.
        $ D_L $ and $ D_R $ represent the left and right detectors respectively.  
        SPD, single photon detector;
        BS, 50:50 beam splitter;
        SNS-source, a system comprising a weak coherent source and modulation devices, satisfying Assumptions~\ref{assu:range}, \ref{assu:vacuum}.  
    }
    \label{fig:Snsqkd-diag}
\end{figure}

Before introducing the protocol, we first outline some fundamental requirements and definitions for its implementation.  
The protocol involves two users, Alice and Bob, as well as an untrusted node, Charlie. 
In each round, Alice and Bob attempt to prepare either a coherent state with a known intensity or a vacuum state, 
following the \textit{sending-or-not-sending} strategy \cite{PhysRevA.98.062323,PhysRevApplied.12.054034,PhysRevApplied.19.064003,PhysRevResearch.6.013266,shan2024improvedpostselectionsecurityanalysis}. 
However, due to the limitations of their sources, they can only generate quantum states that satisfy 
\cref{assu:range} and \ref{assu:vacuum}.
For the untrusted node Charlie, if honest, she performs an interference measurement shown in Fig.~\ref{fig:Snsqkd-diag}. 
Additionally, she must compensate for channel fluctuations so that constructive interference occurs at the left detector and destructive interference at the right detector.  
%
After all transmissions and measurements are completed, 
Alice and Bob negotiate to select a subset of rounds with probability $p_{\text{PE}}$, marked as parameter estimation rounds, for parameter estimation, while the remaining rounds are marked as key extraction rounds.
Furthermore, we need to define certain events,
a $\mathcal{Z}$ event occurs when exactly one of Alice or Bob chooses $r_i = 0$,
an $\mathcal{O}$ event occurs when both Alice and Bob choose $r_i = 0$
and
a $\mathcal{B}$ event occurs when both Alice and Bob choose $r_i = 1$. 
Based on these definitions, our protocol proceeds as follows.
\begin{enumerate}[label=\arabic*., leftmargin=*]
    \item 
    \textbf{State preparation:}{
        In the $i$-th round ($i=1,2,\ldots,N$), Alice and Bob each independently choose a bit setting $r_i \in R = \{0,1\}$.
        Then, based on the value of $r_i$, they send the prepared quantum state with a known lower bound on its vacuum component 
        (satisfying Assumption \ref{assu:range}, \ref{assu:vacuum}) into the quantum channel, 
        where they will be measured by an untrusted nod Charlie.
    }
    \item 
    \textbf{Measurement:}{
        If honest, the untrusted nod Charlie performs an interference measurement on the states sent by Alice and Bob in the $i$-th round. 
        If only the right detector clicks, Charlie records this event as a successful measurement. 
        Charlie public whether each round is successful.
    }
    \item 
    \textbf{Parameter estimation:}{
        Through classical communication, 
        Alice and Bob confirm the number of $\mathcal{Z}$, $\mathcal{O}$ and $\mathcal{B}$ events, 
        denoted as $n_\mathcal{Z}^{\rm PE}$, $n_\mathcal{O}^{\rm PE}$ and $n_\mathcal{B}^{\rm PE}$,
        among all successful and parameter estimation rounds. 
        They then determine the qbit error rate (QBER) $e_{\rm bit}$,
        estimate the upper bound of phase error rate $\bar{n}_{\rm ph}$ 
        and the lower bound of the number of $\mathcal{Z}$ $\underline{n}_{\mathcal{Z}}$ in the key extraction rounds(using method below).
    }
    \item 
    \textbf{Key distillation:}
    Alice and Bob perform error correction and privacy amplification based on the results of the parameter estimation step, and then use the data in the successful rounds to generate the secret keys.
\end{enumerate}

We consider the entanglement-equivalent protocol 
for $N$ rounds.
If there is an ideal source, 
the protocol can be write as
\begin{equation}
    \label{equ:ideal-two-SNS-AB}
    \begin{aligned}
        \ket{\Phi}^{\rm ide}
        =
        \ket{\Phi}_{\rm A}^{\rm ide}
        \otimes
        \ket{\Phi}_{\rm B}^{\rm ide}
        \otimes
        \ket{\Phi}_{\rm PE}
        ,
    \end{aligned}
\end{equation}
where
$
\ket{\Phi}_{\rm PE}
=
        \left[
            \sum_{\mathbf{m}_1^N}
            \left(
                \prod_{i=1}^N \sqrt{p_{m_i}^{{\rm PE}}}
            \right)
            \left(
                \bigotimes_{i=1}^N
                \ket{m_i}_{{{\rm PE}_i}}
            \right)
        \right]
$ denotes a set of auxiliaries used to determine whether a given round is selected for parameter estimation,
$m_i=1$ indicates that the 
$i$-th round is used for parameter estimation, while 
$m_i=0$ means it is not, and
$p_{1}^{{\rm PE}}=1-p_{0}^{{\rm PE}}=p_{{\rm PE}}$;
\begin{equation}
    \label{equ:ideal-two-SNS}
    \begin{aligned}
        \ket{\Phi}_{\rm A}^{\rm ide}
        =
        \left[
            \sum_{\mathbf{r}_1^N}
            \left(
                \prod_{i=1}^N \sqrt{p_{r_i}}
            \right)
            \left(
                \bigotimes_{i=1}^N
                \ket{r_i}_{{A_i}}
                \ket{\psi_{r_i}^{\rm ide}}_{C_i}
            \right)
        \right]
        ,
    \end{aligned}
\end{equation}
where 
$p_{r_i}$ is the the probability of selecting $r_i$,
$\ket{\psi_{r_i}^{\rm ide}}_{C_i}$ denotes the ideal encoded coherent state with \textbf{selected state} $r_i$ to be send into channel,
and $\ket{\Phi}_{\rm B}^{\rm ide}$ is similar to \cref{equ:ideal-two-SNS}.
However, in the following discussion, we omit $\ket{\Phi}_{\rm PE}$, as it is an auxiliary tag negotiated by Alice and Bob after state transmission that introduces no imperfections and is tensor-producted with the remaining transmitted part,
therefore, all transformations do not involve this component.
For the protocol is ideal, we have that 
$\ket{\psi_{0}^{\rm ide}}_{C_i} = \ket{0}$ and $\ket{\psi_{1}^{\rm ide}}_{C_i} = \ket{\mu}$,
where $\ket{0}$ and $\ket{\mu}$ denote the ideal coherent state without any imperfections.
Protocols like \cref{equ:ideal-two-SNS} has been proved secure~\cite{PhysRevApplied.12.054034,PhysRevApplied.19.064003,PhysRevResearch.6.013266,shan2024improvedpostselectionsecurityanalysis}, however
if the source is imperfect, the form will be more complex. 
As we have discussed in \cref{assu:vacuum}, 
instead of $\ket{\psi_{r_i}^{\rm ide}}_{C_i}$, Alice's source send 
$\rho_{
    \mathbf{r}_{i-\xi}^i
    ,
    \mathbf{a}_{i-\xi}^i
}^{\rm A}$.
Consider a protocol that sends the purification 
$\ket{\psi_{
    \mathbf{r}_{i-\xi}^i
    ,
    \mathbf{a}_{i-\xi}^i
}^{\rm imp}}_{C_i}$ of 
$\rho_{
    \mathbf{r}_{i-\xi}^i
    ,
    \mathbf{a}_{i-\xi}^i
}^{\rm A}$ into the channel, the security of this protocol can ensure the security of the protocol that sends 
$\rho_{
    \mathbf{r}_{i-\xi}^i
    ,
    \mathbf{a}_{i-\xi}^i
}^{\rm A}$.
Thus,
if the source is imperfect, the entanglement-equivalent protocol 
of Alice's side \cref{equ:ideal-two-SNS}
will become
\begin{equation}
    \label{equ:unideal-two-SNS}
    \begin{aligned}
        &
        \ket{\Phi}_{\rm A}= 
        \\
        &
        \left[
            \sum_{\mathbf{r}_1^N \mathbf{a}_1^N }
            \left(
                \prod_{i=1}^N \sqrt{
                    p_{r_i}
                    q_{a_i}
                    }
            \right)
            \left(
                \bigotimes_{i=1}^N
                \ket{r_i}_{{A_i}}
                \ket{a_i}_{A_i''}
                \ket{\psi_{
                    \mathbf{r}_{i-\xi}^i
                    ,
                    \mathbf{a}_{i-\xi}^i
                }^{\rm imp}}_{C_i}
            \right)
        \right]
        ,
    \end{aligned}
\end{equation}
where
$q_{a_i}$ is the the probability of selecting $a_i$.

Treat the protocol in \cref{equ:unideal-two-SNS} as the \textit{original protocol} described in section `Security analysis framework addressing corre-
lated sources',
then the \textit{new protocol} without correlation can be expressed as
\begin{equation}
    \label{equ:unideal-two-SNS-new-new}
    \begin{aligned}
        &
        \ket{\Phi}_{\rm A}^{\rm new}
        =
        \Bigg[
            \sum_{\mathbf{r'}_{1}^{\xi N}}
            \sum_{\mathbf{a}_{1}^{(1+\xi)N}}
        \\
        &
            \left(
                \prod_{i=1}^{\xi N} \sqrt{
                    p_{r_i'}
                }
                \prod_{j=1}^{(1+\xi)N} \sqrt{
                    q_{a_j}
                }
            \right)
            \left(
                \bigotimes_{i=1}^{\xi N}
                \ket{r_i'}_{{A_i'}}
                \bigotimes_{j=1}^{(1+\xi) N}
                \ket{a_j}_{{A_j''}}
            \right)
        \\
        &
            \otimes
            \left[
                \sum_{\mathbf{r}_1^N}
                \left(
                    \prod_{i=1}^N \sqrt{p_{r_i}}
                \right)
                \left(
                    \bigotimes_{i=1}^N
                    \ket{r_i}_{{A_i}}
                    \ket{
                        \psi_{
                            r_i,
                            \mathbf{r'}(i,\xi)
                            ,
                            \mathbf{a}(i,\xi)
                        }^{\rm imp'}
                    }_{C_i'}
                \right)
            \right]
        \Bigg]
        ,
    \end{aligned}
\end{equation}
where 
$
\mathbf{r'}(i,\xi)
$
denotes the local ancilla in the previous 
$\xi$ \textit{physical rounds} and the following 
$\xi$ \textit{physical rounds} of the 
$i$-th \textit{key generation round},
$
\mathbf{a}(i,\xi)
$
denotes the system ancilla of the 
up mentioned \textit{physical rounds} and the 
$i$-th \textit{key generation round}, 
and
$
        \ket{
            \psi_{
                r_i,
                \mathbf{r'}(i,\xi),
                \mathbf{a}(i,\xi)
            }^{\rm imp'}
        }_{C_i'}
$
denote the state send into the channel in the $i$-th \textit{key generation round} and the following $\xi$ \textit{physical rounds} (detailed in section `Method' and Supplementary Materials).

As we have discussed, 
the new protocol reveals all \textit{physical rounds} except the \textit{key generation rounds}. 
Furthermore, since the security of the \textit{original protocol} is constrained by that of the \textit{new protocol}, we can relax the assumptions on the \textit{new protocol}. 
Therefore, we further disclose the ancilla in the space 
$A_i''$ for all \textit{physical rounds} and assume that Alice sends additional quantum states into the channel. Thus, for any additional state 
$\ket{
    \psi_{
        r_i,
        \mathbf{r'}(i,\xi)
        ,
        \mathbf{a}(i,\xi)
    }^{\rm add}
}_{C_i''}$ sent into the channel, 
we have a protocol $\ket{\Phi}_{\rm A}^{{\rm new}_2}$
which based on $\ket{\Phi}_{\rm A}^{\rm new}$ sending 
$
        \ket{
            \psi_{
                r_i,
                \mathbf{r'}(i,\xi),
                \mathbf{a}(i,\xi)
            }^{\rm imp'}
        }_{C_i'}
$
in \cref{equ:unideal-two-SNS-new-new}, instead sends 
$
        \ket{
            \psi_{
                r_i,
                \mathbf{r'}(i,\xi),
                \mathbf{a}(i,\xi)
            }^{\rm imp'}
        }_{C_i'}
\ket{
    \psi_{
        r_i,
        \mathbf{r'}(i,\xi)
        ,
        \mathbf{a}(i,\xi)
    }^{\rm add}
}_{C_i''}
$.
Then the security of protocol $\ket{\Phi}_{\rm A}^{\rm new}$ in Eq.~(\ref{equ:unideal-two-SNS-new-new}) can be guaranteed by protocol $\ket{\Phi}_{\rm A}^{{\rm new}_2}$.

Further, we can proof that there exist a set of additional state 
$\ket{
    \psi_{
        r_i,
        \mathbf{r'}(i,\xi)
        ,
        \mathbf{a}(i,\xi)
    }^{\rm add}
}_{C_i''}$ and
a unitary mapping $\mathbf{U}$ acting in space $ \bigotimes_{i=1}^{\xi N}
{A_i'}
\bigotimes_{j=1}^{(1+\xi) N}
{A_j''} 
\bigotimes_{i=1}^N C_i' C_i''$ that makes 
protocol $\ket{\Phi}_{\rm A}^{{\rm new}_2}$ become 
an \textit{equivalent protocol} $\ket{\Phi}_{\rm A}^{\rm equ}$, which is 
i.i.d and
$
            \left(
                \bigotimes_{i=1}^{\xi N}
                \ket{r_i'}_{{A_i'}}
                \bigotimes_{j=1}^{(1+\xi) N}
                \ket{a_j}_{{A_j''}}
            \right)
$ no longer appears entangled within the protocol, but instead appears in a tensor product form.
Thus, because
$r'$ and 
$a$ no longer play any role, we simplify the \textit{equivalent protocol} $\ket{\Phi}_{\rm A}^{\rm equ}$ by removing them. 
The final \textit{equivalent protocol} then satisfies
\begin{equation}
    \label{equ:equal-two-SNS1}
    \begin{aligned}
        \ket{\Phi}_{\rm A}^{\rm equ}= 
        \left[
            \sum_{\mathbf{r}_1^N}
            \left(
                \prod_{i=1}^N \sqrt{p_{r_i}}
            \right)
            \left(
                \bigotimes_{i=1}^N
                \ket{r_i}_{{A_i}}
                \ket{\psi_{r_i}^{\rm equ}}_{C_i'''}
            \right)
        \right]
        ,
    \end{aligned}
\end{equation}
where
$\ket{\psi_{0}^{\rm equ}}_{C_i'''} = \ket{0}$ and $\ket{\psi_{1}^{\rm equ}}_{C_i'''} = \ket{\mu_{\rm equ}}$,
where $\ket{0}$ is the vacuum state and $\ket{\mu_{\rm equ}}$ is the coherent state with an average number of photons equals to $\mu_{\rm equ}$,
satisfies $
{\rm e}^{-\mu_{\rm equ}}
=
\left[
    \sqrt{
        V_{0}^{{\rm A},\xi}
        V_{1}^{{\rm A},\xi}
    }
    -
    \sqrt{
        \left(
            1-V_{0}^{{\rm A},\xi}
        \right)
        \left(
            1-V_{1}^{{\rm A},\xi}
        \right)
    }
\right]^2
$
and
$
    V_{r_i}^{{\rm A},\xi}
=
    V_{r_i}^{\rm A}
\left(
    p_0
    \sqrt{
        V_{0}^{\rm A}
    }
    +
    p_1
    \sqrt{
        V_{1}^{\rm A}
    }
\right)^{2\xi}
$  (detailed in section `Method' and Supplementary Materials).

The above analysis is also applicable to Bob's side. 
In this case, 
we can conclude that a two-state SNS-QKD protocol satisfying 
the \cref{assu:range} and \ref{assu:vacuum}, including the presence of SPF, side-channels, and correlations, has its minimum smooth
entropy constrained by the phase error rate of the \textit{equivalent protocol} under the same measurement outcomes. 
And the \textit{equivalent protocol} ultimately satisfies
\begin{equation}
    \label{equ:equal-two-SNS-final}
    \begin{aligned}
        &
        \ket{\Phi}^{\rm equ}= 
        \left[
            \sum_{\mathbf{r}_1^N}
            \left(
                \prod_{i=1}^N \sqrt{p_{r_i}}
            \right)
            \left(
                \bigotimes_{i=1}^N
                \ket{r_i}_{{A_i}}
                \ket{\psi_{r_i}^{\rm Aequ}}_{C_i^{\rm A}}
            \right)
        \right]
        \\
        &
        \otimes
        \left[
            \sum_{\mathbf{r}_1^N}
            \left(
                \prod_{i=1}^N \sqrt{p_{r_i}}
            \right)
            \left(
                \bigotimes_{i=1}^N
                \ket{r_i}_{{B_i}}
                \ket{\psi_{r_i}^{\rm Bequ}}_{C_i^{\rm B}}
            \right)
        \right]
        \otimes
        \ket{\Phi}_{\rm PE}
        ,
    \end{aligned}
\end{equation}
where
$
        \ket{\psi_{0}^{\rm A(B)equ}}_{C_i^{\rm A(B)}}
        = \ket{0}
$,
$
        \ket{\psi_{1}^{\rm A(B)equ}}_{C_i^{\rm A(B)}}
        =
        \ket{\mu_{\rm equ}^{\rm A(B)}}
$,
and
$
        {\rm e}^{-\mu_{\rm equ}^{\rm A(B)}}
        =
        \left[
            \sqrt{
                V_{0}^{{\rm A(B)},\xi}
                V_{1}^{{\rm A(B)},\xi}
            }
            -
            \sqrt{
                \left(
                    1-V_{0}^{{\rm A(B)},\xi}
                \right)
                \left(
                    1-V_{1}^{{\rm A(B)},\xi}
                \right)
            }
        \right]^2
$,
$
    V_{r_i}^{{\rm A(B)},\xi}
=
    V_{r_i}^{\rm A(B)}
\left(
    p_0
    \sqrt{
        V_{0}^{\rm A(B)}
    }
    +
    p_1
    \sqrt{
        V_{1}^{\rm A(B)}
    }
\right)^{2\xi}
$,
and at this step we recall the auxiliary particle 
$
    \ket{\Phi}_{\rm PE}
$, which indicates whether the rounds are selected for parameter estimation.

After completing the above analysis, from \cref{Proposition:Proposition},
we still need to estimate the phase error rate of the \textit{equivalent protocol} $\ket{\Phi}^{\rm equ}$ to derive our final key rate formula.
Currently, the security of protocols similar to that in \cref{equ:equal-two-SNS-final} has already been proven within the framework of SCS 
QKD~\cite{PhysRevApplied.12.054034,shan2023practicalphasecodingsidechannelsecurequantum,PhysRevApplied.19.064003,PhysRevResearch.6.013266,shan2024improvedpostselectionsecurityanalysis}.
We choose to use postselection security analysis~\cite{Matsuura2024tightconcentration,shan2024improvedpostselectionsecurityanalysis}.
For a two state SNS QKD described in \cref{equ:equal-two-SNS-final}, under collective attack,
we can calculate the upper bound of phase error probability $P_{\rm ph}$, satisfies
\begin{equation}
    \label{equ:ephcakl}
    \begin{aligned}
        &
        P_{\rm ph}\leq \frac{
            p_1p_0
        }{2}
        \Bigg(
            c_0^2\frac{P_{\mathcal{O}}}{(p_0)^2}+c_1^2\frac{P_{\mathcal{B}}}{(p_1)^2}+\bar{c}_2^2
        \\
        &
            +2c_0c_1\sqrt{\frac{P_\mathcal{O}P_\mathcal{B}}{(p_0)^2(p_1)^2}}+c_0\bar{c}_2\sqrt{\frac{P_\mathcal{O}}{(p_0)^2}}+c_1\bar{c}_2\sqrt{\frac{P_\mathcal{B}}{(p_1)^2}}
        \Bigg),
    \end{aligned}
\end{equation}
where $P_{\mathcal{O}}$ and $P_{\mathcal{B}}$ are the probabilities of the $\mathcal{O}$ event and the $\mathcal{B}$ event 
of key extraction rounds
(detailed in section `Method').

Noting that obtaining \cref{equ:ephcakl} requires the assumption that Eve's attack is a collective attack, 
we can conclude that the measurement outcome probabilities are identical for all rounds. 
Moreover, we cannot directly obtain the number $n_\mathcal{O}$ and $n_\mathcal{B}$ of the $\mathcal{O}$ events and the $\mathcal{B}$ events in the key extraction rounds; 
instead, we can only calculate $n_\mathcal{O}^{\rm PE}$ and $n_\mathcal{B}^{\rm PE}$.
And because of here Eve performs a collective attack, $P_{\mathcal{O}}/(1-p_{\rm PE})$ and $P_{\mathcal{B}}/(1-p_{\rm PE})$ are equal to the corresponding portions from the parameter estimation rounds, 
denote as $P_\mathcal{O}^{\rm PE}/p_{\rm PE}$ and $P_\mathcal{B}^{\rm PE}/p_{\rm PE}$. 
Therefore, 
using the Chernoff bound \cite{fa3a69c5-2345-343d-ae4f-de42969ad827,mitzenmacher2017probability}, we can separately estimate their upper bounds such that 
$
    P_\mathcal{O} =  
    \left(
        \left(
            1-p_{\rm PE}
        \right) /
        p_{\rm PE}
    \right)     
    P_\mathcal{O}^{\rm PE}
        \leq
    \left(
        \left(
            1-p_{\rm PE}
        \right) /
        p_{\rm PE}
    \right) 
        {\overline{\text{Cher}}(n_\mathcal{O}^{\rm PE},\epsilon_{\rm ph})}/{N}
$,
$
    P_\mathcal{B} =
    \left(
        \left(
            1-p_{\rm PE}
        \right) /
        p_{\rm PE}
    \right) 
        P_\mathcal{B}^{\rm PE}
        \le
    \left(
        \left(
            1-p_{\rm PE}
        \right) /
        p_{\rm PE}
    \right) 
        {\overline{\text{Cher}}(n_\mathcal{B}^{\rm PE},\epsilon_{\rm ph})}/{N}
$,
where 
$\overline{\text{Cher}}(\cdot,\epsilon)$ is the upper bound of the expectation estimated from the observation with a failure probability $\epsilon$~\cite{shan2024improvedpostselectionsecurityanalysis}, satisfies
$
        \overline{\text{Cher}}(X,\epsilon_x)
        =X+\ln\frac{1}{\epsilon_x}+\sqrt{\ln^2\frac{1}{\epsilon_x}+2X\ln\frac{1}{\epsilon_x}}
$,
$
        \underline{\text{Cher}}(X,\epsilon_x)
        =X+\frac{1}{2}\ln\frac{1}{\epsilon_x}-\frac{1}{2}\sqrt{\ln^2\frac{1}{\epsilon_x}+8X\ln\frac{1}{\epsilon_x}}
$.

Consequently, the phase error number satisfies
$
        n_{\rm ph}\le\bar{n}_{\rm ph}= \overline{\text{cher}}(NP_{\rm ph},\epsilon_{\rm ph})
$,
where $\overline{\text{cher}}(\cdot,\epsilon)$ is the upper bound of the observation estimated from the expectation with a failure probability $\epsilon$~\cite{shan2024improvedpostselectionsecurityanalysis}, satisfies
$
        \overline{\text{cher}}(E,\epsilon_x)=E+\frac{1}{2}\ln\frac{1}{\epsilon_x}+\frac{1}{2}\sqrt{\ln^2\frac{1}{\epsilon_x}+8E\ln\frac{1}{\epsilon_x}}
$,
$
        \underline{\text{cher}}(E,\epsilon_x)=E-\sqrt{2E\ln\frac{1}{\epsilon_x}}.
$
Then, by employing the de Finetti reduction with fixed marginal \cite{PRXQuantum.5.040315}, we can extend the phase error estimation to the case of Eve's coherent attacks 
by appropriately increasing the failure probability of the estimation of $\bar{n}_{\rm ph}$ \cite{shan2024improvedpostselectionsecurityanalysis}.
Specifically, in the case of Eve's coherent attacks, the upper bound of the phase error 
with the failure probability $\epsilon_{\rm ph}$
satisfies
\begin{equation}
    \label{equ:cho}
    \begin{aligned}
        \bar{n}_{\rm ph}^{{\rm c},\epsilon_{\rm ph}}= \overline{\text{cher}}
        \left(
            NP_{\rm ph},
            \frac{\epsilon_{\rm ph}}{3g_{N,64}}
        \right)
        ,
    \end{aligned}
\end{equation}
where
$g_{N,x}=\binom{N+x-1}{N}\le(\frac{e(N+x-1)}{x-1})^{x-1}$~\cite{PRXQuantum.5.040315}, and 
$x$
is the square of
the dimension of Alice, Bob and Charlie, which here we choose 
$x=d_A^2d_B^2d_C^2d_{\rm PE}^2=2^2\times 2^2\times 2^2\times 2^2=256$~\cite{shan2024improvedpostselectionsecurityanalysis}.

Moreover, we cannot directly measure the number $n_{\mathcal{Z}}$ of $\mathcal{Z}$ events and in the key extraction rounds.
Instead, we estimate them using the corresponding occurrences in the parameter estimation rounds. By applying the Chernoff bound~\cite{fa3a69c5-2345-343d-ae4f-de42969ad827,mitzenmacher2017probability}, we can calculate
$p_{\rm PE}(n_{\mathcal{Z}} + n_{\mathcal{Z}}^{\rm PE}) \geq \underline{\text{Cher}}(n_{\mathcal{Z}}^{\rm PE},\epsilon_Z) $.
Thus, the lower bound of $n_{\mathcal{Z}}$ with failure probability $\epsilon_{Z}$ satisfies
\begin{equation}
    \label{equ:cho_132sdvcf}
    \begin{aligned}
\underline{n}_{\mathcal{Z}}^{\epsilon_{Z}} = \frac{\underline{\text{Cher}}(n_{\mathcal{Z}}^{\rm PE},\epsilon_Z) }{p_{\rm PE}} - n_{\mathcal{Z}}^{\rm PE}
.
    \end{aligned}
\end{equation}

Recall the \textit{original protocol},
because of the property of the two-universal hash function \cite{5961850},
if the secure key have a length $l$ with $\epsilon_{\rm tot}$ secure,
then $l=H_\text{min}^\epsilon(\mathbf{Z}|\mathbf{E})-2\log_2\frac{1}{2\bar \epsilon}$ with $\epsilon_{\rm tot}=2\epsilon+\bar \epsilon$.
Next, consider the error correction process. Suppose 
$f n_{\mathbf{Z}} h(e_{\rm bit})$ classical bits are consumed during error correction, where
$f$
is the efficiency of the error correction.
If the key is $\epsilon_{\rm cor}$ correct, 
a hash with length $\log_2\frac{2}{\epsilon_{\rm cor}}$ must be announced for error correction.
Then, the key length become
\begin{equation}
    \label{equ:ltmp}
    \begin{aligned}
    l=H_\text{min}^\epsilon(\mathbf{Z}|\mathbf{E}')-f n_{\mathbf{Z}} h(e_{\rm bit})-\log_2\frac{2}{\epsilon_{\rm cor}}-2\log_2\frac{1}{2\bar \epsilon}
    ,
    \end{aligned}
\end{equation}
with $\epsilon_{\rm tot}=2\epsilon+\bar \epsilon+\epsilon_{\rm cor}$ and $E$ is Eve's system before the error correction \cite{tomamichel2012tight}. 

Due to the conclusion in Eq.~(\ref{equ:ltmp}), the final key rate length is transformed into an estimation problem for the smooth min-entropy of $\mathbf{Z}$ in the \textit{original protocol}. 
We can apply the conclusions from \cref{Proposition:Proposition} to convert this problem into an estimation of phase errors in the \textit{equivalent protocol}, completing the key rate estimation. 
From \cref{equ:cho,equ:cho_132sdvcf}, and
according to 
\cref{equ:lemma2},
we can obtain that
\begin{equation}
    \label{equ:Etmp}
    \begin{aligned}
        H_\text{min}^\epsilon(\mathbf{Z}|\mathbf{E}')
        \geq
        &
        \underline{n}_{\mathcal{Z}}^{\epsilon_0}
        \left(
            1 - h
            \left(
                    \frac{
                        \bar{n}_{\rm ph}^{{\rm c},{
                        (\epsilon_1)^2}}
                    }{
                        \underline{n}_{\mathcal{Z}}^{\epsilon_0}
                    }
            \right)
        \right)
        -
        2
        \xi 
        \log_2 \frac{1}{\epsilon_2}
        \\
        &
        -
        2
        \left(
            \xi+1
        \right)
        \log_2 \frac{1}{\epsilon_3}
    ,
    \end{aligned}
\end{equation}
then we can calculate the  key length $l_{\rm max}$ that satisfies
\begin{equation}
    \label{equ:ltsdc}
    \begin{aligned}
        l_{\rm max}
        =
        &
        \underline{n}_{\mathcal{Z}}^{\epsilon_0}
        \left(
            1 - h\left(
            \frac{
                \bar{n}_{\rm ph}^{{\rm c},{
                    ({\epsilon_1})^2
                }}
            }{
                \underline{n}_{\mathcal{Z}}^{\epsilon_0}
            }
            \right)
        \right)
        -f n_{\mathbf{Z}} h(e_{\rm bit})
        -\log_2\frac{2}{\epsilon_{\rm cor}}
        \\
        &
        -2\log_2\frac{1}{2\bar \epsilon}
        -
        2
        \xi 
        \log_2 \frac{1}{\epsilon_2}
        -
        2
        \left(
            \xi+1
        \right)
        \log_2 \frac{1}{\epsilon_3}
    ,
    \end{aligned}
\end{equation}
where $\bar{n}_{\rm ph}^{{\rm c},{
({\epsilon_1})^2
}}$ satisfies \cref{equ:cho},
$\underline{n}_{\mathcal{Z}}^{\epsilon_0}$ satisfies \cref{equ:cho_132sdvcf},
and
$\epsilon_{\rm tot}=2\epsilon+\bar \epsilon+\epsilon_{\rm cor}+\epsilon_0$
and
$
        {\epsilon_1}
        =
        \left(
            \frac{
                \epsilon - \xi 
                {\epsilon_2}
            }{2 \xi +1}
            -
            {\epsilon_3}
        \right)^{{\xi+1}}
$.

\hfill

\noindent{\bf Simulation result of correlated leakage source secure quantum key distribution.}

Through simulations, we can verify the performance of our protocol. 
Specifically, we set 
the the misalignment error rate to $1\%$,
the detector dark count rate to $p_d = 10^{-9}$ bit per pulse, 
the error correction efficiency to $f=1.16$, the extinction ratio between sending and non-sending intensities to $1/1000$ \cite{Roberts:18,Lu2021},
and the security parameter to $\epsilon_{\rm tot} = 10^{-10}$.
For the specific security parameters, we set
$\epsilon=\bar \epsilon=\epsilon_{\rm cor}=\epsilon_0={\epsilon_{\rm tot}}/{5}$
and
$\epsilon_2 = \epsilon_3 = \epsilon^2$.
Further, we assume that the attenuation from Alice and Bob to the interference node is identical and they have the same upper bound of the intensity, the range of correlations as well as the sending probabilities. 
By optimizing the sending probability and upper bound of intensity, we obtain the Attenuation-Key~Rate curves for different correlation ranges $\xi $.

\begin{figure}[htbp]
    \centering
    \subfigure[
        Secret key rate when
        $N= 10^{12}$.
    ]{
        \includegraphics[width=0.4\textwidth]{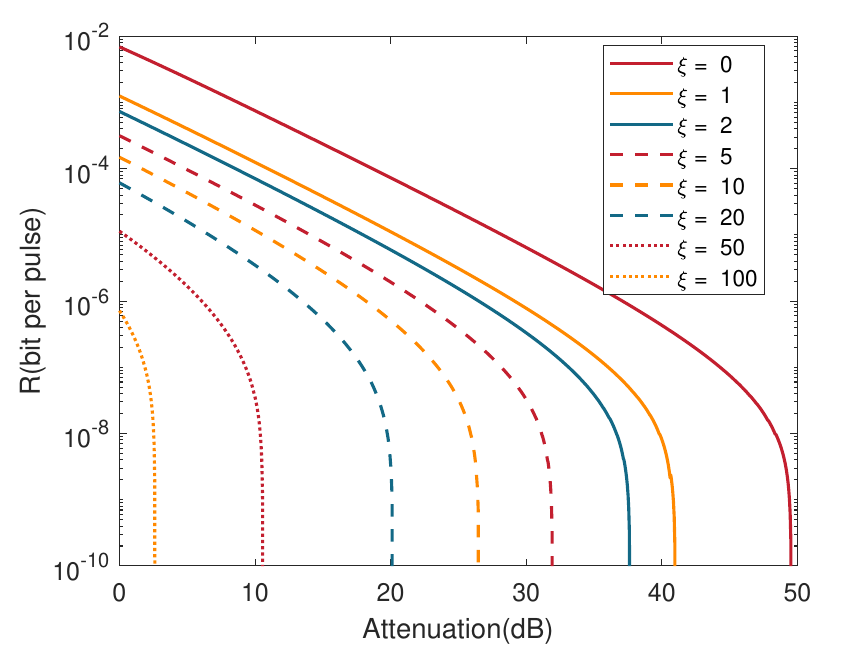}
        \label{fig:sim_n_1e12}
    }
    \subfigure[
        Secret key rate when
        $N= 10^{14}$.
    ]{
        \includegraphics[width=0.4\textwidth]{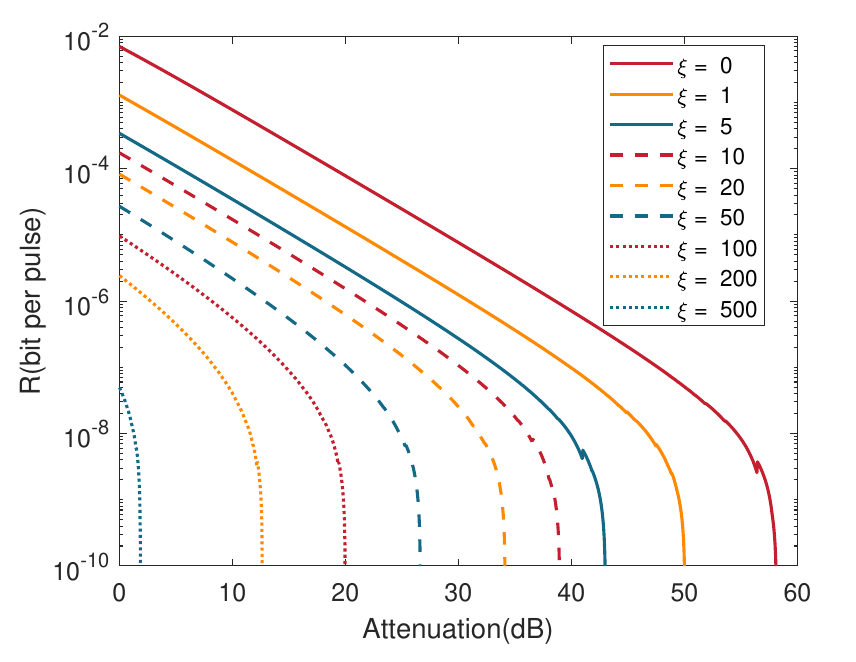}
        \label{fig:sim_n_1e14}
    }
    \caption{
        The simulation result of the performance of our protocol.
        We set  
        the the misalignment error rate to $1\%$,
        detector dark count rate to $p_d = 10^{-9}$ bit per pulse, 
        the error correction efficiency to $f=1.16$, the extinction ratio between sending and non-sending intensities to $1/1000$,
        and the security parameter to $\epsilon_{\rm tot} = 10^{-10}$.
    }
    \label{fig:sim_n}
\end{figure}

\begin{figure}[htbp]
    \includegraphics[width=0.4\textwidth]{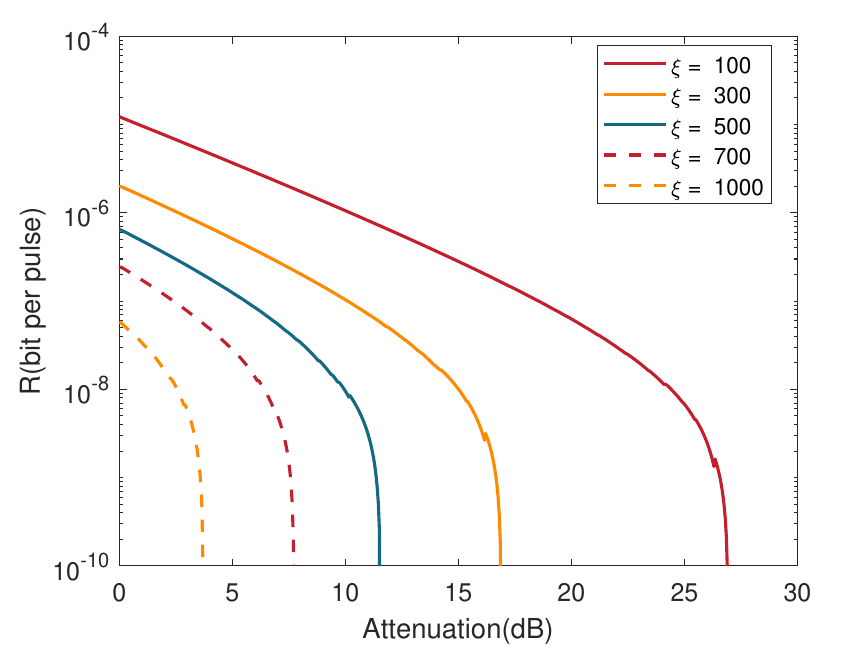}
    \caption{
        The simulation result of the performance of our protocol when $N=2\times 10^{15}$.
        We set  
        the the misalignment error rate to $1\%$,
        detector dark count rate to $p_d = 10^{-9}$ bit per pulse, 
        the error correction efficiency to $f=1.16$, the extinction ratio between sending and non-sending intensities to $1/1000$,
        and the security parameter to $\epsilon_{\rm tot} = 10^{-10}$.
    }
    \label{fig:sim_n_2e15}
\end{figure}

Fig.~\ref{fig:sim_n}~and~\ref{fig:sim_n_2e15} show the numerical simulations results of our protocol.
In Fig.~\ref{fig:sim_n}, we consider two common scenarios where the total number of pulses sent by Alice or Bob, $N$, 
which leads to the finite-length effect, is set to $10^{12}$ and $10^{14}$, 
corresponding to 
\cref{fig:sim_n_1e12,fig:sim_n_1e14} respectively.  
In both cases, we observe that when there is no correlation (i.e. $ \xi = 0 $), our protocol, according to the analysis in our manuscript, 
reduces to the existing SCS-QKD protocol, and the simulation results align with those of SCS-QKD. 
Additionally, we find that the smaller the correlation range $ \xi $, the greater its impact on the key rate and the maximal transmission attenuation.  
For $N=10^{12}$, increasing the correlation range from $\xi=0$ to $1$ results in a maximum attenuation loss of about $9 {\rm dB}$, 
while increasing the correlation range from $\xi=1$ to $5$ also leads to an additional $9 {\rm dB}$ too. 
Furthermore, in the cases of $N=10^{12}$ and $10^{14}$, we simulate correlation ranges up to $\xi=100$ and $500$, respectively.
In Fig.~\ref{fig:sim_n_2e15}, we consider a scenario with a larger $N$. By setting $N=2\times 10^{15}$, 
we find that our protocol can still generate keys even when the correlation range is very large, reaching up to $\xi = 1000$.

\hfill

\noindent{\bf CONCLUSION} 

\hfill

We have addressed a critical security challenge in practical QKD 
with 
correlated leakage source.
We first proposed a security analysis framework that allows for finite-key analysis in the presence of correlations by extending and rearranging QKD rounds 
and applying the generalized chain rule. 
Furthermore, we introduced a secure QKD against 
correlated leakage source,
extending SCS QKD to scenarios where SPFs, side-channels and correlations coexist. 
By characterizing the correlation range $\xi$ and imposing a lower bound on the vacuum component of the transmitted states of the 
correlated leakage source,
we effectively address the security loopholes introduced by source imperfections.

Unlike existing approaches, 
our protocol does not require explicit characterization or magnitudes of correlation and side-channel, 
making it more practical for real-world QKD systems. 
Our simulation results demonstrate the protocol's strong robustness, effectiveness and reliability, especially against very high order correlations.
From the simulation results, we can see that our protocol significantly improves tolerance to correlation range compared to existing protocols. 
Most existing protocols require prior knowledge of not only the correlation range but also the further characterization of both correlation and side-channel 
before conducting security analysis 
\cite{Yoshino2018,aaz4487,zapatero2021security,PhysRevApplied.18.044069,PhysRevResearch.5.023065,Currs-Lorenzo_2024,li2025quantumkeydistributionovercoming}. 
Even for very small correlation levels, the analyzable correlation range typically does not exceed $\xi = 10$.  
Moreover, existing experiments that measure correlation generally suggest that the correlation range $\xi$ is primarily 
concentrated between $ 1 $ and $ 3$
\cite{Yoshino2018,Roberts:18,Lu2021,10050030,9907821,marcomini2024characterisinghigherorderphasecorrelations,li2025quantumkeydistributionovercoming}
, with the worst case scenarios reaching only up to $ \xi = 6 $ \cite{trefilov2024intensitycorrelationsdecoystatebb84}.
Simulation results show that our protocol can handle a correlation range far beyond that of existing protocols and can encompass all currently measured correlation parameters of real devices and systems.  
More importantly, under practical device parameters 
\cite{Yoshino2018,Roberts:18,Lu2021,10050030,9907821,Xie:19,PhysRevApplied.19.014048,marcomini2024characterisinghigherorderphasecorrelations,trefilov2024intensitycorrelationsdecoystatebb84,li2025quantumkeydistributionovercoming}, 
most existing protocols struggle to generate secure keys, 
with only a few achieving this through correlation suppression, 
special encoding structures, or source monitor module~\cite{li2025quantumkeydistributionovercoming}. 
In contrast, our protocol only requires characterization of the correlation range without the need for complex additional steps.  
Furthermore, current correlation analysis protocols have not yet incorporated finite-length effects, whereas our protocol is the first to achieve this.
Compared to existing solutions, our framework and protocol significantly enhance the security and feasibility of QKD under realistic device imperfections.


As modern society's demand for data security continues to grow, 
ensuring the secure deployment of QKD has become an urgent challenge. 
By adopting our security analysis framework, the final security barrier, 
correlations, can now be effectively addressed under finite-key conditions. 
Furthermore, with our 
correlated leakage source secure protocol, 
all security loopholes in QKD can be closed through a simple characterization of the source, 
paving the way for practical and truly secure QKD implementations.
This represents an important advancement toward closing security loopholes in QKD and achieving high performance, 
real world implementations. Future research could further explore optimizing key rates under correlated conditions 
and extending the framework to other QKD protocols to enhance their practical security.

\hfill

\noindent{\bf METHOD} 

\hfill

\noindent{\bf Proof of \cref{lemma:correlated-source-devide} and \ref{lemma:equ-between-have-or-not-corr}.}
%
Recall the definitions in Section `Security analysis framework addressing correlated sources' in `RESULTS',
by the generalized chain-rule result~\cite{6408179}
and data processing inequality, we can obtain that
\begin{equation}
    \label{equ:Zsub2chain-M}
    \begin{aligned}
        &
        H^{\epsilon}_{\rm min}(\mathbf{Z}_A|\mathbf{E}')_{\rho}
        =
        H^{\epsilon}_{\rm min}(\mathbf{Z}_{A_1}
        \mathbf{Z}_{A_2}
        \ldots
        \mathbf{Z}_{A_{\xi+1}}|\mathbf{E}')_{\rho}
        \\
        \geq
        &
        H^{\epsilon_1}_{\rm min}(\mathbf{Z}_{A_1}|\mathbf{Z}_{A_2}
        \mathbf{Z}_{A_3}
        \ldots
        \mathbf{Z}_{A_{\xi+1}}\mathbf{E}')_{\rho}
            \\&
        +
        H^{\epsilon_1'}_{\rm min}(
        \mathbf{Z}_{A_2}
        \ldots
        \mathbf{Z}_{A_{\xi+1}}|\mathbf{E}')_{\rho}
        -f_1
        \\
        \geq
        &
        H^{\epsilon_1}_{\rm min}(\mathbf{Z}_{A_1}|\mathbf{Z}_{A_2}
        \mathbf{Z}_{A_3}
        \ldots
        \mathbf{Z}_{A_{\xi+1}}\mathbf{E}')_{\rho}
            \\&
        +
        H^{\epsilon_1'}_{\rm min}\left(
        \mathbf{Z}_{A_2}
        \ldots
        \mathbf{Z}_{A_{\xi+1}}|\mathbf{Z}_{A_1}\mathbf{E}'\right)_{\rho}
        -f_1
        \\
        \geq
        &
        H^{\epsilon_1}_{\rm min}\left(\mathbf{Z}_{A_1}|\mathbf{Z}_{A_2}
        \mathbf{Z}_{A_3}
        \ldots
        \mathbf{Z}_{A_{\xi+1}}\mathbf{E}'\right)_{\rho}
        -f_1
        \\
        &+
        H^{\epsilon_2}_{\rm min}\left(\mathbf{Z}_{A_2}|\mathbf{Z}_{A_1}
        \mathbf{Z}_{A_3}
        \mathbf{Z}_{A4}
        \ldots
        \mathbf{Z}_{A_{\xi+1}}\mathbf{E}'\right)_{\rho}
        -f_2
        \\ &
        +
        H^{\epsilon_2'}_{\rm min}\left(
        \mathbf{Z}_{A_3}
        \mathbf{Z}_{A4}
        \ldots
        \mathbf{Z}_{A_{\xi+1}}|\mathbf{Z}_{A_1}\mathbf{Z}_{A_2}\mathbf{E}'\right)_{\rho}
        \\ \geq & \ldots
        \\ \geq
        &
        \sum_{i=1}^{\xi+1} \left[
            H^{\epsilon_i}_{\rm min}\left(\mathbf{Z}_{A_i}|
            \left(
                \complement_{\mathbf{Z}_{A}} \mathbf{Z}_{A_i}
            \right)
            \mathbf{E}'
            \right)_{\rho}
        \right]
        -
        \sum_{i=1}^{\xi} (
            f_i
        )
        ,
    \end{aligned}
\end{equation}
where
$
        f_1 
        = 
        2
        \log_2 \frac{1}{\epsilon-2\epsilon_1-\epsilon_1'}
$
        ,
$
        f_i
        = 
        2
        \log_2 \frac{1}{\epsilon_{i-1}'-2\epsilon_i-\epsilon_i'}
$
if $i \in [2,\xi]$,
$
        \epsilon_{\xi+1} 
        = \epsilon_{\xi}'
$.
Because 
$
        \left(
            \complement_{\mathbf{Z}_{A}} \mathbf{Z}_{A_i}
        \right)
        \in
        \left(
            \complement_{\mathbf{D}_{A}} \mathbf{D}_{A_i}
        \right)
$,
using data processing inequality, we can calculate that
\begin{equation}
    \label{equ:Dsubinssunsdawesc}
    \begin{aligned}
        H^{\epsilon_i}_{\rm min}\left(\mathbf{Z}_{A_i}|
        \left(
            \complement_{\mathbf{Z}_{A}} \mathbf{Z}_{A_i}
        \right)
        \mathbf{E}'
        \right)_{\rho}
        &
        =
        H^{\epsilon_i}_{\rm min}\left(\mathbf{Z}_{A_i}|
        \left(
            \complement_{\mathbf{Z}_{A}} \mathbf{Z}_{A_i}
        \right)
        \mathbf{E}'
        \right)_{\rho'_i}
        \\
        &
        \geq
        H^{\epsilon_i}_{\rm min}\left(\mathbf{Z}_{A_i}|
            \left(
                \complement_{\mathbf{D}_{A}} \mathbf{D}_{A_i}
            \right)
            \mathbf{E}'
            \right)_{\rho'_i}
        ,
    \end{aligned}
\end{equation}
where $\rho'_i$ denotes the quantum state in the protocol that include the space
$\mathbf{Z}_{A_i}$, $
\left(
    \complement_{\mathbf{D}_{A}} \mathbf{D}_{A_i}
\right)
$
and
$
\mathbf{E}'$, and $\rho = {\rm Tr}_{
\left(
    \complement_{
        \left(
            \complement_{\mathbf{D}_{A}} \mathbf{D}_{A_i}
        \right)
    }
    \left(
        \complement_{\mathbf{Z}_{A}} \mathbf{Z}_{A_i}
    \right)
\right)}
\rho'_i
$.
Thus, we can prove that in the original protocol, there is
\begin{equation}
    \label{equ:Dsub2chain}
    \begin{aligned}
        &
        H^{\epsilon}_{\rm min}(\mathbf{Z}_A|\mathbf{E}')_{\rho}
        \\
        \geq
        &
        \sum_{i=1}^{\xi+1} \left[
            H^{\epsilon_i}_{\rm min}\left(\mathbf{Z}_{A_i}|
            \left(
                \complement_{\mathbf{D}_{A}} \mathbf{D}_{A_i}
            \right)
            \mathbf{E}'
            \right)_{\rho'_i}
        \right]
        -
        \sum_{i=1}^{\xi} (
            f_i
        )
        .
    \end{aligned}
\end{equation}
And this leads us to \cref{lemma:correlated-source-devide}.

Combine with the fact that $
H^{\tilde{\epsilon}_i}_{\rm min}(\mathbf{Z}_{A_i}'|\mathbf{D}_{A_i}'\mathbf{E}')_{\rho_i''}
        \geq
        H^{\tilde{\epsilon}_i}_{\rm min}(\mathbf{Z}_{A_i}'|\mathbf{E}')_{\rho'}
$,
we can calculate
\begin{equation}
    \label{equ:lemma1-copy}
    \begin{aligned}
        &
        H^{\epsilon}_{\rm min}(\mathbf{Z}_A|\mathbf{E}')_{\rho}
        \geq
        &
        \sum_{i=1}^{\xi+1} \left[
            H^{{\epsilon}_i}_{\rm min}(\mathbf{Z}_{A_i}'|\mathbf{E}')_{\rho'}
        \right]
        -
        \sum_{i=1}^{\xi} (
            f_i
        )
        ,
    \end{aligned}
\end{equation}
where
$
    f_1 
    = 
    2
    \log_2 \frac{1}{\epsilon-2\epsilon_1-\epsilon_1'}
$
    ,
$
    f_i
    = 
    2
    \log_2 \frac{1}{\epsilon_{i-1}'-2\epsilon_i-\epsilon_i'}
$
if $i \in [2,\xi]$,
$
    \epsilon_{\xi+1} 
    = \epsilon_{\xi}'
$.
And this leads us to \cref{lemma:equ-between-have-or-not-corr}.

\hfill

\noindent{\bf Proof of \cref{Proposition:Proposition}.}
As we have discussed in Section `Security analysis framework addressing correlated sources' in `RESULTS',
for most protocols, the smooth min-entropy can be estimated only for specific events in 
$\mathbf{Z}$. 
In this case, by applying the chain-rule result~\cite{6408179} once again, we can obtain
\begin{equation}
    \label{equ:chain234}
    \begin{aligned}
        & \quad \, \,
        H^{\tilde{\epsilon}_i}_{\rm min}(\mathbf{Z}_{A_i}'|\mathbf{E}')_{\rho'}
        \\
        &
        \geq
        H^{\tilde{\epsilon}_i'}_{\rm min}(\mathbf{Z}_{\mathcal{Z},A_i}'|\mathbf{E}')_{\rho'}
        +
        H^{\tilde{\epsilon}_i''}_{\rm min}(\mathbf{Z}_{\mathcal{X},A_i}'|\mathbf{Z}_{\mathcal{Z},A_i}'\mathbf{E}')_{\rho'}
        +
        f_i'
        \\
        &
        \geq
        H^{\tilde{\epsilon}_i'}_{\rm min}(\mathbf{Z}_{\mathcal{Z},A_i}'|\mathbf{E}')_{\rho'}
        +
        f_i'
        ,
    \end{aligned}
\end{equation}
where 
$ \mathbf{Z}_{A_i}' = \mathbf{Z}_{\mathcal{Z},A_i}' \mathbf{Z}_{\mathcal{X},A_i}'$,
$ \mathbf{Z}_{\mathcal{Z},A_i}'$ denote the parts that can estimated,
$ \mathbf{Z}_{\mathcal{X},A_i}'$ denote the parts that can not,
then
setting $\tilde{\epsilon}_i''=0$ and then $f_i'$ satisfies
$f_i'=2\log_2 \frac{1}{\tilde{\epsilon}_i-\tilde{\epsilon}_i'}$.
From the uncertainty
relation and the parameter estimation of the QKD process \cite{Tomamichel2012}, it is known that
\begin{equation}
    \label{equ:uncertaintyrelation}
    \begin{aligned}
        H^{\tilde{\epsilon}_i'}_{\rm min}(\mathbf{Z}_{\mathcal{Z},A_i}'|\mathbf{E}')_{\rho'}
        \geq
        n_i 
        \left(
            1 - h\left(
                \overline{e_i}^{\rm U}_{\tilde{\epsilon}_i'}
            \right)
        \right)
        ,
    \end{aligned}
\end{equation}
where $h(x) = -x {\rm log}_2 (x)-(1-x) {\rm log}_2 (1-x)$, $n_i$ is the number of bits in $\mathbf{Z}_{A_i}'$ 
and
$\overline{e_i}^{\rm U}_{{\epsilon}_i}$ is the upper bound of the phase error rate $e_i$
with the probability 
$p_{\rm fail}^i$ of phase errors exceeding this bound satisfies
$p_{\rm fail}^i = \left(\tilde{\epsilon}_i'\right)^2$.
For 
$\mathbf{Z}_{A}'$ , let it contain 
$n=\sum_{i=1}^{\xi+1} n_i$ bits, and let the phase error rate be 
$e$, satisfying 
$e = 
\left(
{\sum_{i=1}^{\xi+1} n_i {e_i}}
\right)
/
\left(
{\sum_{i=1}^{\xi+1} n_i}
\right)
$.
Also, let $p_{\rm fail}$ be the  the probability
of phase error rate $e$ exceeding the bound
$
\left(
{\sum_{i=1}^{\xi+1} n_i \overline{e_i}^{\rm U}_{\tilde{\epsilon}_i'}}
\right)
/
\left(
{\sum_{i=1}^{\xi+1} n_i}
\right)
$.
Through discussions related to probability theory, we find that if each 
$e_i$ exceeds its upper bound, then 
$e$ must exceed its upper bound; conversely, if none of the 
$e_i$'s exceed their upper bound, then 
$e$ must also not exceed its upper bound. Therefore, we can derive the relationship between these probabilities as
$
\prod p_{\rm fail}^i 
\leq
p_{\rm fail}
\leq
1-\prod 
\left(
    1-
    p_{\rm fail}^i 
\right)
$.
Therefore, since $\overline{e}^{\rm U}_{{\epsilon}}$ is monotonically decreasing with respect to $\epsilon$, we can conclude that
\begin{equation}
    \label{equ:eGuJi}
    \begin{aligned}
        \overline{e}^{\rm U}_{\prod_{i=1}^{\xi+1} \tilde{\epsilon}_i'}
        \geq
        \frac{\sum_{i=1}^{\xi+1} n_i \overline{e_i}^{\rm U}_{\tilde{\epsilon}_i'}}
        {\sum_{i=1}^{\xi+1} n_i}
        ,
    \end{aligned}
\end{equation}
Thus, combine with the the concavity of the 
$h(x)$ function, we can obtain that
\begin{equation}
    \label{equ:sumuncertaintyrelation}
    \begin{aligned}
        \sum_{i=1}^{\xi+1}
        H^{\tilde{\epsilon}_i'}_{\rm min}(\mathbf{Z}_{\mathcal{Z},A_i}'|\mathbf{E}')_{\rho'}
        &
        \geq
        \left(
            \sum_{i=1}^{\xi+1} n_i
        \right)
        \left(
            1-h\left(
                \frac{\sum_{i=1}^{\xi+1} n_i \overline{e_i}^{\rm U}_{{\epsilon}_i'}}
                {\sum_{i=1}^{\xi+1} n_i}
            \right)
        \right)
        \\
        &
        \geq
        n
        \left(
            1 - h\left(
                \overline{e}^{\rm U}_{{\epsilon}'}
            \right)
        \right)
        ,
    \end{aligned}
\end{equation}
where ${\epsilon}' = \prod_{i=1}^{\xi+1} \tilde{\epsilon}_i'$.

Combine 
\cref{lemma:equ-between-have-or-not-corr,equ:chain234,equ:sumuncertaintyrelation},
and without loss of generality taking 
$\tilde{\epsilon}_i = {\epsilon}_i$,
$f_i = f_j$,
$\tilde{\epsilon}_i'=\tilde{\epsilon}_j'$
and ${\epsilon}_i = {\epsilon}_j$, 
we can conclude that the minimum smooth entropy of the original protocol and the upper bound of the phase error estimated in the new protocol satisfy
\begin{equation}
    \label{equ:aHmin-old-and-new}
    \begin{aligned}
        H^{\epsilon}_{\rm min}(\mathbf{Z}_A|\mathbf{E}')_{\rho}
        \geq
        n
        \left(
            1 - h\left(
                \overline{e}^{\rm U}_{\widehat{\epsilon}}
            \right)
        \right)
        -
        \xi f
        -
        \left(
            \xi+1
        \right)
        f'
        ,
    \end{aligned}
\end{equation}
where $\epsilon$, $\widehat{\epsilon}$ and $f$ satisfy
$
{\widehat{\epsilon}} =
        \left(
            \frac{
                \epsilon - \xi \frac{1}{2^{f/2}}
            }{2 \xi +1}
            -
            \frac{1}{2^{f'/2}}
        \right)^{{\xi+1}}
$.
And this leads us to \cref{Proposition:Proposition}.

\hfill

\noindent{\bf Proof that $\ket{\Phi}_{\rm A}^{\rm equ}$ ensures the security of $\ket{\Phi}_{\rm A}$.}
The key of the secure proof of the secure QKD against 
correlated leakage source is to proof that $\ket{\Phi}_{\rm A}^{\rm equ}$ ensures the security of $\ket{\Phi}_{\rm A}$.
We begin by considering the simplest case, where the correlation range 
$\xi = 1$, and the total number of rounds 
$N$ is even. The more general case is discussed in the Supplementary Materials.

First, we will remove the correlation with the conclusion mentioned in \cref{Proposition:Proposition}.
Treat the protocol in Eq.~(\ref{equ:unideal-two-SNS}) as the \textit{original protocol} described in \cref{Proposition:Proposition},
then the \textit{new protocol} without correlation can be expressed as
\begin{equation}
    \label{equ:unideal-two-SNS-new-xi-1}
    \begin{aligned}
        \ket{\Phi}_{\rm A}^{\rm new}
        =
                \ket{\Phi}_{{\rm A}^1}
                \otimes
                \ket{\Phi}_{{\rm A}^2}
        ,
    \end{aligned}
\end{equation}
    where
    $
            \ket{\Phi}_{{\rm A}^1}
            =
            \left[
                \sum_{\mathbf{r}_1^{\frac{N}{2}} \mathbf{r'}_1^{\frac{N}{2}} \mathbf{a}_1^N }
                \left(
                    \prod_{i=1}^{\frac{N}{2}} \sqrt{
                        p_{r_i}
                        p_{r_i'}
                        }
                    \prod_{j=1}^N \sqrt{
                        q_{a_j}
                        }
                \right)
                \left(
                    \bigotimes_{i=1}^{\frac{N}{2}}
                    \ket{r_i'}_{{A_i}'}
                    \bigotimes_{j=1}^N
                    \ket{a_i}_{A_i''}
                \right)
                \left(
                    \bigotimes_{i=1}^{\frac{N}{2}}
                    \ket{r_i}_{{A_i}}
                    \ket{\psi_{
                        r_i,
                        \mathbf{r'}(i,1),
                        \mathbf{a}(i,1)
                    }^{\rm imp'}}_{C_i'}
                \right)
            \right]
    $
    denotes the entanglement-equivalent protocol for the first time of $N$ quantum key distribution rounds
    and
    $
            \ket{\Phi}_{{\rm A}^2}
            =
            \left[
                \sum_{\mathbf{r}_{\frac{N}{2}+1}^{N} \mathbf{r'}_{\frac{N}{2}+1}^{N} \mathbf{a}_{n+1}^{2N} }
                \left(
                    \prod_{i=\frac{N}{2}+1}^{N} \sqrt{
                        p_{r_i}
                        p_{r_i'}
                        }
                    \prod_{j=N+1}^{2N} \sqrt{
                        q_{a_j}
                        }
                \right)
                \left(
                    \bigotimes_{\frac{N}{2}+1}^{N}
                    \ket{r_i'}_{{A_i}'}
                    \bigotimes_{j=N+1}^{2N}
                    \ket{a_i}_{A_i''}
                \right)
                \left(
                    \bigotimes_{\frac{N}{2}+1}^{N}
                    \ket{r_i}_{{A_i}}
                    \ket{\psi_{
                        r_i,
                        \mathbf{r'}(i,1),
                        \mathbf{a}(i,1)
                    }^{\rm imp'}}_{C_i'}
                \right)
            \right]
    $
    denotes the entanglement-equivalent protocol for the next time of $N$ quantum key distribution rounds,
    where
    $r_i$ denotes the encoding ancilla in $i$-th \textit{key generation rounds},
    $r_i'$ denotes the encoding ancilla in $i$-th \textit{leakage rounds},
    $a_i$ denotes the the purified ancilla in the $i$-th \textit{physical rounds},
    and
    $
            \ket{
                \psi_{
                    r_i,
                    \mathbf{r'}(i,\xi),
                    \mathbf{a}(i,\xi)
                }^{\rm imp'}
            }_{C_i'}
    $
    denote the state send into the channel in the $i$-th \textit{key generation round} and the following $\xi$ \textit{physical rounds},
    where in case $\ket{\Phi}_{{\rm A}^1}$, $\mathbf{r'}(i,1) = \mathbf{r'}_{i-1}^{i}$, $\mathbf{a}(i,1) = \mathbf{a}_{2i-2}^{2i}$
    and in case $\ket{\Phi}_{{\rm A}^2}$, $\mathbf{r'}(i,1) = \mathbf{r'}_{i}^{i+1}$, $\mathbf{a}(i,1) = \mathbf{a}_{2i-1}^{2i+1}$,
    satisfies
    \begin{equation}
        \label{equ:define of ket phi 1}
        \begin{aligned}
            \ket{\psi_{
                r_i,
                \mathbf{r'}(i,1),
                \mathbf{a}(i,1)
            }^{\rm imp'}}_{C_i'}
            =
            \left\{
            \begin{aligned}
                &
                \ket{\psi_{
                        r_i {r'}_{i-1}
                        ,
                        \mathbf{a}_{2i-2}^{2i-1}
                    }^{\rm imp}
                }_{C_{2i-1}}
                \ket{\psi_{
                        {r'}_{i} r_i 
                        ,
                        \mathbf{a}_{2i-1}^{2i}
                    }^{\rm imp}
                }_{C_{2i}}
                ,
                \quad
                &
                {
                    \rm
                    in
                    \,
                    case\,
                }
                \ket{\Phi}_{{\rm A}^1}
                ,
                \\
                &
                \ket{\psi_{
                        r_i {r'}_{i}
                        ,
                        \mathbf{a}_{2i-1}^{2i}
                    }^{\rm imp}
                }_{C_{2i}}
                \ket{\psi_{
                        {r'}_{i+1} r_i 
                        ,
                        \mathbf{a}_{2i}^{2i+1}
                    }^{\rm imp}
                }_{C_{2i+1}}
                ,
                \quad
                &
                {
                    \rm
                    in
                    \,
                    case\,
                }
                \ket{\Phi}_{{\rm A}^2}
                ,
            \end{aligned}
            \right.
        \end{aligned}
    \end{equation}
    and where $\ket{\psi_{
        \mathbf{r}_{i-\xi}^i
        ,
        \mathbf{a}_{i-\xi}^i
    }^{\rm imp}}_{C_i}$
    is denoted in Eq.~(\ref{equ:unideal-two-SNS}).

As discussed in \cref{Proposition:Proposition}, 
the \textit{new protocol} can reveal all $r_i'$ and $a_j$ to Eve.
To further ensure that the \textit{new protocol} is not only independently distributed but also i.i.d., 
we additionally require Alice to send an auxiliary quantum state 
$\ket{
    \psi_{
        r_i,
        \mathbf{r'}(i,1)
        ,
        \mathbf{a}(i,1)
    }^{\rm add}
}_{C_i''}$ 
in the $C_i''$ space for each $r_i$. 
We denote this modified protocol as $\ket{\Phi}_{\rm A}^{{\rm new}_2}$. 
For any 
$\ket{
    \psi_{
        r_i,
        \mathbf{r'}(i,1)
        ,
        \mathbf{a}(i,1)
    }^{\rm add}
}_{C_i''}$, 
protocol $\ket{\Phi}_{\rm A}^{{\rm new}_2}$ guarantees the security of $\ket{\Phi}_{\rm A}^{{\rm new}}$ in \cref{equ:unideal-two-SNS-new-xi-1}. 
Thus, combine with \cref{equ:unideal-two-SNS-new-xi-1},
$\ket{\Phi}_{\rm A}^{{\rm new}_2}$ satisfies
\begin{equation}
    \label{equ:unideal-two-SNS-new-new-2new-xi-1}
    \begin{aligned}
        \ket{\Phi}_{\rm A}^{{\rm new}_2}
        =
        &
        \Bigg[
            \sum_{\mathbf{r'}_{1}^{N}}
            \sum_{\mathbf{a}_{1}^{2N}}
            \left(
                \prod_{i=1}^{N} \sqrt{
                    p_{r_i'}
                }
                \prod_{j=1}^{2N} \sqrt{
                    q_{a_j}
                }
            \right)
            \left(
                \bigotimes_{i=1}^{N}
                \ket{r_i'}_{{A_i'}}
                \bigotimes_{j=1}^{2N}
                \ket{a_j}_{{A_j''}}
            \right)
        \\
        &
            \otimes
            \left[
                \sum_{\mathbf{r}_1^N}
                \left(
                    \prod_{i=1}^N \sqrt{p_{r_i}}
                \right)
                \left(
                    \bigotimes_{i=1}^N
                    \ket{r_i}_{{A_i}}
                    \ket{
                        \psi_{
                            r_i,
                            \mathbf{r'}(i,1)
                            ,
                            \mathbf{a}(i,1)
                        }^{\rm imp'}
                    }_{C_i'}
                    \ket{
                        \psi_{
                            r_i,
                            \mathbf{r'}(i,1)
                            ,
                            \mathbf{a}(i,1)
                        }^{\rm add}
                    }_{C_i''}
                \right)
            \right]
        \Bigg]
        .
    \end{aligned}
\end{equation}

After adding additional states
$
                    \ket{
                        \psi_{
                            r_i,
                            \mathbf{r'}(i,1)
                            ,
                            \mathbf{a}(i,1)
                        }^{\rm add}
                    }_{C_i''}
$, 
our goal is to prove that there exist a choice of those states that can make the protocol i.i.d after acting an unitary mapping on 
$\ket{\Phi}_{\rm A}^{{\rm new}_2}$.
To achieve this, first we should isolate the terms related to the $i$-th coding ancilla choice $r_i$ in \cref{equ:unideal-two-SNS-new-new-2new-xi-1},
and denote as $
        \ket{\Phi}_{{\rm A,iso\,}r_i}^{{\rm new}_2,i}
$, satisfies
    \begin{equation}
        \label{equ:iso-ri-define-2}
        \begin{aligned}
            \ket{\Phi}_{{\rm A,iso\,}r_i}^{{\rm new}_2,i}
            =
            \left\{
            \begin{aligned}
                &
                \sum_{\mathbf{r'}(i,\xi)}
                \sum_{\mathbf{a}(i,\xi)}
                \left(
                    \prod_{
                        j=
                        i-1
                    }^{
                        i
                    }
                    \sqrt{
                        p_{r_j'}
                    }
                    \prod_{
                        k=
                        2i-2
                    }^{
                        2i
                    }
                    \sqrt{
                        p_{a_k}
                    }
                \right)
                \left(
                    \bigotimes_{
                        j=
                        i-1
                    }^{
                        i
                    }
                    \ket{r_j'}_{{A_j'}}
                    \bigotimes_{
                        k=
                        2i-2
                    }^{
                        2i
                    }
                    \ket{a_k}_{{A_k''}}
                \right)
                \\
                &
                \quad
                    \otimes
                            \ket{
                                \psi_{
                                    r_i,
                                    \mathbf{r'}(i,\xi)
                                    ,
                                    \mathbf{a}(i,\xi)
                                }^{\rm imp'}
                            }_{C_i'}
                            \ket{
                                \psi_{
                                    r_i,
                                    \mathbf{r'}(i,\xi)
                                    ,
                                    \mathbf{a}(i,\xi)
                                }^{\rm add}
                            }_{C_i''}
                ,
                \quad
                i \leq \frac N 2
                ,
                \\
                &
                \sum_{\mathbf{r'}(i,\xi)}
                \sum_{\mathbf{a}(i,\xi)}
                \left(
                    \prod_{
                        j=
                        i
                    }^{
                        i+1
                    }
                    \sqrt{
                        p_{r_j'}
                    }
                    \prod_{
                        k=
                        2i-1
                    }^{
                        2i+1
                    }
                    \sqrt{
                        p_{a_k}
                    }
                \right)
                \left(
                    \bigotimes_{
                        j=
                        i
                    }^{
                        i+1
                    }
                    \ket{r_j'}_{{A_j'}}
                    \bigotimes_{
                        k=
                        2i-1
                    }^{
                        2i+1
                    }
                    \ket{a_k}_{{A_k''}}
                \right)
                \\
                &
                \quad
                    \otimes
                            \ket{
                                \psi_{
                                    r_i,
                                    \mathbf{r'}(i,\xi)
                                    ,
                                    \mathbf{a}(i,\xi)
                                }^{\rm imp'}
                            }_{C_i'}
                            \ket{
                                \psi_{
                                    r_i,
                                    \mathbf{r'}(i,\xi)
                                    ,
                                    \mathbf{a}(i,\xi)
                                }^{\rm add}
                            }_{C_i''}
                ,
                \quad
                i > \frac N 2
                .
            \end{aligned}
            \right.
        \end{aligned}
    \end{equation}
    In the following discussion, we will take the case of $i\leq N/2$, namely the case $\ket{\Phi}_{{\rm A}^1}$, as an example. The analysis for the other case proceeds analogously.
    Since 
    $\ket{
        \psi_{
            r_i,
            \mathbf{r'}(i,1)
            ,
            \mathbf{a}(i,1)
        }^{\rm add}
    }_{C_i''}$ is arbitrary, we can transfer part of its phase to 
    $\ket{
        \psi_{
            r_i,
            \mathbf{r'}(i,1)
            ,
            \mathbf{a}(i,1)
        }^{\rm imp'}
    }_{C_i'}$   
    to form 
    $\ket{
        \psi_{
            r_i,
            \mathbf{r'}(i,1)
            ,
            \mathbf{a}(i,1)
        }^{\rm imp''}
    }_{C_i'}$   
    and
    $\ket{
        \psi_{
            r_i,
            \mathbf{r'}(i,1)
            ,
            \mathbf{a}(i,1)
        }^{\rm add'}
    }_{C_i''}$,
    such that 
    $
    \bra{0}
    \ket{
        \psi_{
            r_i,
            \mathbf{r'}(i,1)
            ,
            \mathbf{a}(i,1)
        }^{\rm imp''}
    }_{C_i'}$  
    is real and positive,
    where
    $
            \ket{0}_{C_i'} 
            =
            \ket{0}_{C_{2i-1}}
            \ket{0}_{C_{2i}}
    $
    and
    $\ket{0}_{C_{j}}$ is the vacuum state in the $j$-th \textit{physical rounds}.
    For further analysis, we additionally define two states respectively satisfying
    \begin{equation}
        \label{equ:vacandnew-ri-define-2}
        \begin{aligned}
            \ket{\Phi}_{{\rm A,iso\,}r_i}^{{\rm new},i}
            =
            &
            \Bigg[
                \sum_{\mathbf{r'}(i,\xi)}
                \sum_{\mathbf{a}(i,\xi)}
                \left(
                    \prod_{
                        j=
                        i-1
                    }^{
                        i
                    }
                    \sqrt{
                        p_{r_j'}
                    }
                    \prod_{
                        k=
                        2i-2
                    }^{
                        2i
                    }
                    \sqrt{
                        p_{a_k}
                    }
                \right)
                \left(
                    \bigotimes_{
                        j=
                        i-1
                    }^{
                        i
                    }
                    \ket{r_j'}_{{A_j'}}
                    \bigotimes_{
                        k=
                        2i-2
                    }^{
                        2i
                    }
                    \ket{a_k}_{{A_k''}}
                \right)
                        \ket{
                            \psi_{
                                r_i,
                                \mathbf{r'}(i,\xi)
                                ,
                                \mathbf{a}(i,\xi)
                            }^{\rm imp''}
                        }_{C_i'}
            \Bigg]
            ,
            \\
            \ket{\Phi}_{{\rm A,vac}}^{{\rm new},i}
            =
            &
            \left[
                \sum_{\mathbf{r'}(i,\xi)}
                \sum_{\mathbf{a}(i,\xi)}
                \left(
                    \prod_{
                        j=
                        i-1
                    }^{
                        i
                    }
                    \sqrt{
                        p_{r_j'}
                    }
                    \prod_{
                        k=
                        2i-2
                    }^{
                        2i
                    }
                    \sqrt{
                        p_{a_k}
                    }
                \right)
                \left(
                    \bigotimes_{
                        j=
                        i-1
                    }^{
                        i
                    }
                    \ket{r_j'}_{{A_j'}}
                    \bigotimes_{
                        k=
                        2i-2
                    }^{
                        2i
                    }
                    \ket{a_k}_{{A_k''}}
                \right)
                \ket{0}_{C_i'}
            \right]
            .
        \end{aligned}
    \end{equation}
    Thus, due to \cref{assu:vacuum}, which further leads to 
    $
            \mathop{\rm min}\limits_{\mathbf{r}_{i-\xi}^{i-1}}
            \mathop{\rm min}\limits_{\mathbf{a}_{i-\xi}^{i}}
            \left(
            \bra{0}
                    \ket{\psi_{
                        \mathbf{r}_{i-\xi}^i
                        ,
                        \mathbf{a}_{i-\xi}^i
                    }^{\rm imp}}
                    \bra{\psi_{
                        \mathbf{r}_{i-\xi}^i
                        ,
                        \mathbf{a}_{i-\xi}^i
                    }^{\rm imp}}_{C_i}
            \ket{0}
            \right)
            \geq
            V_{r_i}^{\rm A}
    $,
    the two intermediate states defined in \cref{equ:vacandnew-ri-define-2} satisfies
    \begin{equation}
        \label{equ:inner_of_zhongjian-2}
        \begin{aligned}
            \bra{\Phi}_{{\rm A,iso\,}r_i}^{{\rm new},i}
            \ket{\Phi}_{{\rm A,vac}}^{{\rm new},i}
            =
            &
        \left|
            \sum_{\mathbf{r'}(i,\xi)}
            \sum_{\mathbf{a}(i,\xi)}
            \left(
                \prod_{
                    j=
                    i-1
                }^{
                    i
                }
                {
                    p_{r_j'}
                }
                \prod_{
                    k=
                    2i-2
                }^{
                    2i
                }
                {
                    p_{a_k}
                }
            \right)
            \braket{
                \psi_{
                    r_i,
                    \mathbf{r'}(i,\xi)
                    ,
                    \mathbf{a}(i,\xi)
                }^{\rm imp''}
            }{
                0
            }_{C_i'}
        \right|
            \\
            \geq
            &
            \sum_{\mathbf{r'}(i,\xi)}
            \sum_{\mathbf{a}(i,\xi)}
            \left(
                \prod_{
                    j=
                    i-1
                }^{
                    i
                }
                {
                    p_{r_j'}
                }
                \prod_{
                    k=
                    2i-2
                }^{
                    2i
                }
                {
                    p_{a_k}
                }
            \right)
            \left(
                \sqrt{
                    V_{r_i}^{\rm A}
                }
                \prod_{
                    j=
                    i-1
                }^{
                    i
                }
                \sqrt{
                    V_{r_i'}^{\rm A}
                }
            \right)
            \\
                =
            &
                \sqrt{
                    V_{r_i}^{\rm A}
                }
                \left(
                    p_0
                    \sqrt{
                        V_{0}^{\rm A}
                    }
                    +
                    p_1
                    \sqrt{
                        V_{1}^{\rm A}
                    }
                \right)
                =:
                \sqrt{
                    V_{r_i}^{{\rm A},1}
                }
                .
        \end{aligned}
    \end{equation}
    Because \cref{equ:inner_of_zhongjian-2} satisfies for both $r_i = 0$ and $1$, we can find that
    \begin{equation}
        \label{equ:inner_of_two_new-2}
        \begin{aligned}
            \left|
                \bra{\Phi}_{{\rm A,iso\,}0}^{{\rm new},i}
                \ket{\Phi}_{{\rm A,iso\,}1}^{{\rm new},i}
            \right|
            \geq
            \sqrt{
                V_{0}^{{\rm A},1}
                V_{1}^{{\rm A},1}
            }
            -
            \sqrt{
                \left(
                    1-V_{0}^{{\rm A},1}
                \right)
                \left(
                    1-V_{1}^{{\rm A},1}
                \right)
            }
            .
        \end{aligned}
    \end{equation}
    And because from \cref{equ:iso-ri-define-2,equ:inner_of_zhongjian-2}, we have that
    \begin{equation}
        \label{equ:inner-of-new2-2}
        \begin{aligned}
            &
            \bra{\Phi}_{{\rm A,iso\,}0}^{{\rm new}_2,i}
            \ket{\Phi}_{{\rm A,iso\,}1}^{{\rm new}_2,i}
            \\
            =
            &
            \sum_{\mathbf{r'}(i,\xi)}
            \sum_{\mathbf{a}(i,\xi)}
            \left(
                \prod_{
                    j=
                    i-1
                }^{
                    i
                }
                {
                    p_{r_j'}
                }
                \prod_{
                    k=
                    2i-2
                }^{
                    2i
                }
                {
                    p_{a_k}
                }
            \right)
            \braket{
                \psi_{
                    0,
                    \mathbf{r'}(i,\xi)
                    ,
                    \mathbf{a}(i,\xi)
                }^{\rm imp''}
            }{
                \psi_{
                    1,
                    \mathbf{r'}(i,\xi)
                    ,
                    \mathbf{a}(i,\xi)
                }^{\rm imp''}
            }_{C_i'}
            \braket{
                \psi_{
                    0,
                    \mathbf{r'}(i,\xi)
                    ,
                    \mathbf{a}(i,\xi)
                }^{\rm add'}
            }{
                \psi_{
                    1,
                    \mathbf{r'}(i,\xi)
                    ,
                    \mathbf{a}(i,\xi)
                }^{\rm add'}
            }_{C_i''}
        ,
    \\
            &
            \bra{\Phi}_{{\rm A,iso\,}0}^{{\rm new},i}
            \ket{\Phi}_{{\rm A,iso\,}1}^{{\rm new},i}
            =
            \sum_{\mathbf{r'}(i,\xi)}
            \sum_{\mathbf{a}(i,\xi)}
            \left(
                \prod_{
                    j=
                    i-1
                }^{
                    i
                }
                {
                    p_{r_j'}
                }
                \prod_{
                    k=
                    2i-2
                }^{
                    2i
                }
                {
                    p_{a_k}
                }
            \right)
            \braket{
                \psi_{
                    0,
                    \mathbf{r'}(i,\xi)
                    ,
                    \mathbf{a}(i,\xi)
                }^{\rm imp''}
            }{
                \psi_{
                    1,
                    \mathbf{r'}(i,\xi)
                    ,
                    \mathbf{a}(i,\xi)
                }^{\rm imp''}
            }_{C_i'}
        .
        \end{aligned}
    \end{equation}
Since 
$
\left|
\braket{
    \psi_{
        0,
        \mathbf{r'}(i,\xi)
        ,
        \mathbf{a}(i,\xi)
    }^{\rm add'}
}{
    \psi_{
        1,
        \mathbf{r'}(i,\xi)
        ,
        \mathbf{a}(i,\xi)
    }^{\rm add'}
}_{C_i''}
\right|
\leq
1$,
combine with \cref{equ:inner_of_two_new-2}
we can select a specific set of
$\ket{
    \psi_{
        r_i,
        \mathbf{r'}(i,\xi)
        ,
        \mathbf{a}(i,\xi)
    }^{\rm add'}
}_{C_i''}$ such that
\begin{equation}
    \label{equ:inner_of_two_new2-2}
    \begin{aligned}
            \bra{\Phi}_{{\rm A,iso\,}0}^{{\rm new}_2,i}
            \ket{\Phi}_{{\rm A,iso\,}1}^{{\rm new}_2,i}
        =
        \sqrt{
            V_{0}^{{\rm A},1}
            V_{1}^{{\rm A},1}
        }
        -
        \sqrt{
            \left(
                1-V_{0}^{{\rm A},1}
            \right)
            \left(
                1-V_{1}^{{\rm A},1}
            \right)
        }
        ,
    \end{aligned}
\end{equation}
and same prof satisfies when $i >N/2$.
And then we will prof this set of 
$\ket{
    \psi_{
        r_i,
        \mathbf{r'}(i,\xi)
        ,
        \mathbf{a}(i,\xi)
    }^{\rm add'}
}_{C_i''}$
is that can make the protocol i.i.d.

Specifically, we construct this i.i.d protocol as \textit{equivalent protocol} $\ket{\Phi}_{\rm A}^{\rm equ}$, satisfies
\begin{equation}
    \label{equ:1unideal-two-SNS-new-new-2-2}
    \begin{aligned}
        &
        \ket{\Phi}_{\rm A}^{\rm equ}
        =
        \left(
            \prod_{j=1}^{N}
            \mathbf{U}_j
        \right)
        \ket{\Phi}_{\rm A}^{{\rm new}_2}
        \\
        =
        &
        \Bigg[
            \sum_{\mathbf{r'}_{1}^{\xi N}}
            \sum_{\mathbf{a}_{1}^{(1+\xi)N}}
            \left(
                \prod_{i=1}^{ N} \sqrt{
                    p_{r_i'}
                }
                \prod_{j=1}^{2N} \sqrt{
                    q_{a_j}
                }
            \right)
            \left(
                \bigotimes_{i=1}^{N}
                \ket{r_i'}_{{A_i'}}
                \bigotimes_{i=1}^{2 N}
                \ket{a_i}_{{A_i''}}
            \right)
        \\
        &
            \otimes
            \left[
                \sum_{\mathbf{r}_1^N}
                \left(
                    \prod_{i=1}^N \sqrt{p_{r_i}}
                \right)
                \left(
                    \bigotimes_{i=1}^N
                    \ket{r_i}_{{A_i}}
                \ket{\psi_{r_i}^{\rm equ}}_{C_i'''}
                \right)
            \right]
        \Bigg]
        ,
    \end{aligned}
\end{equation}
where $\mathbf{U}_i$ si an unitary mapping from the space
$
\bigcup_{
    j=
    i-1
}^{
    i
} A_j'
\bigcup_{
    2i-2
}^{
    2i
}
A_k''
\bigcup
{C_i'}
\bigcup
{C_i''}
$
into
$
\bigcup_{
    j=
    i-1
}^{
    i
} A_j'
\bigcup_{
    2i-2
}^{
    2i
}
A_k''
\bigcup
{C_i'''}
$
when $i\leq N/2$
and
unitary mapping from the space
$
\bigcup_{
    j=
    i
}^{
    i+1
} A_j'
\bigcup_{
    2i-1
}^{
    2i+1
}
A_k''
\bigcup
{C_i'}
\bigcup
{C_i''}
$
into
$
\bigcup_{
    j=
    i
}^{
    i+1
} A_j'
\bigcup_{
    2i-1
}^{
    2i+1
}
A_k''
\bigcup
{C_i'''}
$
when $i > N/2$,
satisfies
\begin{equation}
    \label{equ:Uafter-ri-define-2}
    \begin{aligned}
        \mathbf{U}_i
        \ket{\Phi}_{{\rm A,iso\,}r_i}^{{\rm new}_2,i}
        =
        \left\{
        \begin{aligned}
            &
            \sum_{\mathbf{r'}(i,\xi)}
            \sum_{\mathbf{a}(i,\xi)}
            \left(
                \prod_{
                    j=
                    i-1
                }^{
                    i
                }
                \sqrt{
                    p_{r_j'}
                }
                \prod_{
                    k=
                    2i-2
                }^{
                    2i
                }
                \sqrt{
                    p_{a_k}
                }
            \right)
            \left(
                \bigotimes_{
                    j=
                    i-1
                }^{
                    i
                }
                \ket{r_j'}_{{A_j'}}
                \bigotimes_{
                    k=
                    2i-2
                }^{
                    2i
                }
                \ket{a_k}_{{A_k''}}
            \right)
                \ket{\psi_{r_i}^{\rm equ}}_{C_i'''}
            ,
            \quad
            i \leq \frac N 2
            ,
            \\
            &
            \sum_{\mathbf{r'}(i,\xi)}
            \sum_{\mathbf{a}(i,\xi)}
            \left(
                \prod_{
                    j=
                    i
                }^{
                    i+1
                }
                \sqrt{
                    p_{r_j'}
                }
                \prod_{
                    k=
                    2i-1
                }^{
                    2i+1
                }
                \sqrt{
                    p_{a_k}
                }
            \right)
            \left(
                \bigotimes_{
                    j=
                    i
                }^{
                    i+1
                }
                \ket{r_j'}_{{A_j'}}
                \bigotimes_{
                    k=
                    2i-1
                }^{
                    2i+1
                }
                \ket{a_k}_{{A_k''}}
            \right)
                \ket{\psi_{r_i}^{\rm equ}}_{C_i'''}
            ,
            \quad
            i > \frac N 2
            .
        \end{aligned}
        \right.
    \end{aligned}
\end{equation}
In both two cases, $\ket{\psi_{r_i}^{\rm equ}}_{C_i'''}$ in \cref{equ:Uafter-ri-define-2} have no relation with $r'_j$ and $a_k$,
so, it is clear that 
$
\bra{\Phi}_{{\rm A,iso\,}0}^{{\rm new}_2,i}
{\mathbf{U}_i}^{\dagger}
\mathbf{U}_i
\ket{\Phi}_{{\rm A,iso\,}1}^{{\rm new}_2,i}
=
\bra{\psi_{0}^{\rm equ}}
\ket{\psi_{1}^{\rm equ}}_{C_i'''}
$.
Thus, $\mathbf{U}_i$ exists if and only if
$
        \bra{\Phi}_{{\rm A,iso\,}0}^{{\rm new}_2,i}
        \ket{\Phi}_{{\rm A,iso\,}1}^{{\rm new}_2,i}
        =
        \braket
        {\psi_{0}^{\rm equ}}
        {\psi_{1}^{\rm equ}}_{C_i'''}
        .
$
Without loss of generality, we assume that
$
        \ket{\psi_{0}^{\rm equ}}_{C_i'''} = \ket{0}, \,
        \ket{\psi_{1}^{\rm equ}}_{C_i'''} = \ket{\mu_{\rm equ}},
$
where $\ket{0}$ is the vacuum state and $\ket{\mu_{\rm equ}}$ is the coherent state with an average number of photons equals to $\mu_{\rm equ}$.
Combine 
\cref{equ:inner_of_two_new2-2},
we can calculate that $\mu_{\rm equ}$ satisfies
\begin{equation}
    \label{equ:equ-two-SNS-mu}
    \begin{aligned}
        {\rm e}^{-\mu_{\rm equ}}
        =&
        \left[
            \sqrt{
                V_{0}^{{\rm A},1}
                V_{1}^{{\rm A},1}
            }
            -
            \sqrt{
                \left(
                    1-V_{0}^{{\rm A},1}
                \right)
                \left(
                    1-V_{1}^{{\rm A},1}
                \right)
            }
        \right]^2
    .
    \end{aligned}
\end{equation}
Furthermore, observing that in protocol $\ket{\Phi}_{\rm A}^{\rm equ}$, 
$r'$ and 
$a$ no longer play any role, we simplify the \textit{equivalent protocol} $\ket{\Phi}_{\rm A}^{\rm equ}$ by removing them. 
The final \textit{equivalent protocol} then satisfies
\begin{equation}
    \label{equ:equal-two-SNS1-2}
    \begin{aligned}
        \ket{\Phi}_{\rm A}^{\rm equ}= 
        \left[
            \sum_{\mathbf{r}_1^N}
            \left(
                \prod_{i=1}^N \sqrt{p_{r_i}}
            \right)
            \left(
                \bigotimes_{i=1}^N
                \ket{r_i}_{{A_i}}
                \ket{\psi_{r_i}^{\rm equ}}_{C_i'''}
            \right)
        \right]
        .
    \end{aligned}
\end{equation}
The above analysis similarly applies to more general values of $\xi$, and the detailed procedure can be found in the Supplementary Materials. 
Ultimately, this leads to a conclusion of the form presented in \cref{equ:equal-two-SNS1}.

\hfill

\noindent{\bf Phase error estimation of secure QKD against imperfect source.}
\noindent{\bf Phase error estimation of secure QKD against correlated leakage source.}

In the protocol described by Eq.~(\ref{equ:equal-two-SNS-final}), similar to the \textit{original protocol}, 
a $\mathcal{Z}$ event occurs when exactly one of Alice or Bob chooses $r_i = 0$,
an $\mathcal{O}$ event occurs when both Alice and Bob choose $r_i = 0$
and
a $\mathcal{B}$ event occurs when both Alice and Bob choose $r_i = 1$.
In this case, we define the phase error for a $\mathcal{Z}$ event as 
$
\ket{++}_{A_i B_i}
=
\left(
    \frac{\ket{0}_{A_i}+\ket{1}_{A_i}}{\sqrt{2}}
\right)
\otimes
\left(
    \frac{\ket{0}_{B_i}+\ket{1}_{B_i}}{\sqrt{2}}
\right)
$
and
$
\ket{--}_{A_i B_i}
=
\left(
    \frac{\ket{0}_{A_i}-\ket{1}_{A_i}}{\sqrt{2}}
\right)
\otimes
\left(
    \frac{\ket{0}_{B_i}-\ket{1}_{B_i}}{\sqrt{2}}
\right)
$. Specifically, a phase error is said to occur when Alice and Bob measure the state
$
\frac{\ket{++}_{A_i B_i}-\ket{--}_{A_i B_i}}{\sqrt{2}}
=
\frac{\ket{0}_{A_i}\ket{1}_{B_i}+\ket{1}_{A_i}\ket{0}_{B_i}}{\sqrt{2}}
$~\cite{PhysRevResearch.6.013266,shan2024improvedpostselectionsecurityanalysis}.



Since we do not assume Charlie to be trusted, we can further consider the case where Charlie is entirely controlled by Eve. 
In this scenario, Eve's attack and Charlie's measurement can be treated jointly. 
For simplicity, we refer to this combined operation as Eve's positive operator-valued measurement (PVOM) 
$\mathcal{M}^{\rm E}_{c_0^N}$
in the following discussion,
where $c_0^N = c_0 c_1 \ldots c_N$ denotes the measurement result announced by Charlie over total $n$ rounds,
$c_i \in \{0,1\}$ 
represents the measurement result in the $i$-th round, 
where 
$1$ indicates a successful measurement 
and $0$ indicates other measurement results, including successful measurement and conclusive results.
First, consider a collective attack. 
In this scenario, Eve's measurement can be represented as 
$
\mathcal{M}^{\rm E}_{c_0^N} 
= \bigotimes_{i=1}^{N}\mathcal{M}^{{\rm E},i}_{c_i} 
$,
where
$\mathcal{M}^{{\rm E},i}_{c_i}$ is a Eve's PVOM in the $i$-th round,
and $\forall i,j \in[0,N], \, {\rm if}\, c_i=c_j, \, \mathcal{M}^{{\rm E},i}_{c_i}=\mathcal{M}^{{\rm E},j}_{c_j}$.
It is worth noting that Eve does not have access to the encoding ancilla spaces of Alice and Bob. 
Therefore, the phase error probability $P_{\rm ph}$ can be shown as
\begin{equation}
    \label{equ:pph1}
    \begin{aligned}
        P_{\rm ph}
        &
        =
        \Tr
    \left(
        \left(
            \mathcal{P}
            \left[
                \frac{\ket{0}_{A_i}\ket{1}_{B_i}+\ket{1}_{A_i}\ket{0}_{B_i}}{\sqrt{2}}
            \right]
            \otimes
            \mathcal{M}^{{\rm E},i}_{1}
        \right)
            \otimes
            \mathcal{P}
            \left[
                \ket{\Phi}^{\rm equ}_i
            \right]
            \otimes
            \ket{0}\bra{0}_{{{\rm PE}_i}}
    \right)
        \\
        &
        =
        \frac{p_0 p_1 (1-p_{\rm PE})}{2}
        \Tr
        \left(
            \mathcal{M}^{{\rm E},i}_{1}
            \mathcal{P}
            \left[
                \ket{\psi_{0}^{\rm Aequ}}_{C_i^{\rm A}}
                \ket{\psi_{1}^{\rm Bequ}}_{C_i^{\rm B}}
                +
                \ket{\psi_{1}^{\rm Aequ}}_{C_i^{\rm A}}
                \ket{\psi_{0}^{\rm Bequ}}_{C_i^{\rm B}}
            \right]
        \right)
        ,
    \end{aligned}
\end{equation}
where $\mathcal{P}[\ket{\cdot}]=\ket{\cdot}\bra{\cdot}$ and
$
        \ket{\Phi}^{\rm equ}_i= 
        \left[
            \sum_{r_i\in \{0,1\}}
            \sqrt{p_{r_i}}
                \ket{r_i}_{{A_i}}
                \ket{\psi_{r_i}^{\rm Aequ}}_{C_i^{\rm A}}
        \right]
        \otimes
        \left[
            \sum_{r_i\in \{0,1\}}
                \sqrt{p_{r_i}}
                \ket{r_i}_{{B_i}}
                \ket{\psi_{r_i}^{\rm Bequ}}_{C_i^{\rm B}}
        \right]
        \otimes
        \left[
            \sum_{\mathbf{m}_i}
                \sqrt{p_{m_i}^{{\rm PE}}}
                \ket{m_i}_{{{\rm PE}_i}}
        \right]
$
denotes the entanglement-equivalent protocol in the $i$-th round, satisfies
$\mathcal{P}[\ket{\Phi}^{\rm equ}] = \bigotimes_{i=1}^N \mathcal{P}[\ket{\Phi}^{\rm equ}_i]$.
Similarly, we can calculate the probabilities of the $\mathcal{O}$ event and the $\mathcal{B}$ event, denote as $P_{\mathcal{O}}$ and $P_{\mathcal{B}}$, satisfy
\begin{equation}
    \label{equ:po}
    \begin{aligned}
        P_{\mathcal{O}}
        &
        =
        \Tr
    \left(
        \left(
            \mathcal{P}
            \left[
                \ket{0}_{A_i}\ket{0}_{B_i}
            \right]
            \otimes
            \mathcal{M}^{{\rm E},i}_{1}
        \right)
            \otimes
            \mathcal{P}
            \left[
                \ket{\Phi}^{\rm equ}_i
            \right]
    \right)
        \\
        &
        =
        (1-p_{\rm PE})
        \left(
            p_0
        \right)^2
        \Tr
        \left(
            \mathcal{M}^{{\rm E},i}_{1}
            \mathcal{P}
            \left[
                \ket{\psi_{0}^{\rm Aequ}}_{C_i^{\rm A}}
                \ket{\psi_{0}^{\rm Bequ}}_{C_i^{\rm B}}
            \right]
        \right)
        ,
\\
        P_{\mathcal{B}}
        &
        =
        \Tr
    \left(
        \left(
            \mathcal{P}
            \left[
                \ket{1}_{A_i}\ket{1}_{B_i}
            \right]
            \otimes
            \mathcal{M}^{{\rm E},i}_{1}
        \right)
            \otimes
            \mathcal{P}
            \left[
                \ket{\Phi}^{\rm equ}_i
            \right]
    \right)
        \\
        &
        =
        (1-p_{\rm PE})
        \left(
            p_1
        \right)^2
        \Tr
        \left(
            \mathcal{M}^{{\rm E},i}_{1}
            \mathcal{P}
            \left[
                \ket{\psi_{1}^{\rm Aequ}}_{C_i^{\rm A}}
                \ket{\psi_{1}^{\rm Bequ}}_{C_i^{\rm B}}
            \right]
        \right)
        .
    \end{aligned}
\end{equation}

As calculated in existing works \cite{PhysRevApplied.12.054034,PhysRevApplied.19.064003,PhysRevResearch.6.013266,shan2024improvedpostselectionsecurityanalysis}, 
the actually transmitted state satisfies
\begin{equation}
    \label{equ:equofstratesad}
    \begin{aligned}
        &
        \ket{\psi_{0}^{\rm Aequ}}_{C_i^{\rm A}}
        \ket{\psi_{1}^{\rm Bequ}}_{C_i^{\rm B}}
        +
        \ket{\psi_{1}^{\rm Aequ}}_{C_i^{\rm A}}
        \ket{\psi_{0}^{\rm Bequ}}_{C_i^{\rm B}}
        \\
        =
        &
        c_0
        \ket{\psi_{0}^{\rm Aequ}}_{C_i^{\rm A}}
        \ket{\psi_{0}^{\rm Bequ}}_{C_i^{\rm B}}
        +
        c_1
        \ket{\psi_{1}^{\rm Aequ}}_{C_i^{\rm A}}
        \ket{\psi_{1}^{\rm Bequ}}_{C_i^{\rm B}}
        +
        \bar c_2\ket{\phi_2}
        ,
    \end{aligned}
\end{equation}
from \cref{equ:equal-two-SNS-final}
where $c_0,c_1>0$, $c_0c_1=1$,
$
        \bar c_2
        =
        \sqrt{
            \left(
                c_0 + c_1
                -2 {\rm e}^{
                    -\frac{\mu_{\rm equ}^{\rm A}}{2}
                } 
            \right)
            \left(
                c_0 + c_1
                -2 {\rm e}^{
                    -\frac{\mu_{\rm equ}^{\rm B}}{2}
                } 
            \right)
        }
$
and $\ket{\phi_2}$ is a normalized state that keep Eq.~(\ref{equ:equofstratesad}) holds.
Although the values of 
$c_0$ and 
$c_1$ can be optimized to improve the protocol's performance, we assume 
$
c_0 = {\rm e}^{
    -\left(\mu_{\rm equ}^{\rm A}+\mu_{\rm equ}^{\rm B}\right)/{4}
} 
$ 
and
$
c_1 = 
1/{c_0}
$
to simplify the analysis.
For $\mathcal{M}^{{\rm E},i}_{1}$ is positive, we can expand its eigenvalues and rewrite as
$\mathcal{M}^{{\rm E},i}_{1} = \sum_{j} \mathcal{P}[\ket{1}^{{\rm E},i}_{j}] $,
where $\ket{1}^{{\rm E},i}_{j}$ is a non-normalized eigenvector and 
$\Tr \left(\mathcal{P}[\ket{1}^{{\rm E},i}_{j}]\right)$ is the corresponding eigenvalue,
satisfy $ \sum_j \Tr \left(\mathcal{P}[\ket{1}^{{\rm E},i}_{j}]\right) = \Tr \left(\mathcal{M}^{{\rm E},i}_{1}\right) \leq 1$.
Then
\cref{equ:po}
can be rewrite as 
\begin{equation}
    \label{equ:sdaeghjuysdadcw}
    \begin{aligned}
P_{\mathcal{O}}
=
(1-p_{\rm PE})
\left(
    p_0
\right)^2
\sum_j
\left|
    \bra{1}^{{\rm E},i}_{j}
        \ket{\psi_{0}^{\rm Aequ}}_{C_i^{\rm A}}
        \ket{\psi_{0}^{\rm Bequ}}_{C_i^{\rm B}}
\right|^2
        ,
        P_{\mathcal{B}}
        =
        (1-p_{\rm PE})
        \left(
            p_1
        \right)^2
        \sum_j
        \left|
            \bra{1}^{{\rm E},i}_{j}
                \ket{\psi_{1}^{\rm Aequ}}_{C_i^{\rm A}}
                \ket{\psi_{1}^{\rm Bequ}}_{C_i^{\rm B}}
        \right|^2
        .
    \end{aligned}
\end{equation}
And combine \cref{equ:pph1,equ:equofstratesad},
we can calculate that
\begin{equation}
    \label{equ:sdaeghjuy}
    \begin{aligned}
        \frac{
            2P_{\rm ph}
        }{
            p_0 p_1 (1-p_{\rm PE})
        }
        =
        &
        \sum_j
        \left|
            \bra{1}^{{\rm E},i}_{j}
            \left(
                c_0
                \ket{\psi_{0}^{\rm Aequ}}_{C_i^{\rm A}}
                \ket{\psi_{0}^{\rm Bequ}}_{C_i^{\rm B}}
                +
                c_1
                \ket{\psi_{1}^{\rm Aequ}}_{C_i^{\rm A}}
                \ket{\psi_{1}^{\rm Bequ}}_{C_i^{\rm B}}
                +
                \bar c_2\ket{\phi_2}
            \right)
        \right|^2
        \\
        \leq
        &
        \sum_j
        \Bigg(
            c_0^2
            \left|
                \bra{1}^{{\rm E},i}_{j}
                    \ket{\psi_{0}^{\rm Aequ}}_{C_i^{\rm A}}
                    \ket{\psi_{0}^{\rm Bequ}}_{C_i^{\rm B}}
            \right|^2
            +
            c_1^2
            \left|
                \bra{1}^{{\rm E},i}_{j}
                    \ket{\psi_{1}^{\rm Aequ}}_{C_i^{\rm A}}
                    \ket{\psi_{1}^{\rm Bequ}}_{C_i^{\rm B}}
            \right|^2
            +
            {\bar{c}_2}^2
            \left|
                \bra{1}^{{\rm E},i}_{j}
                    \bar c_2\ket{\phi_2}
            \right|^2
            \\
            & 
            \qquad \, \, \,
            + 2 c_0 c_1
            \left|
                \bra{1}^{{\rm E},i}_{j}
                    \ket{\psi_{0}^{\rm Aequ}}_{C_i^{\rm A}}
                    \ket{\psi_{0}^{\rm Bequ}}_{C_i^{\rm B}}
            \right|
            \left|
                \bra{1}^{{\rm E},i}_{j}
                    \ket{\psi_{1}^{\rm Aequ}}_{C_i^{\rm A}}
                    \ket{\psi_{1}^{\rm Bequ}}_{C_i^{\rm B}}
            \right|
            \\
            & 
            \qquad \, \, \,
            + 2 c_0 {\bar{c}_2}
            \left|
                \bra{1}^{{\rm E},i}_{j}
                    \ket{\psi_{0}^{\rm Aequ}}_{C_i^{\rm A}}
                    \ket{\psi_{0}^{\rm Bequ}}_{C_i^{\rm B}}
            \right|
            \left|
                \bra{1}^{{\rm E},i}_{j}
                    \bar c_2\ket{\phi_2}
            \right|
            \\
            & 
            \qquad \, \, \,
            + 2 c_1 {\bar{c}_2}
            \left|
                \bra{1}^{{\rm E},i}_{j}
                    \ket{\psi_{1}^{\rm Aequ}}_{C_i^{\rm A}}
                    \ket{\psi_{1}^{\rm Bequ}}_{C_i^{\rm B}}
            \right|
            \left|
                \bra{1}^{{\rm E},i}_{j}
                    \bar c_2\ket{\phi_2}
            \right|
        \Bigg)
        .
    \end{aligned}
\end{equation}
From Cauchy-Schwarz inequality, we can see that $(\sum_i x_i y_i)^2 \leq (\sum_i x_i)^2 (\sum_i y_i)^2 $,
then, combine with the fact that 
$
\sum_j
\left|
    \bra{1}^{{\rm E},i}_{j}
    \bar c_2\ket{\phi_2}
\right|^2
\leq
1
$,
from 
\cref{equ:sdaeghjuysdadcw,equ:sdaeghjuy},
we can calculate the upper bound of $P_{\rm ph}$, satisfies
\cite{PhysRevApplied.12.054034,shan2024improvedpostselectionsecurityanalysis}
\begin{equation}
    \label{equ:ephcakl2}
    \begin{aligned}
        P_{\rm ph}\leq& \frac{p_1p_0}{2}
        \left(
            c_0^2\frac{P_{\mathcal{O}}}{(p_0)^2}+c_1^2\frac{P_{\mathcal{B}}}{(p_1)^2}+\bar{c}_2^2
            +2c_0c_1\sqrt{\frac{P_\mathcal{O}P_\mathcal{B}}{(p_0)^2(p_1)^2}}+c_0\bar{c}_2\sqrt{\frac{P_\mathcal{O}}{(p_0)^2}}+c_1\bar{c}_2\sqrt{\frac{P_\mathcal{B}}{(p_1)^2}}
        \right),
    \end{aligned}
\end{equation}
and this lead us to \cref{equ:ephcakl}.

\hfill

\noindent{\bf DATA AVAILABILITY} 

\hfill

The data that support the findings of this study are available from the corresponding author upon reasonable request.

\hfill

\noindent{\bf ACKNOWLEDGEMENTS} 

\hfill

This work has been supported by the National Natural Science Foundation of China ( Grant No. 62271463, 62301524, 62105318, 61961136004, 62171424 ), 
the Fundamental Research Funds for the Central Universities, the China Postdoctoral Science Foundation ( Grant No. 2022M723064 ), 
Natural Science Foundation of Anhui ( No. 2308085QF216 ), the Innovation Program for Quantum Science and Technology ( Grant No. 2021ZD0300700 ),
and the Youth innovation Fundof the University of Science and Technology of China.
R. W is supported by the Hangzhou Dianzi University start-up grant.

\hfill

\noindent{\bf COMPETING INTERESTS} 

\hfill

The authors declare that they have no competing interests.

\hfill

\noindent{\bf AUTHOR CONTRIBUTIONS} 

\hfill

Jia-Xuan Li and Zhen-Qiang Yin start the project. 
Jia-Xuan Li, Rong Wang and Yang-Guang Shan provide the theoretical analysis of the security analysis framework.
Jia-Xuan Li and Yang-Guang Shan provide the security analysis of secure quantum key distribution against correlated leakage source.
Jia-Xuan Li and Zhen-Qiang Yin mainly prepared the manuscript.
Feng-Yu Lu, Zhen-Qiang Yin, Shuang Wang, Wei Chen, Guang-Can Guo and Zheng-Fu Han supervise the project.

\hfill

\noindent{\bf REFERENCES} 

\bibliography{1}

\begin{thebibliography}{59}%
\makeatletter
\providecommand \@ifxundefined [1]{%
 \@ifx{#1\undefined}
}%
\providecommand \@ifnum [1]{%
 \ifnum #1\expandafter \@firstoftwo
 \else \expandafter \@secondoftwo
 \fi
}%
\providecommand \@ifx [1]{%
 \ifx #1\expandafter \@firstoftwo
 \else \expandafter \@secondoftwo
 \fi
}%
\providecommand \natexlab [1]{#1}%
\providecommand \enquote  [1]{``#1''}%
\providecommand \bibnamefont  [1]{#1}%
\providecommand \bibfnamefont [1]{#1}%
\providecommand \citenamefont [1]{#1}%
\providecommand \href@noop [0]{\@secondoftwo}%
\providecommand \href [0]{\begingroup \@sanitize@url \@href}%
\providecommand \@href[1]{\@@startlink{#1}\@@href}%
\providecommand \@@href[1]{\endgroup#1\@@endlink}%
\providecommand \@sanitize@url [0]{\catcode `\\12\catcode `\$12\catcode
  `\&12\catcode `\#12\catcode `\^12\catcode `\_12\catcode `\%12\relax}%
\providecommand \@@startlink[1]{}%
\providecommand \@@endlink[0]{}%
\providecommand \url  [0]{\begingroup\@sanitize@url \@url }%
\providecommand \@url [1]{\endgroup\@href {#1}{\urlprefix }}%
\providecommand \urlprefix  [0]{URL }%
\providecommand \Eprint [0]{\href }%
\providecommand \doibase [0]{https://doi.org/}%
\providecommand \selectlanguage [0]{\@gobble}%
\providecommand \bibinfo  [0]{\@secondoftwo}%
\providecommand \bibfield  [0]{\@secondoftwo}%
\providecommand \translation [1]{[#1]}%
\providecommand \BibitemOpen [0]{}%
\providecommand \bibitemStop [0]{}%
\providecommand \bibitemNoStop [0]{.\EOS\space}%
\providecommand \EOS [0]{\spacefactor3000\relax}%
\providecommand \BibitemShut  [1]{\csname bibitem#1\endcsname}%
\let\auto@bib@innerbib\@empty
\bibitem [{\citenamefont {Shannon}(1948)}]{6773024}%
  \BibitemOpen
  \bibfield  {author} {\bibinfo {author} {\bibfnamefont {C.~E.}\ \bibnamefont
  {Shannon}},\ }\bibfield  {title} {\bibinfo {title} {A mathematical theory of
  communication},\ }\href {https://doi.org/10.1002/j.1538-7305.1948.tb01338.x}
  {\bibfield  {journal} {\bibinfo  {journal} {The Bell System Technical
  Journal}\ }\textbf {\bibinfo {volume} {27}},\ \bibinfo {pages} {379}
  (\bibinfo {year} {1948})}\BibitemShut {NoStop}%
\bibitem [{\citenamefont {Shor}(1994)}]{365700}%
  \BibitemOpen
  \bibfield  {author} {\bibinfo {author} {\bibfnamefont {P.}~\bibnamefont
  {Shor}},\ }\bibfield  {title} {\bibinfo {title} {Algorithms for quantum
  computation: discrete logarithms and factoring},\ }in\ \href
  {https://doi.org/10.1109/SFCS.1994.365700} {\emph {\bibinfo {booktitle}
  {Proceedings 35th Annual Symposium on Foundations of Computer Science}}}\
  (\bibinfo {year} {1994})\ pp.\ \bibinfo {pages} {124--134}\BibitemShut
  {NoStop}%
\bibitem [{\citenamefont {Stevens}\ \emph {et~al.}(2017)\citenamefont
  {Stevens}, \citenamefont {Bursztein}, \citenamefont {Karpman}, \citenamefont
  {Albertini},\ and\ \citenamefont {Markov}}]{cryptoeprint:2017/190}%
  \BibitemOpen
  \bibfield  {author} {\bibinfo {author} {\bibfnamefont {M.}~\bibnamefont
  {Stevens}}, \bibinfo {author} {\bibfnamefont {E.}~\bibnamefont {Bursztein}},
  \bibinfo {author} {\bibfnamefont {P.}~\bibnamefont {Karpman}}, \bibinfo
  {author} {\bibfnamefont {A.}~\bibnamefont {Albertini}},\ and\ \bibinfo
  {author} {\bibfnamefont {Y.}~\bibnamefont {Markov}},\ }\href
  {https://eprint.iacr.org/2017/190} {\bibinfo {title} {The first collision for
  full {SHA}-1}},\ \bibinfo {howpublished} {Cryptology {ePrint} Archive, Paper
  2017/190} (\bibinfo {year} {2017})\BibitemShut {NoStop}%
\bibitem [{\citenamefont {Bennett}\ and\ \citenamefont
  {Brassard}(1984)}]{BB84}%
  \BibitemOpen
  \bibfield  {author} {\bibinfo {author} {\bibfnamefont {C.~H.}\ \bibnamefont
  {Bennett}}\ and\ \bibinfo {author} {\bibfnamefont {G.}~\bibnamefont
  {Brassard}},\ }\bibfield  {title} {\bibinfo {title} {Quantum cryptography:
  public key distribution and coin tossing},\ }in\ \href@noop {} {\emph
  {\bibinfo {booktitle} {Conf. on Computers, Systems and Signal Processing}}}\
  (\bibinfo {address} {Bangalore},\ \bibinfo {year} {1984})\ p.\ \bibinfo
  {pages} {175}\BibitemShut {NoStop}%
\bibitem [{\citenamefont {Lo}\ and\ \citenamefont
  {Chau}(1999)}]{doi:10.1126/science.283.5410.2050}%
  \BibitemOpen
  \bibfield  {author} {\bibinfo {author} {\bibfnamefont {H.-K.}\ \bibnamefont
  {Lo}}\ and\ \bibinfo {author} {\bibfnamefont {H.~F.}\ \bibnamefont {Chau}},\
  }\bibfield  {title} {\bibinfo {title} {Unconditional security of quantum key
  distribution over arbitrarily long distances},\ }\href
  {https://doi.org/10.1126/science.283.5410.2050} {\bibfield  {journal}
  {\bibinfo  {journal} {Science}\ }\textbf {\bibinfo {volume} {283}},\ \bibinfo
  {pages} {2050} (\bibinfo {year} {1999})}\BibitemShut {NoStop}%
\bibitem [{\citenamefont {Shor}\ and\ \citenamefont
  {Preskill}(2000)}]{PhysRevLett.85.441}%
  \BibitemOpen
  \bibfield  {author} {\bibinfo {author} {\bibfnamefont {P.~W.}\ \bibnamefont
  {Shor}}\ and\ \bibinfo {author} {\bibfnamefont {J.}~\bibnamefont
  {Preskill}},\ }\bibfield  {title} {\bibinfo {title} {Simple proof of security
  of the {BB}84 quantum key distribution protocol},\ }\href
  {https://doi.org/10.1103/PhysRevLett.85.441} {\bibfield  {journal} {\bibinfo
  {journal} {Phys. Rev. Lett.}\ }\textbf {\bibinfo {volume} {85}},\ \bibinfo
  {pages} {441} (\bibinfo {year} {2000})}\BibitemShut {NoStop}%
\bibitem [{\citenamefont {Scarani}\ \emph {et~al.}(2009)\citenamefont
  {Scarani}, \citenamefont {Bechmann-Pasquinucci}, \citenamefont {Cerf},
  \citenamefont {Du\ifmmode~\check{s}\else \v{s}\fi{}ek}, \citenamefont
  {L\"utkenhaus},\ and\ \citenamefont {Peev}}]{RevModPhys.81.1301}%
  \BibitemOpen
  \bibfield  {author} {\bibinfo {author} {\bibfnamefont {V.}~\bibnamefont
  {Scarani}}, \bibinfo {author} {\bibfnamefont {H.}~\bibnamefont
  {Bechmann-Pasquinucci}}, \bibinfo {author} {\bibfnamefont {N.~J.}\
  \bibnamefont {Cerf}}, \bibinfo {author} {\bibfnamefont {M.}~\bibnamefont
  {Du\ifmmode~\check{s}\else \v{s}\fi{}ek}}, \bibinfo {author} {\bibfnamefont
  {N.}~\bibnamefont {L\"utkenhaus}},\ and\ \bibinfo {author} {\bibfnamefont
  {M.}~\bibnamefont {Peev}},\ }\bibfield  {title} {\bibinfo {title} {The
  security of practical quantum key distribution},\ }\href
  {https://doi.org/10.1103/RevModPhys.81.1301} {\bibfield  {journal} {\bibinfo
  {journal} {Rev. Mod. Phys.}\ }\textbf {\bibinfo {volume} {81}},\ \bibinfo
  {pages} {1301} (\bibinfo {year} {2009})}\BibitemShut {NoStop}%
\bibitem [{\citenamefont {Renner}(2008)}]{doi:10.1142/S0219749908003256}%
  \BibitemOpen
  \bibfield  {author} {\bibinfo {author} {\bibfnamefont {R.}~\bibnamefont
  {Renner}},\ }\bibfield  {title} {\bibinfo {title} {Security of quantum key
  distribution},\ }\href {https://doi.org/10.1142/S0219749908003256} {\bibfield
   {journal} {\bibinfo  {journal} {International Journal of Quantum
  Information}\ }\textbf {\bibinfo {volume} {06}},\ \bibinfo {pages} {1}
  (\bibinfo {year} {2008})}\BibitemShut {NoStop}%
\bibitem [{\citenamefont {Braunstein}\ and\ \citenamefont
  {Pirandola}(2012)}]{PhysRevLett.108.130502}%
  \BibitemOpen
  \bibfield  {author} {\bibinfo {author} {\bibfnamefont {S.~L.}\ \bibnamefont
  {Braunstein}}\ and\ \bibinfo {author} {\bibfnamefont {S.}~\bibnamefont
  {Pirandola}},\ }\bibfield  {title} {\bibinfo {title} {Side-channel-free
  quantum key distribution},\ }\href
  {https://doi.org/10.1103/PhysRevLett.108.130502} {\bibfield  {journal}
  {\bibinfo  {journal} {Phys. Rev. Lett.}\ }\textbf {\bibinfo {volume} {108}},\
  \bibinfo {pages} {130502} (\bibinfo {year} {2012})}\BibitemShut {NoStop}%
\bibitem [{\citenamefont {Lo}\ \emph {et~al.}(2012)\citenamefont {Lo},
  \citenamefont {Curty},\ and\ \citenamefont {Qi}}]{PhysRevLett.108.130503}%
  \BibitemOpen
  \bibfield  {author} {\bibinfo {author} {\bibfnamefont {H.-K.}\ \bibnamefont
  {Lo}}, \bibinfo {author} {\bibfnamefont {M.}~\bibnamefont {Curty}},\ and\
  \bibinfo {author} {\bibfnamefont {B.}~\bibnamefont {Qi}},\ }\bibfield
  {title} {\bibinfo {title} {Measurement-device-independent quantum key
  distribution},\ }\href {https://doi.org/10.1103/PhysRevLett.108.130503}
  {\bibfield  {journal} {\bibinfo  {journal} {Phys. Rev. Lett.}\ }\textbf
  {\bibinfo {volume} {108}},\ \bibinfo {pages} {130503} (\bibinfo {year}
  {2012})}\BibitemShut {NoStop}%
\bibitem [{\citenamefont {Curty}\ \emph {et~al.}(2014)\citenamefont {Curty},
  \citenamefont {Xu}, \citenamefont {Cui}, \citenamefont {Lim}, \citenamefont
  {Tamaki},\ and\ \citenamefont {Lo}}]{Curty2014}%
  \BibitemOpen
  \bibfield  {author} {\bibinfo {author} {\bibfnamefont {M.}~\bibnamefont
  {Curty}}, \bibinfo {author} {\bibfnamefont {F.}~\bibnamefont {Xu}}, \bibinfo
  {author} {\bibfnamefont {W.}~\bibnamefont {Cui}}, \bibinfo {author}
  {\bibfnamefont {C.~C.~W.}\ \bibnamefont {Lim}}, \bibinfo {author}
  {\bibfnamefont {K.}~\bibnamefont {Tamaki}},\ and\ \bibinfo {author}
  {\bibfnamefont {H.-K.}\ \bibnamefont {Lo}},\ }\bibfield  {title} {\bibinfo
  {title} {Finite-key analysis for measurement-device-independent quantum key
  distribution},\ }\href {https://doi.org/10.1038/ncomms4732} {\bibfield
  {journal} {\bibinfo  {journal} {Nature Communications}\ }\textbf {\bibinfo
  {volume} {5}},\ \bibinfo {pages} {3732} (\bibinfo {year} {2014})}\BibitemShut
  {NoStop}%
\bibitem [{\citenamefont {Lucamarini}\ \emph {et~al.}(2018)\citenamefont
  {Lucamarini}, \citenamefont {Yuan}, \citenamefont {Dynes},\ and\
  \citenamefont {Shields}}]{Lucamarini2018}%
  \BibitemOpen
  \bibfield  {author} {\bibinfo {author} {\bibfnamefont {M.}~\bibnamefont
  {Lucamarini}}, \bibinfo {author} {\bibfnamefont {Z.~L.}\ \bibnamefont
  {Yuan}}, \bibinfo {author} {\bibfnamefont {J.~F.}\ \bibnamefont {Dynes}},\
  and\ \bibinfo {author} {\bibfnamefont {A.~J.}\ \bibnamefont {Shields}},\
  }\bibfield  {title} {\bibinfo {title} {Overcoming the rate--distance limit of
  quantum key distribution without quantum repeaters},\ }\href
  {https://doi.org/10.1038/s41586-018-0066-6} {\bibfield  {journal} {\bibinfo
  {journal} {Nature}\ }\textbf {\bibinfo {volume} {557}},\ \bibinfo {pages}
  {400} (\bibinfo {year} {2018})}\BibitemShut {NoStop}%
\bibitem [{\citenamefont {Ma}\ \emph {et~al.}(2018)\citenamefont {Ma},
  \citenamefont {Zeng},\ and\ \citenamefont {Zhou}}]{PhysRevX.8.031043}%
  \BibitemOpen
  \bibfield  {author} {\bibinfo {author} {\bibfnamefont {X.}~\bibnamefont
  {Ma}}, \bibinfo {author} {\bibfnamefont {P.}~\bibnamefont {Zeng}},\ and\
  \bibinfo {author} {\bibfnamefont {H.}~\bibnamefont {Zhou}},\ }\bibfield
  {title} {\bibinfo {title} {Phase-matching quantum key distribution},\ }\href
  {https://doi.org/10.1103/PhysRevX.8.031043} {\bibfield  {journal} {\bibinfo
  {journal} {Phys. Rev. X}\ }\textbf {\bibinfo {volume} {8}},\ \bibinfo {pages}
  {031043} (\bibinfo {year} {2018})}\BibitemShut {NoStop}%
\bibitem [{\citenamefont {Wang}\ \emph {et~al.}(2018)\citenamefont {Wang},
  \citenamefont {Yu},\ and\ \citenamefont {Hu}}]{PhysRevA.98.062323}%
  \BibitemOpen
  \bibfield  {author} {\bibinfo {author} {\bibfnamefont {X.-B.}\ \bibnamefont
  {Wang}}, \bibinfo {author} {\bibfnamefont {Z.-W.}\ \bibnamefont {Yu}},\ and\
  \bibinfo {author} {\bibfnamefont {X.-L.}\ \bibnamefont {Hu}},\ }\bibfield
  {title} {\bibinfo {title} {Twin-field quantum key distribution with large
  misalignment error},\ }\href {https://doi.org/10.1103/PhysRevA.98.062323}
  {\bibfield  {journal} {\bibinfo  {journal} {Phys. Rev. A}\ }\textbf {\bibinfo
  {volume} {98}},\ \bibinfo {pages} {062323} (\bibinfo {year}
  {2018})}\BibitemShut {NoStop}%
\bibitem [{\citenamefont {Cui}\ \emph {et~al.}(2019)\citenamefont {Cui},
  \citenamefont {Yin}, \citenamefont {Wang}, \citenamefont {Chen},
  \citenamefont {Wang}, \citenamefont {Guo},\ and\ \citenamefont
  {Han}}]{PhysRevApplied.11.034053}%
  \BibitemOpen
  \bibfield  {author} {\bibinfo {author} {\bibfnamefont {C.}~\bibnamefont
  {Cui}}, \bibinfo {author} {\bibfnamefont {Z.-Q.}\ \bibnamefont {Yin}},
  \bibinfo {author} {\bibfnamefont {R.}~\bibnamefont {Wang}}, \bibinfo {author}
  {\bibfnamefont {W.}~\bibnamefont {Chen}}, \bibinfo {author} {\bibfnamefont
  {S.}~\bibnamefont {Wang}}, \bibinfo {author} {\bibfnamefont {G.-C.}\
  \bibnamefont {Guo}},\ and\ \bibinfo {author} {\bibfnamefont {Z.-F.}\
  \bibnamefont {Han}},\ }\bibfield  {title} {\bibinfo {title} {Twin-field
  quantum key distribution without phase postselection},\ }\href
  {https://doi.org/10.1103/PhysRevApplied.11.034053} {\bibfield  {journal}
  {\bibinfo  {journal} {Phys. Rev. Applied}\ }\textbf {\bibinfo {volume}
  {11}},\ \bibinfo {pages} {034053} (\bibinfo {year} {2019})}\BibitemShut
  {NoStop}%
\bibitem [{\citenamefont {Zeng}\ \emph {et~al.}(2022)\citenamefont {Zeng},
  \citenamefont {Zhou}, \citenamefont {Wu},\ and\ \citenamefont
  {Ma}}]{Zeng2022}%
  \BibitemOpen
  \bibfield  {author} {\bibinfo {author} {\bibfnamefont {P.}~\bibnamefont
  {Zeng}}, \bibinfo {author} {\bibfnamefont {H.}~\bibnamefont {Zhou}}, \bibinfo
  {author} {\bibfnamefont {W.}~\bibnamefont {Wu}},\ and\ \bibinfo {author}
  {\bibfnamefont {X.}~\bibnamefont {Ma}},\ }\bibfield  {title} {\bibinfo
  {title} {Mode-pairing quantum key distribution},\ }\href
  {https://doi.org/10.1038/s41467-022-31534-7} {\bibfield  {journal} {\bibinfo
  {journal} {Nature Communications}\ }\textbf {\bibinfo {volume} {13}},\
  \bibinfo {pages} {3903} (\bibinfo {year} {2022})}\BibitemShut {NoStop}%
\bibitem [{\citenamefont {Pereira}\ \emph {et~al.}(2020)\citenamefont
  {Pereira}, \citenamefont {Kato}, \citenamefont {Mizutani}, \citenamefont
  {Curty},\ and\ \citenamefont {Tamaki}}]{aaz4487}%
  \BibitemOpen
  \bibfield  {author} {\bibinfo {author} {\bibfnamefont {M.}~\bibnamefont
  {Pereira}}, \bibinfo {author} {\bibfnamefont {G.}~\bibnamefont {Kato}},
  \bibinfo {author} {\bibfnamefont {A.}~\bibnamefont {Mizutani}}, \bibinfo
  {author} {\bibfnamefont {M.}~\bibnamefont {Curty}},\ and\ \bibinfo {author}
  {\bibfnamefont {K.}~\bibnamefont {Tamaki}},\ }\bibfield  {title} {\bibinfo
  {title} {Quantum key distribution with correlated sources},\ }\href
  {https://doi.org/10.1126/sciadv.aaz4487} {\bibfield  {journal} {\bibinfo
  {journal} {Science Advances}\ }\textbf {\bibinfo {volume} {6}},\ \bibinfo
  {pages} {eaaz4487} (\bibinfo {year} {2020})},\ \Eprint
  {https://arxiv.org/abs/https://www.science.org/doi/pdf/10.1126/sciadv.aaz4487}
  {https://www.science.org/doi/pdf/10.1126/sciadv.aaz4487} \BibitemShut
  {NoStop}%
\bibitem [{\citenamefont {Tamaki}\ \emph {et~al.}(2014)\citenamefont {Tamaki},
  \citenamefont {Curty}, \citenamefont {Kato}, \citenamefont {Lo},\ and\
  \citenamefont {Azuma}}]{PhysRevA.90.052314}%
  \BibitemOpen
  \bibfield  {author} {\bibinfo {author} {\bibfnamefont {K.}~\bibnamefont
  {Tamaki}}, \bibinfo {author} {\bibfnamefont {M.}~\bibnamefont {Curty}},
  \bibinfo {author} {\bibfnamefont {G.}~\bibnamefont {Kato}}, \bibinfo {author}
  {\bibfnamefont {H.-K.}\ \bibnamefont {Lo}},\ and\ \bibinfo {author}
  {\bibfnamefont {K.}~\bibnamefont {Azuma}},\ }\bibfield  {title} {\bibinfo
  {title} {Loss-tolerant quantum cryptography with imperfect sources},\ }\href
  {https://doi.org/10.1103/PhysRevA.90.052314} {\bibfield  {journal} {\bibinfo
  {journal} {Phys. Rev. A}\ }\textbf {\bibinfo {volume} {90}},\ \bibinfo
  {pages} {052314} (\bibinfo {year} {2014})}\BibitemShut {NoStop}%
\bibitem [{\citenamefont {Yin}\ \emph {et~al.}(2014)\citenamefont {Yin},
  \citenamefont {Fung}, \citenamefont {Ma}, \citenamefont {Zhang},
  \citenamefont {Li}, \citenamefont {Chen}, \citenamefont {Wang}, \citenamefont
  {Guo},\ and\ \citenamefont {Han}}]{PhysRevA.90.052319}%
  \BibitemOpen
  \bibfield  {author} {\bibinfo {author} {\bibfnamefont {Z.-Q.}\ \bibnamefont
  {Yin}}, \bibinfo {author} {\bibfnamefont {C.-H.~F.}\ \bibnamefont {Fung}},
  \bibinfo {author} {\bibfnamefont {X.}~\bibnamefont {Ma}}, \bibinfo {author}
  {\bibfnamefont {C.-M.}\ \bibnamefont {Zhang}}, \bibinfo {author}
  {\bibfnamefont {H.-W.}\ \bibnamefont {Li}}, \bibinfo {author} {\bibfnamefont
  {W.}~\bibnamefont {Chen}}, \bibinfo {author} {\bibfnamefont {S.}~\bibnamefont
  {Wang}}, \bibinfo {author} {\bibfnamefont {G.-C.}\ \bibnamefont {Guo}},\ and\
  \bibinfo {author} {\bibfnamefont {Z.-F.}\ \bibnamefont {Han}},\ }\bibfield
  {title} {\bibinfo {title} {Mismatched-basis statistics enable quantum key
  distribution with uncharacterized qubit sources},\ }\href
  {https://doi.org/10.1103/PhysRevA.90.052319} {\bibfield  {journal} {\bibinfo
  {journal} {Phys. Rev. A}\ }\textbf {\bibinfo {volume} {90}},\ \bibinfo
  {pages} {052319} (\bibinfo {year} {2014})}\BibitemShut {NoStop}%
\bibitem [{\citenamefont {Gottesman}\ \emph {et~al.}(2004)\citenamefont
  {Gottesman}, \citenamefont {Lo}, \citenamefont {Lutkenhaus},\ and\
  \citenamefont {Preskill}}]{1365172}%
  \BibitemOpen
  \bibfield  {author} {\bibinfo {author} {\bibfnamefont {D.}~\bibnamefont
  {Gottesman}}, \bibinfo {author} {\bibfnamefont {H.-K.}\ \bibnamefont {Lo}},
  \bibinfo {author} {\bibfnamefont {N.}~\bibnamefont {Lutkenhaus}},\ and\
  \bibinfo {author} {\bibfnamefont {J.}~\bibnamefont {Preskill}},\ }\bibfield
  {title} {\bibinfo {title} {Security of quantum key distribution with
  imperfect devices},\ }in\ \href {https://doi.org/10.1109/ISIT.2004.1365172}
  {\emph {\bibinfo {booktitle} {International Symposium onInformation Theory,
  2004. ISIT 2004. Proceedings.}}}\ (\bibinfo {year} {2004})\ pp.\ \bibinfo
  {pages} {136--}\BibitemShut {NoStop}%
\bibitem [{\citenamefont {Lo}\ and\ \citenamefont
  {Preskill}(2007)}]{10.5555/2011832.2011834}%
  \BibitemOpen
  \bibfield  {author} {\bibinfo {author} {\bibfnamefont {H.-K.}\ \bibnamefont
  {Lo}}\ and\ \bibinfo {author} {\bibfnamefont {J.}~\bibnamefont {Preskill}},\
  }\bibfield  {title} {\bibinfo {title} {Security of quantum key distribution
  using weak coherent states with nonrandom phases},\ }\href@noop {} {\bibfield
   {journal} {\bibinfo  {journal} {Quantum Info. Comput.}\ }\textbf {\bibinfo
  {volume} {7}},\ \bibinfo {pages} {431–458} (\bibinfo {year}
  {2007})}\BibitemShut {NoStop}%
\bibitem [{\citenamefont {Koashi}(2009)}]{Koashi_2009}%
  \BibitemOpen
  \bibfield  {author} {\bibinfo {author} {\bibfnamefont {M.}~\bibnamefont
  {Koashi}},\ }\bibfield  {title} {\bibinfo {title} {Simple security proof of
  quantum key distribution based on complementarity},\ }\href
  {https://doi.org/10.1088/1367-2630/11/4/045018} {\bibfield  {journal}
  {\bibinfo  {journal} {New Journal of Physics}\ }\textbf {\bibinfo {volume}
  {11}},\ \bibinfo {pages} {045018} (\bibinfo {year} {2009})}\BibitemShut
  {NoStop}%
\bibitem [{\citenamefont {Coles}\ \emph {et~al.}(2016)\citenamefont {Coles},
  \citenamefont {Metodiev},\ and\ \citenamefont {L{\"u}tkenhaus}}]{Coles2016}%
  \BibitemOpen
  \bibfield  {author} {\bibinfo {author} {\bibfnamefont {P.~J.}\ \bibnamefont
  {Coles}}, \bibinfo {author} {\bibfnamefont {E.~M.}\ \bibnamefont
  {Metodiev}},\ and\ \bibinfo {author} {\bibfnamefont {N.}~\bibnamefont
  {L{\"u}tkenhaus}},\ }\bibfield  {title} {\bibinfo {title} {Numerical approach
  for unstructured quantum key distribution},\ }\href
  {https://doi.org/10.1038/ncomms11712} {\bibfield  {journal} {\bibinfo
  {journal} {Nature Communications}\ }\textbf {\bibinfo {volume} {7}},\
  \bibinfo {pages} {11712} (\bibinfo {year} {2016})}\BibitemShut {NoStop}%
\bibitem [{\citenamefont {Wang}\ \emph
  {et~al.}(2019{\natexlab{a}})\citenamefont {Wang}, \citenamefont
  {Primaatmaja}, \citenamefont {Lavie}, \citenamefont {Varvitsiotis},\ and\
  \citenamefont {Lim}}]{Wang2019}%
  \BibitemOpen
  \bibfield  {author} {\bibinfo {author} {\bibfnamefont {Y.}~\bibnamefont
  {Wang}}, \bibinfo {author} {\bibfnamefont {I.~W.}\ \bibnamefont
  {Primaatmaja}}, \bibinfo {author} {\bibfnamefont {E.}~\bibnamefont {Lavie}},
  \bibinfo {author} {\bibfnamefont {A.}~\bibnamefont {Varvitsiotis}},\ and\
  \bibinfo {author} {\bibfnamefont {C.~C.~W.}\ \bibnamefont {Lim}},\ }\bibfield
   {title} {\bibinfo {title} {Characterising the correlations of
  prepare-and-measure quantum networks},\ }\href
  {https://doi.org/10.1038/s41534-019-0133-3} {\bibfield  {journal} {\bibinfo
  {journal} {npj Quantum Information}\ }\textbf {\bibinfo {volume} {5}},\
  \bibinfo {pages} {17} (\bibinfo {year} {2019}{\natexlab{a}})}\BibitemShut
  {NoStop}%
\bibitem [{\citenamefont {Winick}\ \emph {et~al.}(2018)\citenamefont {Winick},
  \citenamefont {L{\"{u}}tkenhaus},\ and\ \citenamefont
  {Coles}}]{Winick2018reliablenumerical}%
  \BibitemOpen
  \bibfield  {author} {\bibinfo {author} {\bibfnamefont {A.}~\bibnamefont
  {Winick}}, \bibinfo {author} {\bibfnamefont {N.}~\bibnamefont
  {L{\"{u}}tkenhaus}},\ and\ \bibinfo {author} {\bibfnamefont {P.~J.}\
  \bibnamefont {Coles}},\ }\bibfield  {title} {\bibinfo {title} {Reliable
  numerical key rates for quantum key distribution},\ }\href
  {https://doi.org/10.22331/q-2018-07-26-77} {\bibfield  {journal} {\bibinfo
  {journal} {{Quantum}}\ }\textbf {\bibinfo {volume} {2}},\ \bibinfo {pages}
  {77} (\bibinfo {year} {2018})}\BibitemShut {NoStop}%
\bibitem [{\citenamefont {Pereira}\ \emph {et~al.}(2019)\citenamefont
  {Pereira}, \citenamefont {Curty},\ and\ \citenamefont
  {Tamaki}}]{Pereira2019}%
  \BibitemOpen
  \bibfield  {author} {\bibinfo {author} {\bibfnamefont {M.}~\bibnamefont
  {Pereira}}, \bibinfo {author} {\bibfnamefont {M.}~\bibnamefont {Curty}},\
  and\ \bibinfo {author} {\bibfnamefont {K.}~\bibnamefont {Tamaki}},\
  }\bibfield  {title} {\bibinfo {title} {Quantum key distribution with flawed
  and leaky sources},\ }\href {https://doi.org/10.1038/s41534-019-0180-9}
  {\bibfield  {journal} {\bibinfo  {journal} {npj Quantum Information}\
  }\textbf {\bibinfo {volume} {5}},\ \bibinfo {pages} {62} (\bibinfo {year}
  {2019})}\BibitemShut {NoStop}%
\bibitem [{\citenamefont {Sixto}\ \emph {et~al.}(2025)\citenamefont {Sixto},
  \citenamefont {Álvaro Navarrete}, \citenamefont {Pereira}, \citenamefont
  {Currás-Lorenzo}, \citenamefont {Tamaki},\ and\ \citenamefont
  {Curty}}]{sixto2025quantumkeydistributionimperfectly}%
  \BibitemOpen
  \bibfield  {author} {\bibinfo {author} {\bibfnamefont {X.}~\bibnamefont
  {Sixto}}, \bibinfo {author} {\bibnamefont {Álvaro Navarrete}}, \bibinfo
  {author} {\bibfnamefont {M.}~\bibnamefont {Pereira}}, \bibinfo {author}
  {\bibfnamefont {G.}~\bibnamefont {Currás-Lorenzo}}, \bibinfo {author}
  {\bibfnamefont {K.}~\bibnamefont {Tamaki}},\ and\ \bibinfo {author}
  {\bibfnamefont {M.}~\bibnamefont {Curty}},\ }\href
  {https://arxiv.org/abs/2411.13948} {\bibinfo {title} {Quantum key
  distribution with imperfectly isolated devices}} (\bibinfo {year} {2025}),\
  \Eprint {https://arxiv.org/abs/2411.13948} {arXiv:2411.13948 [quant-ph]}
  \BibitemShut {NoStop}%
\bibitem [{\citenamefont {Wang}\ \emph
  {et~al.}(2019{\natexlab{b}})\citenamefont {Wang}, \citenamefont {Hu},\ and\
  \citenamefont {Yu}}]{PhysRevApplied.12.054034}%
  \BibitemOpen
  \bibfield  {author} {\bibinfo {author} {\bibfnamefont {X.-B.}\ \bibnamefont
  {Wang}}, \bibinfo {author} {\bibfnamefont {X.-L.}\ \bibnamefont {Hu}},\ and\
  \bibinfo {author} {\bibfnamefont {Z.-W.}\ \bibnamefont {Yu}},\ }\bibfield
  {title} {\bibinfo {title} {Practical long-distance side-channel-free quantum
  key distribution},\ }\href {https://doi.org/10.1103/PhysRevApplied.12.054034}
  {\bibfield  {journal} {\bibinfo  {journal} {Phys. Rev. Appl.}\ }\textbf
  {\bibinfo {volume} {12}},\ \bibinfo {pages} {054034} (\bibinfo {year}
  {2019}{\natexlab{b}})}\BibitemShut {NoStop}%
\bibitem [{\citenamefont {Jiang}\ \emph {et~al.}(2023)\citenamefont {Jiang},
  \citenamefont {Yu}, \citenamefont {Hu},\ and\ \citenamefont
  {Wang}}]{PhysRevApplied.19.064003}%
  \BibitemOpen
  \bibfield  {author} {\bibinfo {author} {\bibfnamefont {C.}~\bibnamefont
  {Jiang}}, \bibinfo {author} {\bibfnamefont {Z.-W.}\ \bibnamefont {Yu}},
  \bibinfo {author} {\bibfnamefont {X.-L.}\ \bibnamefont {Hu}},\ and\ \bibinfo
  {author} {\bibfnamefont {X.-B.}\ \bibnamefont {Wang}},\ }\bibfield  {title}
  {\bibinfo {title} {Side-channel-secure quantum key distribution with
  imperfect vacuum sources},\ }\href
  {https://doi.org/10.1103/PhysRevApplied.19.064003} {\bibfield  {journal}
  {\bibinfo  {journal} {Phys. Rev. Appl.}\ }\textbf {\bibinfo {volume} {19}},\
  \bibinfo {pages} {064003} (\bibinfo {year} {2023})}\BibitemShut {NoStop}%
\bibitem [{\citenamefont {Jiang}\ \emph {et~al.}(2024)\citenamefont {Jiang},
  \citenamefont {Hu}, \citenamefont {Yu},\ and\ \citenamefont
  {Wang}}]{PhysRevResearch.6.013266}%
  \BibitemOpen
  \bibfield  {author} {\bibinfo {author} {\bibfnamefont {C.}~\bibnamefont
  {Jiang}}, \bibinfo {author} {\bibfnamefont {X.-L.}\ \bibnamefont {Hu}},
  \bibinfo {author} {\bibfnamefont {Z.-W.}\ \bibnamefont {Yu}},\ and\ \bibinfo
  {author} {\bibfnamefont {X.-B.}\ \bibnamefont {Wang}},\ }\bibfield  {title}
  {\bibinfo {title} {Side-channel security of practical quantum key
  distribution},\ }\href {https://doi.org/10.1103/PhysRevResearch.6.013266}
  {\bibfield  {journal} {\bibinfo  {journal} {Phys. Rev. Res.}\ }\textbf
  {\bibinfo {volume} {6}},\ \bibinfo {pages} {013266} (\bibinfo {year}
  {2024})}\BibitemShut {NoStop}%
\bibitem [{\citenamefont {Shan}\ \emph {et~al.}(2024)\citenamefont {Shan},
  \citenamefont {Yin}, \citenamefont {Wang}, \citenamefont {Chen},
  \citenamefont {He}, \citenamefont {Guo},\ and\ \citenamefont
  {Han}}]{shan2024improvedpostselectionsecurityanalysis}%
  \BibitemOpen
  \bibfield  {author} {\bibinfo {author} {\bibfnamefont {Y.-G.}\ \bibnamefont
  {Shan}}, \bibinfo {author} {\bibfnamefont {Z.-Q.}\ \bibnamefont {Yin}},
  \bibinfo {author} {\bibfnamefont {S.}~\bibnamefont {Wang}}, \bibinfo {author}
  {\bibfnamefont {W.}~\bibnamefont {Chen}}, \bibinfo {author} {\bibfnamefont
  {D.-Y.}\ \bibnamefont {He}}, \bibinfo {author} {\bibfnamefont {G.-C.}\
  \bibnamefont {Guo}},\ and\ \bibinfo {author} {\bibfnamefont {Z.-F.}\
  \bibnamefont {Han}},\ }\href {https://arxiv.org/abs/2409.19538} {\bibinfo
  {title} {Improved postselection security analysis of phase error estimation
  in quantum key distribution}} (\bibinfo {year} {2024}),\ \Eprint
  {https://arxiv.org/abs/2409.19538} {arXiv:2409.19538 [quant-ph]} \BibitemShut
  {NoStop}%
\bibitem [{\citenamefont {Ac\'{\i}n}\ \emph {et~al.}(2007)\citenamefont
  {Ac\'{\i}n}, \citenamefont {Brunner}, \citenamefont {Gisin}, \citenamefont
  {Massar}, \citenamefont {Pironio},\ and\ \citenamefont
  {Scarani}}]{PhysRevLett.98.230501}%
  \BibitemOpen
  \bibfield  {author} {\bibinfo {author} {\bibfnamefont {A.}~\bibnamefont
  {Ac\'{\i}n}}, \bibinfo {author} {\bibfnamefont {N.}~\bibnamefont {Brunner}},
  \bibinfo {author} {\bibfnamefont {N.}~\bibnamefont {Gisin}}, \bibinfo
  {author} {\bibfnamefont {S.}~\bibnamefont {Massar}}, \bibinfo {author}
  {\bibfnamefont {S.}~\bibnamefont {Pironio}},\ and\ \bibinfo {author}
  {\bibfnamefont {V.}~\bibnamefont {Scarani}},\ }\bibfield  {title} {\bibinfo
  {title} {Device-independent security of quantum cryptography against
  collective attacks},\ }\href {https://doi.org/10.1103/PhysRevLett.98.230501}
  {\bibfield  {journal} {\bibinfo  {journal} {Phys. Rev. Lett.}\ }\textbf
  {\bibinfo {volume} {98}},\ \bibinfo {pages} {230501} (\bibinfo {year}
  {2007})}\BibitemShut {NoStop}%
\bibitem [{\citenamefont {Zhang}\ \emph {et~al.}(2022)\citenamefont {Zhang},
  \citenamefont {Hu}, \citenamefont {Jiang}, \citenamefont {Chen},
  \citenamefont {Liu}, \citenamefont {Zhang}, \citenamefont {Yu}, \citenamefont
  {Li}, \citenamefont {You}, \citenamefont {Wang}, \citenamefont {Wang},
  \citenamefont {Zhang},\ and\ \citenamefont {Pan}}]{PhysRevLett.128.190503}%
  \BibitemOpen
  \bibfield  {author} {\bibinfo {author} {\bibfnamefont {C.}~\bibnamefont
  {Zhang}}, \bibinfo {author} {\bibfnamefont {X.-L.}\ \bibnamefont {Hu}},
  \bibinfo {author} {\bibfnamefont {C.}~\bibnamefont {Jiang}}, \bibinfo
  {author} {\bibfnamefont {J.-P.}\ \bibnamefont {Chen}}, \bibinfo {author}
  {\bibfnamefont {Y.}~\bibnamefont {Liu}}, \bibinfo {author} {\bibfnamefont
  {W.}~\bibnamefont {Zhang}}, \bibinfo {author} {\bibfnamefont {Z.-W.}\
  \bibnamefont {Yu}}, \bibinfo {author} {\bibfnamefont {H.}~\bibnamefont {Li}},
  \bibinfo {author} {\bibfnamefont {L.}~\bibnamefont {You}}, \bibinfo {author}
  {\bibfnamefont {Z.}~\bibnamefont {Wang}}, \bibinfo {author} {\bibfnamefont
  {X.-B.}\ \bibnamefont {Wang}}, \bibinfo {author} {\bibfnamefont
  {Q.}~\bibnamefont {Zhang}},\ and\ \bibinfo {author} {\bibfnamefont {J.-W.}\
  \bibnamefont {Pan}},\ }\bibfield  {title} {\bibinfo {title} {Experimental
  side-channel-secure quantum key distribution},\ }\href
  {https://doi.org/10.1103/PhysRevLett.128.190503} {\bibfield  {journal}
  {\bibinfo  {journal} {Phys. Rev. Lett.}\ }\textbf {\bibinfo {volume} {128}},\
  \bibinfo {pages} {190503} (\bibinfo {year} {2022})}\BibitemShut {NoStop}%
\bibitem [{\citenamefont {Nagamatsu}\ \emph {et~al.}(2016)\citenamefont
  {Nagamatsu}, \citenamefont {Mizutani}, \citenamefont {Ikuta}, \citenamefont
  {Yamamoto}, \citenamefont {Imoto},\ and\ \citenamefont
  {Tamaki}}]{PhysRevA.93.042325}%
  \BibitemOpen
  \bibfield  {author} {\bibinfo {author} {\bibfnamefont {Y.}~\bibnamefont
  {Nagamatsu}}, \bibinfo {author} {\bibfnamefont {A.}~\bibnamefont {Mizutani}},
  \bibinfo {author} {\bibfnamefont {R.}~\bibnamefont {Ikuta}}, \bibinfo
  {author} {\bibfnamefont {T.}~\bibnamefont {Yamamoto}}, \bibinfo {author}
  {\bibfnamefont {N.}~\bibnamefont {Imoto}},\ and\ \bibinfo {author}
  {\bibfnamefont {K.}~\bibnamefont {Tamaki}},\ }\bibfield  {title} {\bibinfo
  {title} {Security of quantum key distribution with light sources that are not
  independently and identically distributed},\ }\href
  {https://doi.org/10.1103/PhysRevA.93.042325} {\bibfield  {journal} {\bibinfo
  {journal} {Phys. Rev. A}\ }\textbf {\bibinfo {volume} {93}},\ \bibinfo
  {pages} {042325} (\bibinfo {year} {2016})}\BibitemShut {NoStop}%
\bibitem [{\citenamefont {Mizutani}\ \emph {et~al.}(2019)\citenamefont
  {Mizutani}, \citenamefont {Kato}, \citenamefont {Azuma}, \citenamefont
  {Curty}, \citenamefont {Ikuta}, \citenamefont {Yamamoto}, \citenamefont
  {Imoto}, \citenamefont {Lo},\ and\ \citenamefont {Tamaki}}]{Mizutani2019}%
  \BibitemOpen
  \bibfield  {author} {\bibinfo {author} {\bibfnamefont {A.}~\bibnamefont
  {Mizutani}}, \bibinfo {author} {\bibfnamefont {G.}~\bibnamefont {Kato}},
  \bibinfo {author} {\bibfnamefont {K.}~\bibnamefont {Azuma}}, \bibinfo
  {author} {\bibfnamefont {M.}~\bibnamefont {Curty}}, \bibinfo {author}
  {\bibfnamefont {R.}~\bibnamefont {Ikuta}}, \bibinfo {author} {\bibfnamefont
  {T.}~\bibnamefont {Yamamoto}}, \bibinfo {author} {\bibfnamefont
  {N.}~\bibnamefont {Imoto}}, \bibinfo {author} {\bibfnamefont {H.-K.}\
  \bibnamefont {Lo}},\ and\ \bibinfo {author} {\bibfnamefont {K.}~\bibnamefont
  {Tamaki}},\ }\bibfield  {title} {\bibinfo {title} {Quantum key distribution
  with setting-choice-independently correlated light sources},\ }\href
  {https://doi.org/10.1038/s41534-018-0122-y} {\bibfield  {journal} {\bibinfo
  {journal} {npj Quantum Information}\ }\textbf {\bibinfo {volume} {5}},\
  \bibinfo {pages} {8} (\bibinfo {year} {2019})}\BibitemShut {NoStop}%
\bibitem [{\citenamefont {Yoshino}\ \emph {et~al.}(2018)\citenamefont
  {Yoshino}, \citenamefont {Fujiwara}, \citenamefont {Nakata}, \citenamefont
  {Sumiya}, \citenamefont {Sasaki}, \citenamefont {Takeoka}, \citenamefont
  {Sasaki}, \citenamefont {Tajima}, \citenamefont {Koashi},\ and\ \citenamefont
  {Tomita}}]{Yoshino2018}%
  \BibitemOpen
  \bibfield  {author} {\bibinfo {author} {\bibfnamefont {K.-i.}\ \bibnamefont
  {Yoshino}}, \bibinfo {author} {\bibfnamefont {M.}~\bibnamefont {Fujiwara}},
  \bibinfo {author} {\bibfnamefont {K.}~\bibnamefont {Nakata}}, \bibinfo
  {author} {\bibfnamefont {T.}~\bibnamefont {Sumiya}}, \bibinfo {author}
  {\bibfnamefont {T.}~\bibnamefont {Sasaki}}, \bibinfo {author} {\bibfnamefont
  {M.}~\bibnamefont {Takeoka}}, \bibinfo {author} {\bibfnamefont
  {M.}~\bibnamefont {Sasaki}}, \bibinfo {author} {\bibfnamefont
  {A.}~\bibnamefont {Tajima}}, \bibinfo {author} {\bibfnamefont
  {M.}~\bibnamefont {Koashi}},\ and\ \bibinfo {author} {\bibfnamefont
  {A.}~\bibnamefont {Tomita}},\ }\bibfield  {title} {\bibinfo {title} {Quantum
  key distribution with an efficient countermeasure against correlated
  intensity fluctuations in optical pulses},\ }\href
  {https://doi.org/10.1038/s41534-017-0057-8} {\bibfield  {journal} {\bibinfo
  {journal} {npj Quantum Information}\ }\textbf {\bibinfo {volume} {4}},\
  \bibinfo {pages} {8} (\bibinfo {year} {2018})}\BibitemShut {NoStop}%
\bibitem [{\citenamefont {Zapatero}\ \emph {et~al.}(2021)\citenamefont
  {Zapatero}, \citenamefont {Navarrete}, \citenamefont {Tamaki},\ and\
  \citenamefont {Curty}}]{zapatero2021security}%
  \BibitemOpen
  \bibfield  {author} {\bibinfo {author} {\bibfnamefont {V.}~\bibnamefont
  {Zapatero}}, \bibinfo {author} {\bibfnamefont {{\'A}.}~\bibnamefont
  {Navarrete}}, \bibinfo {author} {\bibfnamefont {K.}~\bibnamefont {Tamaki}},\
  and\ \bibinfo {author} {\bibfnamefont {M.}~\bibnamefont {Curty}},\ }\bibfield
   {title} {\bibinfo {title} {Security of quantum key distribution with
  intensity correlations},\ }\href
  {https://quantum-journal.org/papers/q-2021-12-07-602/#} {\bibfield  {journal}
  {\bibinfo  {journal} {Quantum}\ }\textbf {\bibinfo {volume} {5}},\ \bibinfo
  {pages} {602} (\bibinfo {year} {2021})}\BibitemShut {NoStop}%
\bibitem [{\citenamefont {Sixto}\ \emph {et~al.}(2022)\citenamefont {Sixto},
  \citenamefont {Zapatero},\ and\ \citenamefont
  {Curty}}]{PhysRevApplied.18.044069}%
  \BibitemOpen
  \bibfield  {author} {\bibinfo {author} {\bibfnamefont {X.}~\bibnamefont
  {Sixto}}, \bibinfo {author} {\bibfnamefont {V.}~\bibnamefont {Zapatero}},\
  and\ \bibinfo {author} {\bibfnamefont {M.}~\bibnamefont {Curty}},\ }\bibfield
   {title} {\bibinfo {title} {Security of decoy-state quantum key distribution
  with correlated intensity fluctuations},\ }\href
  {https://doi.org/10.1103/PhysRevApplied.18.044069} {\bibfield  {journal}
  {\bibinfo  {journal} {Phys. Rev. Appl.}\ }\textbf {\bibinfo {volume} {18}},\
  \bibinfo {pages} {044069} (\bibinfo {year} {2022})}\BibitemShut {NoStop}%
\bibitem [{\citenamefont {Pereira}\ \emph {et~al.}(2023)\citenamefont
  {Pereira}, \citenamefont {Curr\'as-Lorenzo}, \citenamefont {Navarrete},
  \citenamefont {Mizutani}, \citenamefont {Kato}, \citenamefont {Curty},\ and\
  \citenamefont {Tamaki}}]{PhysRevResearch.5.023065}%
  \BibitemOpen
  \bibfield  {author} {\bibinfo {author} {\bibfnamefont {M.}~\bibnamefont
  {Pereira}}, \bibinfo {author} {\bibfnamefont {G.}~\bibnamefont
  {Curr\'as-Lorenzo}}, \bibinfo {author} {\bibfnamefont {A.}~\bibnamefont
  {Navarrete}}, \bibinfo {author} {\bibfnamefont {A.}~\bibnamefont {Mizutani}},
  \bibinfo {author} {\bibfnamefont {G.}~\bibnamefont {Kato}}, \bibinfo {author}
  {\bibfnamefont {M.}~\bibnamefont {Curty}},\ and\ \bibinfo {author}
  {\bibfnamefont {K.}~\bibnamefont {Tamaki}},\ }\bibfield  {title} {\bibinfo
  {title} {Modified bb84 quantum key distribution protocol robust to source
  imperfections},\ }\href {https://doi.org/10.1103/PhysRevResearch.5.023065}
  {\bibfield  {journal} {\bibinfo  {journal} {Phys. Rev. Res.}\ }\textbf
  {\bibinfo {volume} {5}},\ \bibinfo {pages} {023065} (\bibinfo {year}
  {2023})}\BibitemShut {NoStop}%
\bibitem [{\citenamefont {Currás-Lorenzo}\ \emph {et~al.}(2023)\citenamefont
  {Currás-Lorenzo}, \citenamefont {Nahar}, \citenamefont {Lütkenhaus},
  \citenamefont {Tamaki},\ and\ \citenamefont {Curty}}]{Currs-Lorenzo_2024}%
  \BibitemOpen
  \bibfield  {author} {\bibinfo {author} {\bibfnamefont {G.}~\bibnamefont
  {Currás-Lorenzo}}, \bibinfo {author} {\bibfnamefont {S.}~\bibnamefont
  {Nahar}}, \bibinfo {author} {\bibfnamefont {N.}~\bibnamefont {Lütkenhaus}},
  \bibinfo {author} {\bibfnamefont {K.}~\bibnamefont {Tamaki}},\ and\ \bibinfo
  {author} {\bibfnamefont {M.}~\bibnamefont {Curty}},\ }\bibfield  {title}
  {\bibinfo {title} {Security of quantum key distribution with imperfect phase
  randomisation},\ }\href {https://doi.org/10.1088/2058-9565/ad141c} {\bibfield
   {journal} {\bibinfo  {journal} {Quantum Science and Technology}\ }\textbf
  {\bibinfo {volume} {9}},\ \bibinfo {pages} {015025} (\bibinfo {year}
  {2023})}\BibitemShut {NoStop}%
\bibitem [{\citenamefont {Li}\ \emph {et~al.}(2025)\citenamefont {Li},
  \citenamefont {Lu}, \citenamefont {Wang}, \citenamefont {Zapatero},
  \citenamefont {Curty}, \citenamefont {Wang}, \citenamefont {Yin},
  \citenamefont {Chen}, \citenamefont {He}, \citenamefont {Guo},\ and\
  \citenamefont {Han}}]{li2025quantumkeydistributionovercoming}%
  \BibitemOpen
  \bibfield  {author} {\bibinfo {author} {\bibfnamefont {J.-X.}\ \bibnamefont
  {Li}}, \bibinfo {author} {\bibfnamefont {F.-Y.}\ \bibnamefont {Lu}}, \bibinfo
  {author} {\bibfnamefont {Z.-H.}\ \bibnamefont {Wang}}, \bibinfo {author}
  {\bibfnamefont {V.}~\bibnamefont {Zapatero}}, \bibinfo {author}
  {\bibfnamefont {M.}~\bibnamefont {Curty}}, \bibinfo {author} {\bibfnamefont
  {S.}~\bibnamefont {Wang}}, \bibinfo {author} {\bibfnamefont {Z.-Q.}\
  \bibnamefont {Yin}}, \bibinfo {author} {\bibfnamefont {W.}~\bibnamefont
  {Chen}}, \bibinfo {author} {\bibfnamefont {D.-Y.}\ \bibnamefont {He}},
  \bibinfo {author} {\bibfnamefont {G.-C.}\ \bibnamefont {Guo}},\ and\ \bibinfo
  {author} {\bibfnamefont {Z.-F.}\ \bibnamefont {Han}},\ }\href
  {https://arxiv.org/abs/2501.13482} {\bibinfo {title} {Quantum key
  distribution overcoming practical correlated intensity fluctuations}}
  (\bibinfo {year} {2025}),\ \Eprint {https://arxiv.org/abs/2501.13482}
  {arXiv:2501.13482 [quant-ph]} \BibitemShut {NoStop}%
\bibitem [{\citenamefont {Currás-Lorenzo}\ \emph {et~al.}(2024)\citenamefont
  {Currás-Lorenzo}, \citenamefont {Pereira}, \citenamefont {Kato},
  \citenamefont {Curty},\ and\ \citenamefont
  {Tamaki}}]{currslorenzo2024securityframeworkquantumkey}%
  \BibitemOpen
  \bibfield  {author} {\bibinfo {author} {\bibfnamefont {G.}~\bibnamefont
  {Currás-Lorenzo}}, \bibinfo {author} {\bibfnamefont {M.}~\bibnamefont
  {Pereira}}, \bibinfo {author} {\bibfnamefont {G.}~\bibnamefont {Kato}},
  \bibinfo {author} {\bibfnamefont {M.}~\bibnamefont {Curty}},\ and\ \bibinfo
  {author} {\bibfnamefont {K.}~\bibnamefont {Tamaki}},\ }\href
  {https://arxiv.org/abs/2305.05930} {\bibinfo {title} {A security framework
  for quantum key distribution implementations}} (\bibinfo {year} {2024}),\
  \Eprint {https://arxiv.org/abs/2305.05930} {arXiv:2305.05930 [quant-ph]}
  \BibitemShut {NoStop}%
\bibitem [{\citenamefont {Vitanov}\ \emph {et~al.}(2013)\citenamefont
  {Vitanov}, \citenamefont {Dupuis}, \citenamefont {Tomamichel},\ and\
  \citenamefont {Renner}}]{6408179}%
  \BibitemOpen
  \bibfield  {author} {\bibinfo {author} {\bibfnamefont {A.}~\bibnamefont
  {Vitanov}}, \bibinfo {author} {\bibfnamefont {F.}~\bibnamefont {Dupuis}},
  \bibinfo {author} {\bibfnamefont {M.}~\bibnamefont {Tomamichel}},\ and\
  \bibinfo {author} {\bibfnamefont {R.}~\bibnamefont {Renner}},\ }\bibfield
  {title} {\bibinfo {title} {Chain rules for smooth min- and max-entropies},\
  }\href {https://doi.org/10.1109/TIT.2013.2238656} {\bibfield  {journal}
  {\bibinfo  {journal} {IEEE Transactions on Information Theory}\ }\textbf
  {\bibinfo {volume} {59}},\ \bibinfo {pages} {2603} (\bibinfo {year}
  {2013})}\BibitemShut {NoStop}%
\bibitem [{\citenamefont {Trefilov}\ \emph {et~al.}(2024)\citenamefont
  {Trefilov}, \citenamefont {Sixto}, \citenamefont {Zapatero}, \citenamefont
  {Huang}, \citenamefont {Curty},\ and\ \citenamefont
  {Makarov}}]{trefilov2024intensitycorrelationsdecoystatebb84}%
  \BibitemOpen
  \bibfield  {author} {\bibinfo {author} {\bibfnamefont {D.}~\bibnamefont
  {Trefilov}}, \bibinfo {author} {\bibfnamefont {X.}~\bibnamefont {Sixto}},
  \bibinfo {author} {\bibfnamefont {V.}~\bibnamefont {Zapatero}}, \bibinfo
  {author} {\bibfnamefont {A.}~\bibnamefont {Huang}}, \bibinfo {author}
  {\bibfnamefont {M.}~\bibnamefont {Curty}},\ and\ \bibinfo {author}
  {\bibfnamefont {V.}~\bibnamefont {Makarov}},\ }\href
  {https://arxiv.org/abs/2411.00709} {\bibinfo {title} {Intensity correlations
  in decoy-state bb84 quantum key distribution systems}} (\bibinfo {year}
  {2024}),\ \Eprint {https://arxiv.org/abs/2411.00709} {arXiv:2411.00709
  [quant-ph]} \BibitemShut {NoStop}%
\bibitem [{\citenamefont {Roberts}\ \emph {et~al.}(2018)\citenamefont
  {Roberts}, \citenamefont {Pittaluga}, \citenamefont {Minder}, \citenamefont
  {Lucamarini}, \citenamefont {Dynes}, \citenamefont {Yuan},\ and\
  \citenamefont {Shields}}]{Roberts:18}%
  \BibitemOpen
  \bibfield  {author} {\bibinfo {author} {\bibfnamefont {G.~L.}\ \bibnamefont
  {Roberts}}, \bibinfo {author} {\bibfnamefont {M.}~\bibnamefont {Pittaluga}},
  \bibinfo {author} {\bibfnamefont {M.}~\bibnamefont {Minder}}, \bibinfo
  {author} {\bibfnamefont {M.}~\bibnamefont {Lucamarini}}, \bibinfo {author}
  {\bibfnamefont {J.~F.}\ \bibnamefont {Dynes}}, \bibinfo {author}
  {\bibfnamefont {Z.~L.}\ \bibnamefont {Yuan}},\ and\ \bibinfo {author}
  {\bibfnamefont {A.~J.}\ \bibnamefont {Shields}},\ }\bibfield  {title}
  {\bibinfo {title} {Patterning-effect mitigating intensity modulator for
  secure decoy-state quantum key distribution},\ }\href
  {https://doi.org/10.1364/OL.43.005110} {\bibfield  {journal} {\bibinfo
  {journal} {Opt. Lett.}\ }\textbf {\bibinfo {volume} {43}},\ \bibinfo {pages}
  {5110} (\bibinfo {year} {2018})}\BibitemShut {NoStop}%
\bibitem [{\citenamefont {Lu}\ \emph {et~al.}(2021)\citenamefont {Lu},
  \citenamefont {Lin}, \citenamefont {Wang}, \citenamefont {Fan-Yuan},
  \citenamefont {Ye}, \citenamefont {Wang}, \citenamefont {Yin}, \citenamefont
  {He}, \citenamefont {Chen}, \citenamefont {Guo},\ and\ \citenamefont
  {Han}}]{Lu2021}%
  \BibitemOpen
  \bibfield  {author} {\bibinfo {author} {\bibfnamefont {F.-Y.}\ \bibnamefont
  {Lu}}, \bibinfo {author} {\bibfnamefont {X.}~\bibnamefont {Lin}}, \bibinfo
  {author} {\bibfnamefont {S.}~\bibnamefont {Wang}}, \bibinfo {author}
  {\bibfnamefont {G.-J.}\ \bibnamefont {Fan-Yuan}}, \bibinfo {author}
  {\bibfnamefont {P.}~\bibnamefont {Ye}}, \bibinfo {author} {\bibfnamefont
  {R.}~\bibnamefont {Wang}}, \bibinfo {author} {\bibfnamefont {Z.-Q.}\
  \bibnamefont {Yin}}, \bibinfo {author} {\bibfnamefont {D.-Y.}\ \bibnamefont
  {He}}, \bibinfo {author} {\bibfnamefont {W.}~\bibnamefont {Chen}}, \bibinfo
  {author} {\bibfnamefont {G.-C.}\ \bibnamefont {Guo}},\ and\ \bibinfo {author}
  {\bibfnamefont {Z.-F.}\ \bibnamefont {Han}},\ }\bibfield  {title} {\bibinfo
  {title} {Intensity modulator for secure, stable, and high-performance
  decoy-state quantum key distribution},\ }\href
  {https://doi.org/10.1038/s41534-021-00418-x} {\bibfield  {journal} {\bibinfo
  {journal} {npj Quantum Information}\ }\textbf {\bibinfo {volume} {7}},\
  \bibinfo {pages} {75} (\bibinfo {year} {2021})}\BibitemShut {NoStop}%
\bibitem [{\citenamefont {Lu}\ \emph {et~al.}(2023)\citenamefont {Lu},
  \citenamefont {Wang}, \citenamefont {Wang}, \citenamefont {Yin},
  \citenamefont {Chen}, \citenamefont {Kang}, \citenamefont {He}, \citenamefont
  {Chen}, \citenamefont {Fan-Yuan}, \citenamefont {Guo},\ and\ \citenamefont
  {Han}}]{10050030}%
  \BibitemOpen
  \bibfield  {author} {\bibinfo {author} {\bibfnamefont {F.-Y.}\ \bibnamefont
  {Lu}}, \bibinfo {author} {\bibfnamefont {Z.-H.}\ \bibnamefont {Wang}},
  \bibinfo {author} {\bibfnamefont {S.}~\bibnamefont {Wang}}, \bibinfo {author}
  {\bibfnamefont {Z.-Q.}\ \bibnamefont {Yin}}, \bibinfo {author} {\bibfnamefont
  {J.-L.}\ \bibnamefont {Chen}}, \bibinfo {author} {\bibfnamefont
  {X.}~\bibnamefont {Kang}}, \bibinfo {author} {\bibfnamefont {D.-Y.}\
  \bibnamefont {He}}, \bibinfo {author} {\bibfnamefont {W.}~\bibnamefont
  {Chen}}, \bibinfo {author} {\bibfnamefont {G.-J.}\ \bibnamefont {Fan-Yuan}},
  \bibinfo {author} {\bibfnamefont {G.-C.}\ \bibnamefont {Guo}},\ and\ \bibinfo
  {author} {\bibfnamefont {Z.-F.}\ \bibnamefont {Han}},\ }\bibfield  {title}
  {\bibinfo {title} {Intensity tomography method for secure and
  high-performance quantum key distribution},\ }\href
  {https://doi.org/10.1109/JLT.2023.3247766} {\bibfield  {journal} {\bibinfo
  {journal} {Journal of Lightwave Technology}\ }\textbf {\bibinfo {volume}
  {41}},\ \bibinfo {pages} {4895} (\bibinfo {year} {2023})}\BibitemShut
  {NoStop}%
\bibitem [{\citenamefont {Kang}\ \emph {et~al.}(2023)\citenamefont {Kang},
  \citenamefont {Lu}, \citenamefont {Wang}, \citenamefont {Chen}, \citenamefont
  {Wang}, \citenamefont {Yin}, \citenamefont {He}, \citenamefont {Chen},
  \citenamefont {Fan-Yuan}, \citenamefont {Guo},\ and\ \citenamefont
  {Han}}]{9907821}%
  \BibitemOpen
  \bibfield  {author} {\bibinfo {author} {\bibfnamefont {X.}~\bibnamefont
  {Kang}}, \bibinfo {author} {\bibfnamefont {F.-Y.}\ \bibnamefont {Lu}},
  \bibinfo {author} {\bibfnamefont {S.}~\bibnamefont {Wang}}, \bibinfo {author}
  {\bibfnamefont {J.-L.}\ \bibnamefont {Chen}}, \bibinfo {author}
  {\bibfnamefont {Z.-H.}\ \bibnamefont {Wang}}, \bibinfo {author}
  {\bibfnamefont {Z.-Q.}\ \bibnamefont {Yin}}, \bibinfo {author} {\bibfnamefont
  {D.-Y.}\ \bibnamefont {He}}, \bibinfo {author} {\bibfnamefont
  {W.}~\bibnamefont {Chen}}, \bibinfo {author} {\bibfnamefont {G.-J.}\
  \bibnamefont {Fan-Yuan}}, \bibinfo {author} {\bibfnamefont {G.-C.}\
  \bibnamefont {Guo}},\ and\ \bibinfo {author} {\bibfnamefont {Z.-F.}\
  \bibnamefont {Han}},\ }\bibfield  {title} {\bibinfo {title}
  {Patterning-effect calibration algorithm for secure decoy-state quantum key
  distribution},\ }\href {https://doi.org/10.1109/JLT.2022.3211442} {\bibfield
  {journal} {\bibinfo  {journal} {Journal of Lightwave Technology}\ }\textbf
  {\bibinfo {volume} {41}},\ \bibinfo {pages} {75} (\bibinfo {year}
  {2023})}\BibitemShut {NoStop}%
\bibitem [{\citenamefont {Marcomini}\ \emph {et~al.}(2024)\citenamefont
  {Marcomini}, \citenamefont {Currás-Lorenzo}, \citenamefont {Rusca},
  \citenamefont {Valle}, \citenamefont {Tamaki},\ and\ \citenamefont
  {Curty}}]{marcomini2024characterisinghigherorderphasecorrelations}%
  \BibitemOpen
  \bibfield  {author} {\bibinfo {author} {\bibfnamefont {A.}~\bibnamefont
  {Marcomini}}, \bibinfo {author} {\bibfnamefont {G.}~\bibnamefont
  {Currás-Lorenzo}}, \bibinfo {author} {\bibfnamefont {D.}~\bibnamefont
  {Rusca}}, \bibinfo {author} {\bibfnamefont {A.}~\bibnamefont {Valle}},
  \bibinfo {author} {\bibfnamefont {K.}~\bibnamefont {Tamaki}},\ and\ \bibinfo
  {author} {\bibfnamefont {M.}~\bibnamefont {Curty}},\ }\href
  {https://arxiv.org/abs/2412.03738} {\bibinfo {title} {Characterising
  higher-order phase correlations in gain-switched laser sources with
  application to quantum key distribution}} (\bibinfo {year} {2024}),\ \Eprint
  {https://arxiv.org/abs/2412.03738} {arXiv:2412.03738 [quant-ph]} \BibitemShut
  {NoStop}%
\bibitem [{\citenamefont {Tomamichel}\ \emph
  {et~al.}(2012{\natexlab{a}})\citenamefont {Tomamichel}, \citenamefont {Lim},
  \citenamefont {Gisin},\ and\ \citenamefont {Renner}}]{Tomamichel2012}%
  \BibitemOpen
  \bibfield  {author} {\bibinfo {author} {\bibfnamefont {M.}~\bibnamefont
  {Tomamichel}}, \bibinfo {author} {\bibfnamefont {C.~C.~W.}\ \bibnamefont
  {Lim}}, \bibinfo {author} {\bibfnamefont {N.}~\bibnamefont {Gisin}},\ and\
  \bibinfo {author} {\bibfnamefont {R.}~\bibnamefont {Renner}},\ }\bibfield
  {title} {\bibinfo {title} {Tight finite-key analysis for quantum
  cryptography},\ }\href {https://doi.org/10.1038/ncomms1631} {\bibfield
  {journal} {\bibinfo  {journal} {Nature Communications}\ }\textbf {\bibinfo
  {volume} {3}},\ \bibinfo {pages} {634} (\bibinfo {year}
  {2012}{\natexlab{a}})}\BibitemShut {NoStop}%
\bibitem [{\citenamefont {Shan}\ \emph {et~al.}(2023)\citenamefont {Shan},
  \citenamefont {Yin}, \citenamefont {Wang}, \citenamefont {Chen},
  \citenamefont {He}, \citenamefont {Guo},\ and\ \citenamefont
  {Han}}]{shan2023practicalphasecodingsidechannelsecurequantum}%
  \BibitemOpen
  \bibfield  {author} {\bibinfo {author} {\bibfnamefont {Y.-G.}\ \bibnamefont
  {Shan}}, \bibinfo {author} {\bibfnamefont {Z.-Q.}\ \bibnamefont {Yin}},
  \bibinfo {author} {\bibfnamefont {S.}~\bibnamefont {Wang}}, \bibinfo {author}
  {\bibfnamefont {W.}~\bibnamefont {Chen}}, \bibinfo {author} {\bibfnamefont
  {D.-Y.}\ \bibnamefont {He}}, \bibinfo {author} {\bibfnamefont {G.-C.}\
  \bibnamefont {Guo}},\ and\ \bibinfo {author} {\bibfnamefont {Z.-F.}\
  \bibnamefont {Han}},\ }\href {https://arxiv.org/abs/2305.13861} {\bibinfo
  {title} {Practical phase-coding side-channel-secure quantum key
  distribution}} (\bibinfo {year} {2023}),\ \Eprint
  {https://arxiv.org/abs/2305.13861} {arXiv:2305.13861 [quant-ph]} \BibitemShut
  {NoStop}%
\bibitem [{\citenamefont {Matsuura}\ \emph {et~al.}(2024)\citenamefont
  {Matsuura}, \citenamefont {Yamano}, \citenamefont {Kuramochi}, \citenamefont
  {Sasaki},\ and\ \citenamefont {Koashi}}]{Matsuura2024tightconcentration}%
  \BibitemOpen
  \bibfield  {author} {\bibinfo {author} {\bibfnamefont {T.}~\bibnamefont
  {Matsuura}}, \bibinfo {author} {\bibfnamefont {S.}~\bibnamefont {Yamano}},
  \bibinfo {author} {\bibfnamefont {Y.}~\bibnamefont {Kuramochi}}, \bibinfo
  {author} {\bibfnamefont {T.}~\bibnamefont {Sasaki}},\ and\ \bibinfo {author}
  {\bibfnamefont {M.}~\bibnamefont {Koashi}},\ }\bibfield  {title} {\bibinfo
  {title} {Tight concentration inequalities for quantum adversarial setups
  exploiting permutation symmetry},\ }\href
  {https://doi.org/10.22331/q-2024-11-27-1540} {\bibfield  {journal} {\bibinfo
  {journal} {{Quantum}}\ }\textbf {\bibinfo {volume} {8}},\ \bibinfo {pages}
  {1540} (\bibinfo {year} {2024})}\BibitemShut {NoStop}%
\bibitem [{\citenamefont
  {Chernoff}(1952)}]{fa3a69c5-2345-343d-ae4f-de42969ad827}%
  \BibitemOpen
  \bibfield  {author} {\bibinfo {author} {\bibfnamefont {H.}~\bibnamefont
  {Chernoff}},\ }\bibfield  {title} {\bibinfo {title} {A measure of asymptotic
  efficiency for tests of a hypothesis based on the sum of observations},\
  }\href {http://www.jstor.org/stable/2236576} {\bibfield  {journal} {\bibinfo
  {journal} {The Annals of Mathematical Statistics}\ }\textbf {\bibinfo
  {volume} {23}},\ \bibinfo {pages} {493} (\bibinfo {year} {1952})}\BibitemShut
  {NoStop}%
\bibitem [{\citenamefont {Mitzenmacher}\ and\ \citenamefont
  {Upfal}(2017)}]{mitzenmacher2017probability}%
  \BibitemOpen
  \bibfield  {author} {\bibinfo {author} {\bibfnamefont {M.}~\bibnamefont
  {Mitzenmacher}}\ and\ \bibinfo {author} {\bibfnamefont {E.}~\bibnamefont
  {Upfal}},\ }\href@noop {} {\emph {\bibinfo {title} {Probability and
  computing: Randomization and probabilistic techniques in algorithms and data
  analysis}}}\ (\bibinfo  {publisher} {Cambridge university press},\ \bibinfo
  {year} {2017})\BibitemShut {NoStop}%
\bibitem [{\citenamefont {Nahar}\ \emph {et~al.}(2024)\citenamefont {Nahar},
  \citenamefont {Tupkary}, \citenamefont {Zhao}, \citenamefont {L\"utkenhaus},\
  and\ \citenamefont {Tan}}]{PRXQuantum.5.040315}%
  \BibitemOpen
  \bibfield  {author} {\bibinfo {author} {\bibfnamefont {S.}~\bibnamefont
  {Nahar}}, \bibinfo {author} {\bibfnamefont {D.}~\bibnamefont {Tupkary}},
  \bibinfo {author} {\bibfnamefont {Y.}~\bibnamefont {Zhao}}, \bibinfo {author}
  {\bibfnamefont {N.}~\bibnamefont {L\"utkenhaus}},\ and\ \bibinfo {author}
  {\bibfnamefont {E.~Y.-Z.}\ \bibnamefont {Tan}},\ }\bibfield  {title}
  {\bibinfo {title} {Postselection technique for optical quantum key
  distribution with improved de finetti reductions},\ }\href
  {https://doi.org/10.1103/PRXQuantum.5.040315} {\bibfield  {journal} {\bibinfo
   {journal} {PRX Quantum}\ }\textbf {\bibinfo {volume} {5}},\ \bibinfo {pages}
  {040315} (\bibinfo {year} {2024})}\BibitemShut {NoStop}%
\bibitem [{\citenamefont {Tomamichel}\ \emph {et~al.}(2011)\citenamefont
  {Tomamichel}, \citenamefont {Schaffner}, \citenamefont {Smith},\ and\
  \citenamefont {Renner}}]{5961850}%
  \BibitemOpen
  \bibfield  {author} {\bibinfo {author} {\bibfnamefont {M.}~\bibnamefont
  {Tomamichel}}, \bibinfo {author} {\bibfnamefont {C.}~\bibnamefont
  {Schaffner}}, \bibinfo {author} {\bibfnamefont {A.}~\bibnamefont {Smith}},\
  and\ \bibinfo {author} {\bibfnamefont {R.}~\bibnamefont {Renner}},\
  }\bibfield  {title} {\bibinfo {title} {Leftover hashing against quantum side
  information},\ }\href {https://doi.org/10.1109/TIT.2011.2158473} {\bibfield
  {journal} {\bibinfo  {journal} {IEEE Transactions on Information Theory}\
  }\textbf {\bibinfo {volume} {57}},\ \bibinfo {pages} {5524} (\bibinfo {year}
  {2011})}\BibitemShut {NoStop}%
\bibitem [{\citenamefont {Tomamichel}\ \emph
  {et~al.}(2012{\natexlab{b}})\citenamefont {Tomamichel}, \citenamefont {Lim},
  \citenamefont {Gisin},\ and\ \citenamefont {Renner}}]{tomamichel2012tight}%
  \BibitemOpen
  \bibfield  {author} {\bibinfo {author} {\bibfnamefont {M.}~\bibnamefont
  {Tomamichel}}, \bibinfo {author} {\bibfnamefont {C.~C.~W.}\ \bibnamefont
  {Lim}}, \bibinfo {author} {\bibfnamefont {N.}~\bibnamefont {Gisin}},\ and\
  \bibinfo {author} {\bibfnamefont {R.}~\bibnamefont {Renner}},\ }\bibfield
  {title} {\bibinfo {title} {Tight finite-key analysis for quantum
  cryptography},\ }\href {https://doi.org/10.1038/ncomms1631.} {\bibfield
  {journal} {\bibinfo  {journal} {Nature communications}\ }\textbf {\bibinfo
  {volume} {3}},\ \bibinfo {pages} {634} (\bibinfo {year}
  {2012}{\natexlab{b}})}\BibitemShut {NoStop}%
\bibitem [{\citenamefont {Xie}\ \emph {et~al.}(2019)\citenamefont {Xie},
  \citenamefont {Li}, \citenamefont {Jiang}, \citenamefont {Cai}, \citenamefont
  {Yin}, \citenamefont {Ren}, \citenamefont {Wang}, \citenamefont {Liao},\ and\
  \citenamefont {Peng}}]{Xie:19}%
  \BibitemOpen
  \bibfield  {author} {\bibinfo {author} {\bibfnamefont {H.-B.}\ \bibnamefont
  {Xie}}, \bibinfo {author} {\bibfnamefont {Y.}~\bibnamefont {Li}}, \bibinfo
  {author} {\bibfnamefont {C.}~\bibnamefont {Jiang}}, \bibinfo {author}
  {\bibfnamefont {W.-Q.}\ \bibnamefont {Cai}}, \bibinfo {author} {\bibfnamefont
  {J.}~\bibnamefont {Yin}}, \bibinfo {author} {\bibfnamefont {J.-G.}\
  \bibnamefont {Ren}}, \bibinfo {author} {\bibfnamefont {X.-B.}\ \bibnamefont
  {Wang}}, \bibinfo {author} {\bibfnamefont {S.-K.}\ \bibnamefont {Liao}},\
  and\ \bibinfo {author} {\bibfnamefont {C.-Z.}\ \bibnamefont {Peng}},\
  }\bibfield  {title} {\bibinfo {title} {Optically injected intensity-stable
  pulse source for secure quantum key distribution},\ }\href
  {https://doi.org/10.1364/OE.27.012231} {\bibfield  {journal} {\bibinfo
  {journal} {Opt. Express}\ }\textbf {\bibinfo {volume} {27}},\ \bibinfo
  {pages} {12231} (\bibinfo {year} {2019})}\BibitemShut {NoStop}%
\bibitem [{\citenamefont {Huang}\ \emph {et~al.}(2023)\citenamefont {Huang},
  \citenamefont {Mizutani}, \citenamefont {Lo}, \citenamefont {Makarov},\ and\
  \citenamefont {Tamaki}}]{PhysRevApplied.19.014048}%
  \BibitemOpen
  \bibfield  {author} {\bibinfo {author} {\bibfnamefont {A.}~\bibnamefont
  {Huang}}, \bibinfo {author} {\bibfnamefont {A.}~\bibnamefont {Mizutani}},
  \bibinfo {author} {\bibfnamefont {H.-K.}\ \bibnamefont {Lo}}, \bibinfo
  {author} {\bibfnamefont {V.}~\bibnamefont {Makarov}},\ and\ \bibinfo {author}
  {\bibfnamefont {K.}~\bibnamefont {Tamaki}},\ }\bibfield  {title} {\bibinfo
  {title} {Characterization of state-preparation uncertainty in quantum key
  distribution},\ }\href {https://doi.org/10.1103/PhysRevApplied.19.014048}
  {\bibfield  {journal} {\bibinfo  {journal} {Phys. Rev. Appl.}\ }\textbf
  {\bibinfo {volume} {19}},\ \bibinfo {pages} {014048} (\bibinfo {year}
  {2023})}\BibitemShut {NoStop}%
\end{thebibliography}%

\hfill

\noindent{\bf Supplementary Materials} 

In this supplementary material, we will prove that for more general values of $ \xi $, there still exists a $\ket{\Phi}_{\rm A}^{\rm equ}$ that ensures the security of $\ket{\Phi}_{\rm A}$.

Recalling the main text, we first have \cref{Proposition:Proposition} under \cref{assu:range}, stated as follows.
\begin{assumption}
    \label{assu:range}
    The correlation is constrained within a maximum range $\xi$.
\end{assumption}
\begin{definition}
    \label{def:new-protocol}
    For any original protocol with a maximum correlation range $\xi$, 
    we define a corresponding new protocol by repeating the original protocol $ \xi + 1 $ times. 
    In the $i$-th repetition, only the rounds whose indices satisfy modulo $ \xi + 1 $ congruent to $ i $ (with the remainder $ \xi + 1 $ interpreted as $0$) 
    are used for key generation, while all other rounds are disclosed. 
    The raw key of original protocol is denoted by $\mathbf{Z}_A$, the raw key of the $i$-th repetition of the new protocol is denoted by $\mathbf{Z}_{A_i}'$ and the the raw key of the whole new protocol is denoted by $\mathbf{Z}_{A}'$.
\end{definition}
\begin{proposition}
    \label{Proposition:Proposition}
    The lower bond of the smooth min-entropy $H^{\epsilon}_{\rm min}(\mathbf{Z}_A|\mathbf{E}')_{\rho}$ of an original protocol with a maximum correlation range $\xi$ can be
    bond by the upper bond of the estimation of phase error rate $\overline{e}^{\rm U}_{\widehat{\epsilon}}$ of the new protocol,
    where the definition of new protocol is in \cref{def:new-protocol}.
    This relation satisfies
    \begin{equation}
        \label{equ:lemma2}
        \begin{aligned}
            H^{\epsilon}_{\rm min}(\mathbf{Z}_A|\mathbf{E}')_{\rho}
            \geq
            n
            \left(
                1 - h\left(
                    \overline{e}^{\rm U}_{\widehat{\epsilon}}
                \right)
            \right)
            -
            \xi f
            -
            \left(
                \xi+1
            \right)
            f'
            ,
        \end{aligned}
    \end{equation}
    where $\epsilon$, $\widehat{\epsilon}$ and $f$ satisfy
    $
    {\widehat{\epsilon}} =
            \left(
                \frac{
                    \epsilon - \xi \frac{1}{2^{f/2}}
                }{2 \xi +1}
                -
                \frac{1}{2^{f'/2}}
            \right)^{{\xi+1}}
    $.
\end{proposition}

Base on \cref{Proposition:Proposition},  we demonstrated in the main text the security of the two-state SNS protocol\cite{PhysRevApplied.12.054034,PhysRevApplied.19.064003,PhysRevResearch.6.013266,shan2024improvedpostselectionsecurityanalysis} 
under correlation, assuming that both \cref{assu:range} and \cref{assu:vacuum} are satisfied. \cref{assu:vacuum} is presented below.
\begin{assumption}
    \label{assu:vacuum}
    For the two-state SNS-QKD protocol, the lower bound of the proportion of vacuum states in each round, under both the send and not send scenarios, is known. 
    Specifically, given the $i$-th round and its preceding $\xi$ rounds, the state sent into the channel during the current round 
    $\rho_{
        \mathbf{r}_{i-\xi}^i
        ,
        \mathbf{a}_{i-\xi}^i
    }^{\rm A(B)}$
    satisfies
    \begin{equation}
        \label{equ:ass2}
        \begin{aligned}
            \mathop{\rm min}\limits_{\mathbf{r}_{i-\xi}^{i-1}}
            \left(
                \mathop{\rm min}\limits_{\mathbf{a}_{i-\xi}^{i}}
                \left(
                \left|
                    \bra{0}
                    \rho_{
                        \mathbf{r}_{i-\xi}^i
                        ,
                        \mathbf{a}_{i-\xi}^i
                    }^{\rm A(B)}
                    \ket{0}
                \right|
                \right)
            \right)
            \geq
            V_{r_i}^{\rm A(B)}
            ,
        \end{aligned}
    \end{equation}
    where
    $r_i \in \{0,1\}$ denotes the encoding setting in $i$-th round,
    $a_i$ denotes the set of ancillas in the system that are potentially related to rounds and can influence the transmitted state, which includes controls over SPF, correlation, side-channel and so on,
    $ V_{r_i}^{\rm A(B)}$ denotes the lower bound of the proportion of vacuum states
and 
the sequence from $i$-th to the $j$-th round for  
$a$ and $r$ are defined as 
$
{a}_{i}^j 
:= a_{j}a_{j-1}\ldots a_{i}
$
and
$
{r}_{i}^j
:= r_{j}r_{j-1}\ldots r_{i}
$
respectively.
\end{assumption}

The main text has already provided the proof for the case where $ \xi = 1 $. 
Here, we present the proof for the general case where $ \xi $ takes arbitrary values. 
In this scenario, the entanglement-based equivalent protocol on Alice's side satisfies
\begin{equation}
    \label{equ:unideal-two-SNS}
    \begin{aligned}
        \ket{\Phi}_{\rm A}= 
        \left[
            \sum_{\mathbf{r}_1^N \mathbf{a}_1^N }
            \left(
                \prod_{i=1}^N \sqrt{
                    p_{r_i}
                    q_{a_i}
                    }
            \right)
            \left(
                \bigotimes_{i=1}^N
                \ket{r_i}_{{A_i}}
                \ket{a_i}_{A_i''}
                \ket{\psi_{
                    \mathbf{r}_{i-\xi}^i
                    ,
                    \mathbf{a}_{i-\xi}^i
                }^{\rm imp}}_{C_i}
            \right)
        \right]
        ,
    \end{aligned}
\end{equation}
where
$q_{a_i}$ is the the probability of selecting $a_i$.
Recall the 
\cref{equ:ass2} of
\cref{assu:vacuum}, it can be rewritten 
as
\begin{equation}
    \label{equ:assu223}
    \begin{aligned}
        \mathop{\rm min}\limits_{\mathbf{r}_{i-\xi}^{i-1}}
        \mathop{\rm min}\limits_{\mathbf{a}_{i-\xi}^{i}}
        \left(
        \bra{0}
                \ket{\psi_{
                    \mathbf{r}_{i-\xi}^i
                    ,
                    \mathbf{a}_{i-\xi}^i
                }^{\rm imp}}
                \bra{\psi_{
                    \mathbf{r}_{i-\xi}^i
                    ,
                    \mathbf{a}_{i-\xi}^i
                }^{\rm imp}}_{C_i}
        \ket{0}
        \right)
        \geq
        V_{r_i}^{\rm A}
        .
    \end{aligned}
\end{equation}

Treat the protocol in Eq.~(\ref{equ:unideal-two-SNS}) as the \textit{original protocol} described in \cref{Proposition:Proposition},
then the \textit{new protocol} without correlation can be expressed as
\begin{equation}
    \label{equ:unideal-two-SNS-new3}
    \begin{aligned}
        \ket{\Phi}_{\rm A}^{\rm new}
        =
        \bigotimes_{i=1}^{\xi+1}
            \left(
                \ket{\Phi}_{{\rm A}^i}
            \right)
        ,
    \end{aligned}
\end{equation}
and
\begin{equation}
    \label{equ:unideal-two-SNS-new5}
    \begin{aligned}
        \ket{\Phi}_{{\rm A}^i}= 
        \mathop{\mathbf{U}}\limits^{\rm shift}
        \left[
            \sum_{\mathbf{m}_{(i-1)N+1}^{iN}}
            \sum_{\mathbf{r'}_{(i-1)N+1}^{iN}}
            \sum_{\mathbf{a}_{(i-1)N+1}^{iN}}
            \left(
                \prod_{j=(i-1)N+1}^{iN} \sqrt{
                    p_{r_j'}
                    q_{a_j}
                    }
            \right)
            \left(
                \bigotimes_{j=(i-1)N+1}^{iN}
                \ket{m_j}_{M_j}
                \ket{r_j'}_{{A_j'}}
                \ket{a_j}_{A_j''}
                \ket{\psi_{
                    \mathbf{r'}_{i-\xi}^i
                    ,
                    \mathbf{a}_{i-\xi}^i
                }^{\rm imp}}_{C_i}
            \right)
        \right]
        ,
    \end{aligned}
\end{equation}
where 
the marker
$
        \ket{m_j}_{M_j}
$
satisfies
\begin{equation}
    \label{equ:unideal-two-SNS-new6}
    \begin{aligned}
        \ket{m_j}_{M_j}
=\left\{
\begin{aligned}
    \ket{1}_{M_j} & \quad j\in\left\{
        (i-1) N + i + n (\xi+1)
        |
        i \in \mathbb{N},
        n \in \mathbb{N}_0,
        i + n (\xi+1)\leq N
    \right\} \\
    \ket{0}_{M_j} &  \quad 
    {\rm otherwise}
\end{aligned}
\right.
    \end{aligned}
    ,
\end{equation}
which mark the \textit{key generation rounds} mentioned in \cref{Proposition:Proposition} with state $1$,
and the 
map 
$\mathop{\mathbf{U}}\limits^{\rm shift}$ represents a process that uses the marker 
$\ket{m_j}_{M_j}$ to classify the local ancilla 
$\ket{r_j'}_{{A_j'}}$ either into the encoded qubit space $A_j$ or to remain in the local ancilla space $A_j'$,
specifically
\begin{equation}
    \label{equ:unideal-two-SNS-new7}
    \begin{aligned}
        \mathop{\mathbf{U}}\limits^{\rm shift}
        \ket{1}_{M_j}
        \ket{r_j}_{{A_j'}}
        =
        \ket{r_j}_{{A_j}}
        ,
        \quad
        \mathop{\mathbf{U}}\limits^{\rm shift}
        \ket{0}_{M_j}
        \ket{r_j'}_{{A_j'}}
        =
        \ket{r_j'}_{{A_j'}}
        .
    \end{aligned}
\end{equation}
Thus, we can observe that there are 
$N$ encoded qubits within the 
$\bigcup_{j=1}^{(\xi+1)N} A_j$ space, while there are 
$\xi N$ ancillas remaining in the 
$\bigcup_{j=1}^{(\xi+1)N} A_j'$ space. This observation can also be intuitively inferred from Fig.~\ref{fig:oriANDnewPROT_new}.
Or in another word, before the map $\mathop{\mathbf{U}}\limits^{\rm shift}$,
the 
$\bigcup_{j=1}^{(\xi+1)N} A_j$ space is void and 
the
$\bigcup_{j=1}^{(\xi+1)N} A_j'$ space has $(\xi+1)N$ non-void subspace $A_j$.
And after the map $\mathop{\mathbf{U}}\limits^{\rm shift}$,
the 
$\bigcup_{j=1}^{(\xi+1)N} A_j$ space has $N$ non-void subspaces and 
the
$\bigcup_{j=1}^{(\xi+1)N} A_j'$ space has $N$ void subspace,
and these subspaces transformed by the map is those in the \textit{key generation rounds}, which satisfies
$j\in\left\{
    (i-1) N + i + n (\xi+1)
    |
    i \in \mathbb{N},
    n \in \mathbb{N}_0,
    i + n (\xi+1)\leq N
\right\}$.
Afterward, we perform a rearrangement, discarding the empty subspaces (as they hold no significance). 
This rearrangement results in
$\bigcup_{j=1}^{(\xi+1)N} A_j
\to
\bigcup_{j=1}^{N} A_j$
and
$\bigcup_{j=1}^{(\xi+1)N} A_j'
\to
\bigcup_{j=1}^{\xi N} A_j'$.
Then we denote the mappings and inverse mappings of the indices under these two rearrangements as $f_{\rm rea}(i)$, $f_{\rm rea}'(i)$
and $f_{\rm rea}^{-1}(i)$, 
$f_{\rm rea}^{\prime -1}(i)$.

Even though the \textit{new protocol} given by Eq.~(\ref{equ:unideal-two-SNS-new3}) is an uncorrelated protocol, it still significantly differs from the common two-state SNS protocol. 
Furthermore, due to the lack of critical parameters, estimating its phase error rate remains challenging. 
To address this, we further constrain the security of the \textit{original protocol} by introducing an \textit{equivalent protocol} to analyze its security.
By reorganizing Eq.~(\ref{equ:unideal-two-SNS-new3}) to emphasize and reorder the encoded qubit space of the \textit{key generation rounds}, we obtain that the entanglement-equivalent protocol of the new protocol satisfies
\begin{equation}
    \label{equ:unideal-two-SNS-new-new}
    \begin{aligned}
        \ket{\Phi}_{\rm A}^{\rm new}
        =
        &
        \Bigg[
            \sum_{\mathbf{r'}_{1}^{\xi N}}
            \sum_{\mathbf{a}_{1}^{(1+\xi)N}}
            \left(
                \prod_{i=1}^{\xi N} \sqrt{
                    p_{r_i'}
                }
                \prod_{j=1}^{(1+\xi)N} \sqrt{
                    q_{a_j}
                }
            \right)
            \left(
                \bigotimes_{i=1}^{\xi N}
                \ket{r_i'}_{{A_i'}}
                \bigotimes_{j=1}^{(1+\xi) N}
                \ket{a_j}_{{A_j''}}
            \right)
        \\
        &
            \otimes
            \left[
                \sum_{\mathbf{r}_1^N}
                \left(
                    \prod_{i=1}^N \sqrt{p_{r_i}}
                \right)
                \left(
                    \bigotimes_{i=1}^N
                    \ket{r_i}_{{A_i}}
                    \ket{
                        \psi_{
                            r_i,
                            \mathbf{r'}(i,\xi)
                            ,
                            \mathbf{a}(i,\xi)
                        }^{\rm imp'}
                    }_{C_i'}
                \right)
            \right]
        \Bigg]
        ,
    \end{aligned}
\end{equation}
where 
$
\mathbf{r'}(i,\xi)
=
            \mathbf{r'}_{
                f_{\rm rea}'(
                    f_{\rm rea}^{-1}(i)
                    - \xi
                )
            }^{
                f_{\rm rea}'(
                    f_{\rm rea}^{-1}(i)
                    +\xi
                )
            }
$
denotes the local ancilla in the previous 
$\xi$ \textit{physical rounds} and the following 
$\xi$ \textit{physical rounds} of the 
$i$-th \textit{key generation round},
$
\mathbf{a}(i,\xi)
=
            \mathbf{a}_{
                    f_{\rm rea}^{-1}(i)
                    - \xi
            }^{
                    f_{\rm rea}^{-1}(i)
                    +\xi
            }
$
denotes the system ancilla of the 
up mentioned \textit{physical rounds} and the 
$i$-th \textit{key generation round}, 
and
$
        \ket{
            \psi_{
                r_i,
                \mathbf{r'}(i,\xi),
                \mathbf{a}(i,\xi)
            }^{\rm imp'}
        }_{C_i'}
$
denote the state send into the channel in the $i$-th \textit{key generation round} and the following $\xi$ \textit{physical rounds}, satisfies
\begin{equation}
    \label{equ:unideal-two-SNS-new-new2}
    \begin{aligned}
        \ket{
            \psi_{
                r_i,
                \mathbf{r'}(i,\xi),
                \mathbf{a}(i,\xi)
            }^{\rm imp'}
        }_{C_i'}
        =
        &
        \ket{\psi_{
            r_{f_{\rm rea}^{-1}(i)}
            \mathbf{r'}_{
                f_{\rm rea}'(
                    f_{\rm rea}^{-1}(i)
                    - \xi
                )
            }^{
                f_{\rm rea}'(
                    f_{\rm rea}^{-1}(i)
                    -1
                )
            }
            ,
            \mathbf{a}_{
                    f_{\rm rea}^{-1}(i)
                    - \xi
            }^{
                    f_{\rm rea}^{-1}(i)
            }
        }^{\rm imp}}_{C_{f_{\rm rea}^{-1}(i)}}     
        \\
        &
        \bigotimes_{j=1}^{\xi}
        \ket{\psi_{
            \mathbf{r'}_{
                f_{\rm rea}'(
                    f_{\rm rea}^{-1}(i)
                    +1
                )
            }^{
                f_{\rm rea}'(
                    f_{\rm rea}^{-1}(i)
                    +j
                )
            }
            r_{f_{\rm rea}^{-1}(i)}
            \mathbf{r'}_{
                f_{\rm rea}'(
                    f_{\rm rea}^{-1}(i)
                    - \xi +j
                )
            }^{
                f_{\rm rea}'(
                    f_{\rm rea}^{-1}(i)
                    -1
                )
            }
            ,
            \mathbf{a}_{
                    f_{\rm rea}^{-1}(i)
                    - \xi +j
            }^{
                    f_{\rm rea}^{-1}(i)
                    +j
            }
        }^{\rm imp}}_{C_{f_{\rm rea}^{-1}(i)}+j}     
        ,
    \end{aligned}
\end{equation}
and where $\ket{\psi_{
    \mathbf{r}_{i-\xi}^i
    ,
    \mathbf{a}_{i-\xi}^i
}^{\rm imp}}_{C_i}$
is denoted in Eq.~(\ref{equ:unideal-two-SNS}).

As discussed in \cref{Proposition:Proposition}, 
the new protocol reveals all \textit{physical rounds} except the \textit{key generation rounds}. 
Furthermore, since the security of the \textit{original protocol} is constrained by that of the \textit{new protocol}, we can relax the assumptions on the \textit{new protocol}. 
Therefore, we further disclose the ancilla in the space 
$A_i''$ for all \textit{physical rounds} and assume that Alice sends additional quantum states into the channel. Thus, for any additional state 
$\ket{
    \psi_{
        r_i,
        \mathbf{r'}(i,\xi)
        ,
        \mathbf{a}(i,\xi)
    }^{\rm add}
}_{C_i''}$ sent into the channel, the security of protocol $\ket{\Phi}_{\rm A}^{\rm new}$ in Eq.~(\ref{equ:unideal-two-SNS-new-new}) can be guaranteed by protocol $\ket{\Phi}_{\rm A}^{{\rm new}_2}$, satisfying
\begin{equation}
    \label{equ:unideal-two-SNS-new-new-2new}
    \begin{aligned}
        \ket{\Phi}_{\rm A}^{{\rm new}_2}
        =
        &
        \Bigg[
            \sum_{\mathbf{r'}_{1}^{\xi N}}
            \sum_{\mathbf{a}_{1}^{(1+\xi)N}}
            \left(
                \prod_{i=1}^{\xi N} \sqrt{
                    p_{r_i'}
                }
                \prod_{j=1}^{(1+\xi)N} \sqrt{
                    q_{a_j}
                }
            \right)
            \left(
                \bigotimes_{i=1}^{\xi N}
                \ket{r_i'}_{{A_i'}}
                \bigotimes_{j=1}^{(1+\xi) N}
                \ket{a_j}_{{A_j''}}
            \right)
        \\
        &
            \otimes
            \left[
                \sum_{\mathbf{r}_1^N}
                \left(
                    \prod_{i=1}^N \sqrt{p_{r_i}}
                \right)
                \left(
                    \bigotimes_{i=1}^N
                    \ket{r_i}_{{A_i}}
                    \ket{
                        \psi_{
                            r_i,
                            \mathbf{r'}(i,\xi)
                            ,
                            \mathbf{a}(i,\xi)
                        }^{\rm imp'}
                    }_{C_i'}
                    \ket{
                        \psi_{
                            r_i,
                            \mathbf{r'}(i,\xi)
                            ,
                            \mathbf{a}(i,\xi)
                        }^{\rm add}
                    }_{C_i''}
                \right)
            \right]
        \Bigg]
        .
    \end{aligned}
\end{equation}
In protocol $\ket{\Phi}_{\rm A}^{{\rm new}_2}$, 
we may isolate the terms related to the $i$-th coding ancilla choice
$r_i$ and denote it as 
\begin{equation}
    \label{equ:iso-ri-define}
    \begin{aligned}
        \ket{\Phi}_{{\rm A,iso\,}r_i}^{{\rm new}_2,i}
        =
        &
        \Bigg[
            \sum_{\mathbf{r'}(i,\xi)}
            \sum_{\mathbf{a}(i,\xi)}
            \left(
                \prod_{
                    j=
                    f_{\rm rea}'(
                        f_{\rm rea}^{-1}(i)
                        - \xi
                    )
                }^{
                    j=
                    f_{\rm rea}'(
                        f_{\rm rea}^{-1}(i)
                        + \xi
                    )
                }
                \sqrt{
                    p_{r_j'}
                }
                \prod_{
                    k=
                        f_{\rm rea}^{-1}(i)
                        - \xi
                }^{
                    k=
                        f_{\rm rea}^{-1}(i)
                        + \xi
                }
                \sqrt{
                    p_{a_k}
                }
            \right)
            \left(
                \bigotimes_{
                    j=
                    f_{\rm rea}'(
                        f_{\rm rea}^{-1}(i)
                        - \xi
                    )
                }^{
                    j=
                    f_{\rm rea}'(
                        f_{\rm rea}^{-1}(i)
                        + \xi
                    )
                }
                \ket{r_j'}_{{A_j'}}
                \bigotimes_{
                    k=
                        f_{\rm rea}^{-1}(i)
                        - \xi
                }^{
                    k=
                        f_{\rm rea}^{-1}(i)
                        + \xi
                }
                \ket{a_k}_{{A_k''}}
            \right)
        \\
        &
            \otimes
                    \ket{
                        \psi_{
                            r_i,
                            \mathbf{r'}(i,\xi)
                            ,
                            \mathbf{a}(i,\xi)
                        }^{\rm imp'}
                    }_{C_i'}
                    \ket{
                        \psi_{
                            r_i,
                            \mathbf{r'}(i,\xi)
                            ,
                            \mathbf{a}(i,\xi)
                        }^{\rm add}
                    }_{C_i''}
        \Bigg]
        .  
    \end{aligned}
\end{equation}
According to \cref{equ:unideal-two-SNS-new-new2}, we further define the vacuum state in the 
$C_i'$ space as 
$\ket{0}_{C_i'}$, satisfying
\begin{equation}
    \label{equ:vaccc}
    \begin{aligned}
        \ket{0}_{C_i'} 
        =
        &
        \ket{0}_{C_{f_{\rm rea}^{-1}(i)}}
        \bigotimes_{j=1}^{\xi}
        \ket{0}_{C_{f_{\rm rea}^{-1}(i)}+j}
        .
    \end{aligned}
\end{equation}
Since
$
                    \ket{
                        \psi_{
                            r_i,
                            \mathbf{r'}(i,\xi)
                            ,
                            \mathbf{a}(i,\xi)
                        }^{\rm add}
                    }_{C_i''}
$
is arbitrary, we can transfer some of its phase to
$
                    \ket{
                        \psi_{
                            r_i,
                            \mathbf{r'}(i,\xi)
                            ,
                            \mathbf{a}(i,\xi)
                        }^{\rm imp'}
                    }_{C_i'}
$, 
to form a new state
$
                    \ket{
                        \psi_{
                            r_i,
                            \mathbf{r'}(i,\xi)
                            ,
                            \mathbf{a}(i,\xi)
                        }^{\rm imp''}
                    }_{C_i'}
$
such that 
$
\ket{
    \psi_{
        r_i,
        \mathbf{r'}(i,\xi)
        ,
        \mathbf{a}(i,\xi)
    }^{\rm imp'}
}_{C_i'}                    
                    \ket{
                        \psi_{
                            r_i,
                            \mathbf{r'}(i,\xi)
                            ,
                            \mathbf{a}(i,\xi)
                        }^{\rm add}
                    }_{C_i''}
                    =
                    \ket{
                        \psi_{
                            r_i,
                            \mathbf{r'}(i,\xi)
                            ,
                            \mathbf{a}(i,\xi)
                        }^{\rm imp''}
                    }_{C_i'}                    
                                        \ket{
                                            \psi_{
                                                r_i,
                                                \mathbf{r'}(i,\xi)
                                                ,
                                                \mathbf{a}(i,\xi)
                                            }^{\rm add'}
                                        }_{C_i''}
$,
$
\left|
    \bra{
    \psi_{
        r_i,
        \mathbf{r'}(i,\xi)
        ,
        \mathbf{a}(i,\xi)
    }^{\rm imp'}
}   
\ket{
    \psi_{
        r_i,
        \mathbf{r'}(i,\xi)
        ,
        \mathbf{a}(i,\xi)
    }^{\rm imp''}
}_{C_i'}   
\right|
=
\left|
                    \bra{
                        \psi_{
                            r_i,
                            \mathbf{r'}(i,\xi)
                            ,
                            \mathbf{a}(i,\xi)
                        }^{\rm add}
                    }
                    \ket{
                        \psi_{
                            r_i,
                            \mathbf{r'}(i,\xi)
                            ,
                            \mathbf{a}(i,\xi)
                        }^{\rm add'}
                    }_{C_i''}
\right|
=
1
$,
and
$
\bra{
    0
}   
\ket{
    \psi_{
        r_i,
        \mathbf{r'}(i,\xi)
        ,
        \mathbf{a}(i,\xi)
    }^{\rm imp''}
}_{C_i'}   
$
is real and positive.
It is also worth noting that after this step, 
$
                    \ket{
                        \psi_{
                            r_i,
                            \mathbf{r'}(i,\xi)
                            ,
                            \mathbf{a}(i,\xi)
                        }^{\rm add'}
                    }_{C_i''}
$
still retains its arbitrariness, both in terms of phase and magnitude.
Subsequently, for the sake of computational simplicity, we further define two intermediate states,
$
    \ket{\Phi}_{{\rm A,iso\,}r_i}^{{\rm new},i}
$
and
$
    \ket{\Phi}_{{\rm A,vac}}^{{\rm new},i}
$,
satisfy
\begin{equation}
    \label{equ:vacandnew-ri-define}
    \begin{aligned}
        \ket{\Phi}_{{\rm A,iso\,}r_i}^{{\rm new},i}
        =
        &
        \Bigg[
            \sum_{\mathbf{r'}(i,\xi)}
            \sum_{\mathbf{a}(i,\xi)}
            \left(
                \prod_{
                    j=
                    f_{\rm rea}'(
                        f_{\rm rea}^{-1}(i)
                        - \xi
                    )
                }^{
                    j=
                    f_{\rm rea}'(
                        f_{\rm rea}^{-1}(i)
                        + \xi
                    )
                }
                \sqrt{
                    p_{r_j'}
                }
                \prod_{
                    k=
                        f_{\rm rea}^{-1}(i)
                        - \xi
                }^{
                    k=
                        f_{\rm rea}^{-1}(i)
                        + \xi
                }
                \sqrt{
                    p_{a_k}
                }
            \right)
            \left(
                \bigotimes_{
                    j=
                    f_{\rm rea}'(
                        f_{\rm rea}^{-1}(i)
                        - \xi
                    )
                }^{
                    j=
                    f_{\rm rea}'(
                        f_{\rm rea}^{-1}(i)
                        + \xi
                    )
                }
                \ket{r_j'}_{{A_j'}}
                \bigotimes_{
                    k=
                        f_{\rm rea}^{-1}(i)
                        - \xi
                }^{
                    k=
                        f_{\rm rea}^{-1}(i)
                        + \xi
                }
                \ket{a_k}_{{A_k''}}
            \right)
        \\
        &
            \otimes
                    \ket{
                        \psi_{
                            r_i,
                            \mathbf{r'}(i,\xi)
                            ,
                            \mathbf{a}(i,\xi)
                        }^{\rm imp''}
                    }_{C_i'}
        \Bigg]
        ,
        \\
        \ket{\Phi}_{{\rm A,vac}}^{{\rm new},i}
        =
        &
        \left[
            \sum_{\mathbf{r'}(i,\xi)}
            \sum_{\mathbf{a}(i,\xi)}
            \left(
                \prod_{
                    j=
                    f_{\rm rea}'(
                        f_{\rm rea}^{-1}(i)
                        - \xi
                    )
                }^{
                    j=
                    f_{\rm rea}'(
                        f_{\rm rea}^{-1}(i)
                        + \xi
                    )
                }
                \sqrt{
                    p_{r_j'}
                }
                \prod_{
                    k=
                        f_{\rm rea}^{-1}(i)
                        - \xi
                }^{
                    k=
                        f_{\rm rea}^{-1}(i)
                        + \xi
                }
                \sqrt{
                    p_{a_k}
                }
            \right)
            \left(
                \bigotimes_{
                    j=
                    f_{\rm rea}'(
                        f_{\rm rea}^{-1}(i)
                        - \xi
                    )
                }^{
                    j=
                    f_{\rm rea}'(
                        f_{\rm rea}^{-1}(i)
                        + \xi
                    )
                }
                \ket{r_j'}_{{A_j'}}
                \bigotimes_{
                    k=
                        f_{\rm rea}^{-1}(i)
                        - \xi
                }^{
                    k=
                        f_{\rm rea}^{-1}(i)
                        + \xi
                }
                \ket{a_k}_{{A_k''}}
            \right)
            \ket{0}_{C_i'}
        \right]
        .
    \end{aligned}
\end{equation}
Thus, due to \cref{assu:vacuum}, which further leads to \cref{equ:assu223},
the two intermediate states defined in \cref{equ:vacandnew-ri-define} satisfies
\begin{equation}
    \label{equ:inner_of_zhongjian}
    \begin{aligned}
        \bra{\Phi}_{{\rm A,iso\,}r_i}^{{\rm new},i}
        \ket{\Phi}_{{\rm A,vac}}^{{\rm new},i}
        =
        &
    \left|
        \sum_{\mathbf{r'}(i,\xi)}
        \sum_{\mathbf{a}(i,\xi)}
        \left(
            \prod_{
                j=
                f_{\rm rea}'(
                    f_{\rm rea}^{-1}(i)
                    - \xi
                )
            }^{
                j=
                f_{\rm rea}'(
                    f_{\rm rea}^{-1}(i)
                    + \xi
                )
            }
            {
                p_{r_j'}
            }
            \prod_{
                k=
                    f_{\rm rea}^{-1}(i)
                    - \xi
            }^{
                k=
                    f_{\rm rea}^{-1}(i)
                    + \xi
            }
            {
                p_{a_k}
            }
        \right)
        \braket{
            \psi_{
                r_i,
                \mathbf{r'}(i,\xi)
                ,
                \mathbf{a}(i,\xi)
            }^{\rm imp''}
        }{
            0
        }_{C_i'}
    \right|
        \\
        \geq
        &
        \sum_{\mathbf{r'}(i,\xi)}
        \sum_{\mathbf{a}(i,\xi)}
        \left(
            \prod_{
                j=
                f_{\rm rea}'(
                    f_{\rm rea}^{-1}(i)
                    - \xi
                )
            }^{
                j=
                f_{\rm rea}'(
                    f_{\rm rea}^{-1}(i)
                    + \xi
                )
            }
            {
                p_{r_j'}
            }
            \prod_{
                k=
                    f_{\rm rea}^{-1}(i)
                    - \xi
            }^{
                k=
                    f_{\rm rea}^{-1}(i)
                    + \xi
            }
            {
                p_{a_k}
            }
        \right)
        \left(
            \sqrt{
                V_{r_i}^{\rm A}
            }
            \prod_{
                j=
                f_{\rm rea}'(
                    f_{\rm rea}^{-1}(i)
                    - \xi
                )
            }^{
                j=
                f_{\rm rea}'(
                    f_{\rm rea}^{-1}(i)
                    + \xi
                )
            }
            \sqrt{
                V_{r_i'}^{\rm A}
            }
        \right)
        \\
            =
        &
            \sqrt{
                V_{r_i}^{\rm A}
            }
            \left(
                p_0
                \sqrt{
                    V_{0}^{\rm A}
                }
                +
                p_1
                \sqrt{
                    V_{1}^{\rm A}
                }
            \right)^{\xi}
            =:
            \sqrt{
                V_{r_i}^{{\rm A},\xi}
            }
            .
    \end{aligned}
\end{equation}
Because \cref{equ:inner_of_zhongjian} satisfies for both $r_i = 0$ and $1$, we can find that
\begin{equation}
    \label{equ:inner_of_two_new}
    \begin{aligned}
        \left|
            \bra{\Phi}_{{\rm A,iso\,}0}^{{\rm new},i}
            \ket{\Phi}_{{\rm A,iso\,}1}^{{\rm new},i}
        \right|
        \geq
        \sqrt{
            V_{0}^{{\rm A},\xi}
            V_{1}^{{\rm A},\xi}
        }
        -
        \sqrt{
            \left(
                1-V_{0}^{{\rm A},\xi}
            \right)
            \left(
                1-V_{1}^{{\rm A},\xi}
            \right)
        }
        .
    \end{aligned}
\end{equation}
Further, from \cref{equ:iso-ri-define} and \cref{equ:vacandnew-ri-define}, we can calculate that
\begin{equation}
    \label{equ:inner-of-new2}
    \begin{aligned}
        &
        \bra{\Phi}_{{\rm A,iso\,}0}^{{\rm new}_2,i}
        \ket{\Phi}_{{\rm A,iso\,}1}^{{\rm new}_2,i}
        \\
        =
        &
        \sum_{\mathbf{r'}(i,\xi)}
        \sum_{\mathbf{a}(i,\xi)}
        \left(
            \prod_{
                j=
                f_{\rm rea}'(
                    f_{\rm rea}^{-1}(i)
                    - \xi
                )
            }^{
                j=
                f_{\rm rea}'(
                    f_{\rm rea}^{-1}(i)
                    + \xi
                )
            }
            {
                p_{r_j'}
            }
            \prod_{
                k=
                    f_{\rm rea}^{-1}(i)
                    - \xi
            }^{
                k=
                    f_{\rm rea}^{-1}(i)
                    + \xi
            }
            {
                p_{a_k}
            }
        \right)
        \braket{
            \psi_{
                0,
                \mathbf{r'}(i,\xi)
                ,
                \mathbf{a}(i,\xi)
            }^{\rm imp''}
        }{
            \psi_{
                1,
                \mathbf{r'}(i,\xi)
                ,
                \mathbf{a}(i,\xi)
            }^{\rm imp''}
        }_{C_i'}
        \braket{
            \psi_{
                0,
                \mathbf{r'}(i,\xi)
                ,
                \mathbf{a}(i,\xi)
            }^{\rm add'}
        }{
            \psi_{
                1,
                \mathbf{r'}(i,\xi)
                ,
                \mathbf{a}(i,\xi)
            }^{\rm add'}
        }_{C_i''}
    ,
    \end{aligned}
\end{equation}
and
\begin{equation}
    \label{equ:inner-of-new}
    \begin{aligned}
        &
        \bra{\Phi}_{{\rm A,iso\,}0}^{{\rm new},i}
        \ket{\Phi}_{{\rm A,iso\,}1}^{{\rm new},i}
        =
        &
        \sum_{\mathbf{r'}(i,\xi)}
        \sum_{\mathbf{a}(i,\xi)}
        \left(
            \prod_{
                j=
                f_{\rm rea}'(
                    f_{\rm rea}^{-1}(i)
                    - \xi
                )
            }^{
                j=
                f_{\rm rea}'(
                    f_{\rm rea}^{-1}(i)
                    + \xi
                )
            }
            {
                p_{r_j'}
            }
            \prod_{
                k=
                    f_{\rm rea}^{-1}(i)
                    - \xi
            }^{
                k=
                    f_{\rm rea}^{-1}(i)
                    + \xi
            }
            {
                p_{a_k}
            }
        \right)
        \braket{
            \psi_{
                0,
                \mathbf{r'}(i,\xi)
                ,
                \mathbf{a}(i,\xi)
            }^{\rm imp''}
        }{
            \psi_{
                1,
                \mathbf{r'}(i,\xi)
                ,
                \mathbf{a}(i,\xi)
            }^{\rm imp''}
        }_{C_i'}
    .
    \end{aligned}
\end{equation}
Since 
$
\left|
\braket{
    \psi_{
        0,
        \mathbf{r'}(i,\xi)
        ,
        \mathbf{a}(i,\xi)
    }^{\rm add'}
}{
    \psi_{
        1,
        \mathbf{r'}(i,\xi)
        ,
        \mathbf{a}(i,\xi)
    }^{\rm add'}
}_{C_i''}
\right|
\leq
1$,
combine with \cref{equ:inner_of_two_new}
we can select a specific set of
$\ket{
    \psi_{
        r_i,
        \mathbf{r'}(i,\xi)
        ,
        \mathbf{a}(i,\xi)
    }^{\rm add'}
}_{C_i''}$ such that
\begin{equation}
    \label{equ:inner_of_two_new2}
    \begin{aligned}
            \bra{\Phi}_{{\rm A,iso\,}0}^{{\rm new}_2,i}
            \ket{\Phi}_{{\rm A,iso\,}1}^{{\rm new}_2,i}
        =
        \sqrt{
            V_{0}^{{\rm A},\xi}
            V_{1}^{{\rm A},\xi}
        }
        -
        \sqrt{
            \left(
                1-V_{0}^{{\rm A},\xi}
            \right)
            \left(
                1-V_{1}^{{\rm A},\xi}
            \right)
        }
        .
    \end{aligned}
\end{equation}

To reiterate, in protocol $\ket{\Phi}_{\rm A}^{{\rm new}_2}$, all quantum states in spaces other than the encoded qubit space 
$\bigcup_{j=1}^{N} A_j $, including ancillary particles, are sent into the channel. 
Therefore, performing a unitary mapping in the remaining spaces does not affect the security of the protocol. 
Specifically, we construct an \textit{equivalent protocol} $\ket{\Phi}_{\rm A}^{\rm equ}$, satisfies
\begin{equation}
    \label{equ:1unideal-two-SNS-new-new-2}
    \begin{aligned}
        \ket{\Phi}_{\rm A}^{\rm equ}
        =
        &
        \left(
            \prod_{j=1}^{N}
            \mathbf{U}_j
        \right)
        \ket{\Phi}_{\rm A}^{{\rm new}_2}
        \\
        =
        &
        \Bigg[
            \sum_{\mathbf{r'}_{1}^{\xi N}}
            \sum_{\mathbf{a}_{1}^{(1+\xi)N}}
            \left(
                \prod_{i=1}^{\xi N} \sqrt{
                    p_{r_i'}
                }
                \prod_{j=1}^{(1+\xi)N} \sqrt{
                    q_{a_j}
                }
            \right)
            \left(
                \bigotimes_{i=1}^{\xi N}
                \ket{r_i'}_{{A_i'}}
                \bigotimes_{i=1}^{(1+\xi) N}
                \ket{a_i}_{{A_i''}}
            \right)
        \\
        &
            \otimes
            \left[
                \sum_{\mathbf{r}_1^N}
                \left(
                    \prod_{i=1}^N \sqrt{p_{r_i}}
                \right)
                \left(
                    \bigotimes_{i=1}^N
                    \ket{r_i}_{{A_i}}
                \ket{\psi_{r_i}^{\rm equ}}_{C_i'''}
                \right)
            \right]
        \Bigg]
        ,
    \end{aligned}
\end{equation}
where $\mathbf{U}_i$ si an unitary mapping from the space
$
\bigcup_{
    j=
    f_{\rm rea}'(
        f_{\rm rea}^{-1}(i)
        - \xi
    )
}^{
    j=
    f_{\rm rea}'(
        f_{\rm rea}^{-1}(i)
        + \xi
    )
} A_j'
\bigcup_{
    k=
        f_{\rm rea}^{-1}(i)
        - \xi
}^{
    k=
    f_{\rm rea}^{-1}(i)
    + \xi
}
A_k''
\bigcup
{C_i'}
\bigcup
{C_i''}
$
into
\\
$
\bigcup_{
    j=
    f_{\rm rea}'(
        f_{\rm rea}^{-1}(i)
        - \xi
    )
}^{
    j=
    f_{\rm rea}'(
        f_{\rm rea}^{-1}(i)
        + \xi
    )
} A_j'
\bigcup_{
    k=
        f_{\rm rea}^{-1}(i)
        - \xi
}^{
    k=
    f_{\rm rea}^{-1}(i)
    + \xi
}
A_k''
\bigcup
{C_i'''}
$,
satisfies
\begin{equation}
    \label{equ:1unideal-two-SNS-new-new-21}
    \begin{aligned}
        &
        \mathbf{U}_i
        \ket{\Phi}_{{\rm A,iso\,}r_i}^{{\rm new}_2,i}
        \\
        =
        &
        \Bigg[
            \sum_{\mathbf{r'}(i,\xi)}
            \sum_{\mathbf{a}(i,\xi)}
            \left(
                \prod_{
                    j=
                    f_{\rm rea}'(
                        f_{\rm rea}^{-1}(i)
                        - \xi
                    )
                }^{
                    j=
                    f_{\rm rea}'(
                        f_{\rm rea}^{-1}(i)
                        + \xi
                    )
                }
                \sqrt{
                    p_{r_j'}
                }
                \prod_{
                    k=
                        f_{\rm rea}^{-1}(i)
                        - \xi
                }^{
                    k=
                        f_{\rm rea}^{-1}(i)
                        + \xi
                }
                \sqrt{
                    p_{a_k}
                }
            \right)
            \left(
                \bigotimes_{
                    j=
                    f_{\rm rea}'(
                        f_{\rm rea}^{-1}(i)
                        - \xi
                    )
                }^{
                    j=
                    f_{\rm rea}'(
                        f_{\rm rea}^{-1}(i)
                        + \xi
                    )
                }
                \ket{r_j'}_{{A_j'}}
                \bigotimes_{
                    k=
                        f_{\rm rea}^{-1}(i)
                        - \xi
                }^{
                    k=
                        f_{\rm rea}^{-1}(i)
                        + \xi
                }
                \ket{a_k}_{{A_k''}}
            \right)
            \otimes
                \ket{\psi_{r_i}^{\rm equ}}_{C_i'''}
        \Bigg]
        .
    \end{aligned}
\end{equation}
From Eq.~(\ref{equ:1unideal-two-SNS-new-new-21}), we can see that $\mathbf{U}_i$ exists if and only if
\begin{equation}
    \label{equ:1new-two-SNS-asdce}
    \begin{aligned}
        &
        \bra{\Phi}_{{\rm A,iso\,}0}^{{\rm new}_2,i}
        \ket{\Phi}_{{\rm A,iso\,}1}^{{\rm new}_2,i}
        =
        &
        \sum_{\mathbf{r'}(i,\xi)}
        \sum_{\mathbf{a}(i,\xi)}
        \left(
            \prod_{
                j=
                f_{\rm rea}'(
                    f_{\rm rea}^{-1}(i)
                    - \xi
                )
            }^{
                j=
                f_{\rm rea}'(
                    f_{\rm rea}^{-1}(i)
                    + \xi
                )
            }
            {
                p_{r_j'}
            }
            \prod_{
                k=
                    f_{\rm rea}^{-1}(i)
                    - \xi
            }^{
                k=
                    f_{\rm rea}^{-1}(i)
                    + \xi
            }
            {
                p_{a_k}
            }
        \right)
        \braket
        {\psi_{0}^{\rm equ}}
        {\psi_{1}^{\rm equ}}_{C_i'''}
        =
        \braket
        {\psi_{0}^{\rm equ}}
        {\psi_{1}^{\rm equ}}_{C_i'''}
        .
    \end{aligned}
\end{equation}

Without loss of generality, we assume that
\begin{equation}
    \label{equ:equ-two-SNS-phi1}
    \begin{aligned}
        \ket{\psi_{0}^{\rm equ}}_{C_i'''} = \ket{0}, \,
        \ket{\psi_{1}^{\rm equ}}_{C_i'''} = \ket{\mu_{\rm equ}},
    \end{aligned}
\end{equation}
where $\ket{0}$ is the vacuum state and $\ket{\mu_{\rm equ}}$ is the coherent state with an average number of photons equals to $\mu_{\rm equ}$.
Combine 
\cref{equ:inner_of_two_new2,equ:1new-two-SNS-asdce,equ:equ-two-SNS-phi1},
we can calculate that $\mu_{\rm equ}$ satisfies
\begin{equation}
    \label{equ:equ-two-SNS-mu}
    \begin{aligned}
        {\rm e}^{-\mu_{\rm equ}}
        =&
        \left[
            \sqrt{
                V_{0}^{{\rm A},\xi}
                V_{1}^{{\rm A},\xi}
            }
            -
            \sqrt{
                \left(
                    1-V_{0}^{{\rm A},\xi}
                \right)
                \left(
                    1-V_{1}^{{\rm A},\xi}
                \right)
            }
        \right]^2
    .
    \end{aligned}
\end{equation}
Furthermore, observing that in protocol $\ket{\Phi}_{\rm A}^{\rm equ}$ as defined in Eq.~(\ref{equ:1unideal-two-SNS-new-new-2}), 
$r'$ and 
$a$ no longer play any role, we simplify the \textit{equivalent protocol} $\ket{\Phi}_{\rm A}^{\rm equ}$ by removing them. 
The final \textit{equivalent protocol} then satisfies
\begin{equation}
    \label{equ:equal-two-SNS1}
    \begin{aligned}
        \ket{\Phi}_{\rm A}^{\rm equ}= 
        \left[
            \sum_{\mathbf{r}_1^N}
            \left(
                \prod_{i=1}^N \sqrt{p_{r_i}}
            \right)
            \left(
                \bigotimes_{i=1}^N
                \ket{r_i}_{{A_i}}
                \ket{\psi_{r_i}^{\rm equ}}_{C_i'''}
            \right)
        \right]
        .
    \end{aligned}
\end{equation}

The above analysis provides a proof that, for general values of $ \xi $, protocol $\ket{\Phi}_{\rm A}^{\rm equ}$ satisfies \cref{equ:equal-two-SNS1} ensures the security of $\ket{\Phi}_{\rm A}$.

\end{document}